\documentclass[twoside, twocolumn]{article}

\usepackage[margin=2cm]{geometry}
\usepackage{amsmath,amssymb}
\usepackage{graphicx}
\usepackage{booktabs}
\usepackage{multirow}
\usepackage{hyperref}
\usepackage[numbers,sort&compress]{natbib}
\usepackage{xcolor}
\usepackage{subcaption}
\usepackage{array}
\usepackage{tabularx}
\usepackage{float}
\usepackage{enumitem}
\usepackage{placeins}   
\usepackage{stfloats}   

\usepackage{titlesec}
\titleformat{\section}{\large\bfseries}{\thesection.}{0.5em}{}
\titleformat{\subsection}{\normalsize\bfseries}{\thesubsection}{0.5em}{}
\titleformat{\subsubsection}{\normalsize\itshape}{\thesubsubsection}{0.5em}{}

\hypersetup{
    colorlinks=true,
    linkcolor=blue!60!black,
    citecolor=blue!60!black,
    urlcolor=blue!60!black
}

\newcommand{\Rtwo}{R$^{2}$}

\newcommand{\ntrain}{n_{\text{train}}}

\title{Surrogate-Gated Generation and Foundation-Model Embeddings for Bayesian Materials Design}

\author{
    Sk Md Ahnaf Akif Alvi,$^{a,e,*}$ Jan Janssen,$^{d}$ Danny Perez,$^{f}$ Douglas Allaire,$^{b}$ and Raymundo Arr\'{o}yave$^{a,b,c}$\\[6pt]
    \footnotesize $^{a}$Department of Materials Science and Engineering, Texas A\&M University, College Station, TX 77843, USA\\
    \footnotesize $^{b}$J.\ Mike Walker '66 Department of Mechanical Engineering, Texas A\&M University, College Station, TX 77843, USA\\
    \footnotesize $^{c}$Wm Michael Barnes '64 Department of Industrial and Systems Engineering, Texas A\&M University, College Station, TX 77843, USA\\
    \footnotesize $^{d}$Max-Planck-Institute for Sustainable Materials, D\"{u}sseldorf, Germany 40237\\
    \footnotesize $^{e}$Theoretical Division T-1, Los Alamos National Laboratory, Los Alamos, NM 87544, USA\\
    \footnotesize $^{f}$X Computational Physics XCP-AI4ND, Los Alamos National Laboratory, Los Alamos, NM 87544, USA\\[3pt]
    \footnotesize $^{*}$Corresponding author. E-mail: ahnafalvi@tamu.edu
}

\date{}

\begin{document}

\twocolumn[
  \begin{@twocolumnfalse}
    \maketitle
    \begin{abstract}
Closed-loop materials discovery iterates between proposing candidate structures and evaluating their properties, and property evaluation dominates the cost.
  In the generative variant, a learned prior proposes candidate crystals and a property oracle scores them; we ask whether a cheap probabilistic surrogate can triage the generator's output, and what such a surrogate must do well.
  Across three architecturally distinct pretrained diffusion priors (MatterGen, CrystalFlow, ADiT) and two targets (room-temperature heat capacity and bulk modulus), we insert a Gaussian process acquisition gate between structure generation and the oracle in an RL-steered generative workflow.
  The gate matches or exceeds ungated fine-tuning of the generative model while capping oracle calls at a fixed per-cycle budget.
  Budget-matched ablations isolate the mechanism.
  At an identical four-call budget, ranking-based selection outperforms arbitrary selection, confirming that the gain comes from the surrogate's choice; the gate comes within $\sim$9\% of exhaustive oracle spending at roughly one-fifth of the calls.
  A density-functional-theory check of the bulk-modulus discoveries confirms the learned oracle to within 2.5\% on average and the surrogate's ranking of the generated structures at Spearman $\rho = 0.94$.
  A cross-factorial benchmark of surrogate performance spanning mechanical, electronic, and vibrational properties identifies pretrained ORB embeddings with a Gaussian process as the most reliable combination, which we adopt as the building blocks of the proposed workflow.
  The complete pipeline is released as open-source software.
    \end{abstract}
    \vspace{1cm}
  \end{@twocolumnfalse}
]

\section{Introduction}
\label{sec:introduction}

Generative models are poised to dramatically alter the way  crystals with tailored properties are designed.
Diffusion and flow-based models now sample crystal structures directly from a learned distribution over (near-)stable materials manifold, often conditioned on specific target properties~\cite{generative_review_2025}, as demonstrated with CDVAE~\cite{xie2022cdvae}, DiffCSP~\cite{jiao2023diffcsp}, MatterGen~\cite{zeni2025mattergen}, FlowMM~\cite{miller2024flowmm}, CrystalFlow~\cite{luo2025crystalflow}, and the all-atom diffusion transformer ADiT~\cite{joshi2025adit}.
Reinforcement learning sharpens these priors online by reformulating denoising as a Markov decision process. DDPO~\cite{black2023ddpo} and DPOK~\cite{fan2023dpok} first proposed to fine-tune image diffusion against a reward, and MatInvent~\cite{chen2025matinvent} and CrystalFormer-RL~\cite{cao2025crystalformerrl} carry the idea to crystals.
The result is a closed loop: the prior proposes, an oracle scores, and the policy updates toward higher-scoring regions of chemical space.

Whether it is density functional theory (DFT) or a machine-learning surrogate of it, scoring every proposed structure dominates the cost of the loop. A generator that emits tens of candidates per cycle quickly makes oracle calls the binding constraint.
An obvious remedy is to triage: place a cheap surrogate between the generator and the oracle, and spend oracle budget only on candidates most likely to be high performing.
Yet, this gating step is largely absent from closed-loop crystal generation.
Surveys of generative models for materials~\cite{generative_review_2025} and of reinforcement-learning fine-tuning of diffusion~\cite{uehara2024rldiffusion} map the design space without an online, uncertainty-aware acquisition gate. The closest precedent, LCOM~\cite{qi2023lcom}, applies a variational-autoencoder gate to crystal structure prediction, not generative design.
Self-driving laboratories~\cite{burger2020mobile} show that surrogate-guided selection sharply reduces the number of expensive evaluations.

Such a gate can be cheap because a Bayesian-optimization surrogate~\cite{frazier2018tutorial,jones1998efficient} does not need to predict property values accurately; it needs to \emph{rank} candidates, as
acquisition functions such as Expected Improvement are driven chiefly by how candidates are ordered relative to the incumbent best. A surrogate that puts the right structures at the top therefore remains useful even when its point predictions are poor.
This relaxes the requirements on the surrogate.
Foundation models for atomistic simulation, such as MACE~\cite{batatia2022mace,batatia2024foundation}, ORB~\cite{orb2024}, and UMA/eSEN~\cite{uma2025}, are pretrained on millions of DFT calculations and produce transferable structural embeddings. Integration of these embeddings into a Gaussian process~\cite{rasmussen2006gaussian} may give exactly such a ranking surrogate with no task-specific training.

To test this end to end, we insert a Gaussian-process gate into the published MatInvent loop.
The gate scores stable, novel candidates by Expected Improvement with a diversity term and sends only the top few to the oracle.
We run the gated loop across three architecturally distinct pretrained priors (MatterGen, CrystalFlow, ADiT) and two property targets (room-temperature heat capacity and bulk modulus), comparing gated and ungated policies over five seeds each.
The gate matches or exceeds ungated fine-tuning while capping oracle calls at a fixed per-cycle budget.
A first ablation isolates the mechanism: at an identical four-call budget, ranking-based selection outperforms arbitrary selection, so the gain comes from the gate's choice of candidates, not from spending less.
A second ablation maps the efficiency frontier: in the most competitive case, the gate comes within $\sim$9\% of exhaustive oracle spending at roughly one-fifth of the calls.

To test the same requirement outside the loop, we run a cross-factorial study over four embeddings, three Gaussian-process surrogate families, three dimensionalities, and three training-set sizes across mechanical, electronic, and vibrational property datasets (8{,}640 cross-validated fits).
Existing benchmarks either evaluate one featurizer with one model~\cite{batatia2024foundation} or many models on fixed descriptors~\cite{dunn2020benchmarking}. In both cases they report point-prediction accuracy rather than ranking.
Among the embeddings we tested, pretrained ORB with a Gaussian process was the most reliable across all three domains, and we adopt it as the default in the loop.

Section~\ref{sec:methods} formalizes the gate and the benchmark setup; Section~\ref{sec:results} reports the closed-loop and static studies; Section~\ref{sec:discussion} and Section~\ref{sec:conclusion} interpret the results and outline next steps.

\section{Methods}
\label{sec:methods}

\subsection{Pipeline Overview}
\label{sec:pipeline}

\begin{figure}[H]
    \centering
    \includegraphics[width=\columnwidth]{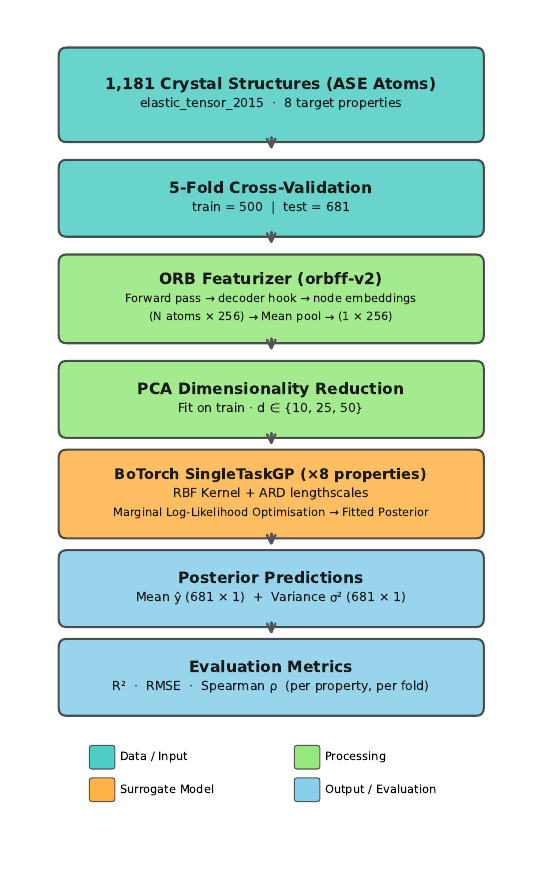}
    \caption{Modular pipeline for foundation-model-based surrogate modeling. Crystal structures are featurized by one of four embedding methods (SOAP, MACE, ORB, UMA/eSEN), projected to lower dimensions via PCA, and fed into a probabilistic surrogate model (GP, MTGP, or DGP). Each component can be independently swapped, enabling systematic cross-factorial evaluation.}
    \label{fig:pipeline}
\end{figure}

Our benchmark uses a modular pipeline with three swappable stages (Figure~\ref{fig:pipeline}).
Given a crystal structure, the pipeline proceeds as follows: (i)~a \emph{featurizer} converts the atomic arrangement into a fixed-length numerical vector; (ii)~principal component analysis (PCA) projects this high-dimensional embedding into a reduced subspace; (iii)~a \emph{probabilistic surrogate model} maps the reduced features to a predictive distribution over the target property, yielding a mean prediction~$\mu(\mathbf{x})$ and uncertainty~$\sigma(\mathbf{x})$. 

\label{sec:pca}
We evaluate three reduced dimensionalities $d \in \{10, 25, 50\}$ via PCA~\cite{jolliffe2002principal}. PCA serves two purposes: it reduces kernel hyperparameter count (one ARD lengthscale per input dimension) and mitigates collinearity among embedding features. The PCA basis is fit on the training set only, then applied to the test set to prevent data leakage.

Any component can be swapped independently, enabling systematic cross-factorial evaluation. We tested 4~featurizers $\times$ 3~PCA settings $\times$ 3~surrogates $\times$ 3~training sizes, yielding 108~configurations per dataset.

\subsection{Datasets}
\label{sec:datasets}

We evaluate our pipeline on three benchmark datasets spanning material property domains: mechanical (elastic) and electronic (dielectric) datasets from the matminer library~\cite{ward2018matminer}, and a vibrational (phonon) dataset from the Materials Project DFPT phonon database~\cite{petretto2018high}.
Table~\ref{tab:datasets} summarizes the key characteristics of each dataset.

\begin{table*}[!htb]
    \centering
    \caption{Summary of the three benchmark datasets. The elastic and dielectric datasets are sourced from matminer; the phonon dataset is the Materials Project DFPT phonon database~\cite{petretto2018high}. All contain DFT-computed properties from the Materials Project~\cite{jain2013materials}.}
    \label{tab:datasets}
    \small
    \begin{tabular}{@{}llcc@{}}
        \toprule
        \textbf{Dataset} & \textbf{Property Type} & \textbf{Structures} & \textbf{Targets} \\
        \midrule
        \texttt{elastic\_tensor\_2015} & Mechanical & $\sim$1{,}181 & 8 \\
        \texttt{dielectric\_constant} & Electronic & $\sim$1{,}056 & 4 \\
        \texttt{mp\_dfpt\_phonon} & Phonon/Thermal & $\sim$1{,}253 & 4 \\
        \bottomrule
    \end{tabular}
\end{table*}

\textbf{Elastic tensor dataset}~\cite{deJong2015elastic} contains 1{,}181 inorganic crystalline structures with DFT-computed elastic constants.
We predict 8~target properties: the Voigt, Voigt--Reuss--Hill (VRH), and Reuss averages of both bulk modulus ($K$) and shear modulus ($G$), together with elastic anisotropy and Poisson's ratio.
These properties range from ``easy'' (bulk modulus variants, depending primarily on bond stiffness and local chemistry) to ``hard'' (elastic anisotropy, requiring directional bonding and crystal symmetry).

\textbf{Dielectric constant dataset}~\cite{petousis2017high} contains 1{,}056 structures with 4~target properties: band gap, refractive index ($n$), and the electronic and total components of the polycrystalline dielectric constant (\texttt{poly\_electronic} and \texttt{poly\_total}).
These properties are sensitive to electronic structure details that are only indirectly captured by geometric embeddings.

\textbf{Phonon thermodynamics dataset} is drawn from the Materials Project DFPT phonon database~\cite{petretto2018high}. It contains 1{,}253 structures with 4~target properties evaluated at 300\,K: the per-atom heat capacity (\texttt{Cv\_300K}), vibrational entropy (\texttt{S\_300K}), Helmholtz free energy (\texttt{F\_300K}), and the maximum phonon frequency (\texttt{max\_phonon\_freq}).

\subsection{Featurizers}
\label{sec:featurizers}

We compare four featurization strategies spanning the spectrum from classical hand-crafted descriptors to foundation model embeddings.
Table~\ref{tab:featurizers} summarizes their key architectural differences.

\begin{table*}[!htb]
    \centering
    \caption{Comparison of the four featurization methods. Foundation model embeddings (MACE, ORB, UMA/eSEN) are extracted from pretrained models without fine-tuning. The raw embedding dimensionality is reduced via PCA before surrogate training.}
    \label{tab:featurizers}
    \small
    \begin{tabular}{@{}lllcl@{}}
        \toprule
        \textbf{Featurizer} & \textbf{Type} & \textbf{Pretraining Data} & \textbf{Raw Dim.} & \textbf{Key Reference} \\
        \midrule
        SOAP & Classical descriptor & None (analytical) & $\sim$10{,}000+ & Bart{\'o}k \emph{et al.}~\cite{bartok2013representing} \\
        MACE & Foundation model & MPTrj ($\sim$1.6M structs., $\sim$146k materials) & $\sim$256 & Batatia \emph{et al.}~\cite{batatia2022mace,batatia2024foundation} \\
        ORB & Foundation model & OMat24 ($>$100M configs.) & $\sim$256 & Orbital Materials~\cite{orb2024} \\
        UMA/eSEN & Foundation model & Multiple sources & $\sim$256 & Meta FAIR~\cite{uma2025} \\
        \bottomrule
    \end{tabular}
\end{table*}

\textbf{SOAP} (Smooth Overlap of Atomic Positions)~\cite{bartok2013representing}  descriptors  encode local atomic environments as the power spectrum of a smooth atomic density expanded in radial basis functions and spherical harmonics.
We use the DScribe implementation~\cite{dscribe2020} with $r_{\text{cut}} = 5$~{\AA}, $n_{\max} = 4$, $\ell_{\max} = 3$, and outer averaging.
The structure-level descriptor is obtained by averaging over all atomic environments.
SOAP serves as the classical baseline: it encodes only local neighborhoods within the cutoff radius.

\textbf{MACE}~\cite{batatia2022mace,batatia2024foundation} is an equivariant message-passing neural network that captures many-body interactions through higher-order tensor products.
We extract pooled node embeddings from the pretrained MACE-MP-0 foundation model~\cite{batatia2024foundation}, which was trained on the MPTrj dataset of $\sim$1.6~million structures (relaxation-trajectory frames drawn from $\sim$146{,}000 Materials Project materials)~\cite{deng2023chgnet}.
We use the model as a frozen feature extractor.

\textbf{ORB}~\cite{orb2024} is a graph neural network pretrained by Orbital Materials. We use the \texttt{orb\_v3\_conservative\_inf\_omat} checkpoint, whose \texttt{omat} suffix denotes training on the OMat24 dataset of over 100~million atomic configurations.
Its graph attention architecture is tuned for periodic crystal structures.
We extract final-layer embeddings and average over atomic sites.

\textbf{UMA/eSEN} (Universal Materials Architecture)~\cite{uma2025} is Meta FAIR's family of interatomic potentials based on the eSCN/eSEN architecture.
Pretrained on heterogeneous materials data, UMA/eSEN is the most recently released of the foundation models we benchmark.
For featurization we use the \texttt{uma-s-1p1} checkpoint with the OMat task head. We extract final-layer embeddings and average over atomic sites, as for MACE and ORB.
The closed-loop probe oracle in Section~\ref{sec:closed_loop_methods} uses a separate \texttt{eSEN-30M-OAM} checkpoint from the same family.

\subsection{Surrogate Models}
\label{sec:surrogates}

All surrogate models are implemented in BoTorch~\cite{balandat2020botorch}, built on GPyTorch~\cite{gardner2018gpytorch}.

\textbf{Gaussian Process (GP).}
We use BoTorch's single-task Gaussian process at its current default: an RBF kernel with automatic relevance determination (ARD) under a dimension-scaled prior.
Hyperparameters are optimized by maximizing the marginal log-likelihood.
The GP is the standard surrogate for Bayesian optimization: given a set of observations, it provides a closed-form posterior over the target function with calibrated uncertainty estimates~\cite{rasmussen2006gaussian}.
Each target property is modeled independently.

\textbf{Multi-Task GP (MTGP).}
We use BoTorch's multi-task Gaussian process at its current defaults: an RBF data kernel with ARD and an intrinsic coregionalization model (ICM)~\cite{bonilla2007multi} task structure at full rank (rank = number of tasks).
The ICM kernel decomposes the multi-output covariance as a Kronecker product of a task correlation matrix and a shared spatial kernel, allowing information to flow between related properties.
The hypothesis is that correlated properties (e.g., $K_{\text{Voigt}}$, $K_{\text{VRH}}$, $K_{\text{Reuss}}$) should benefit from joint modeling.

\textbf{Deep GP (DGP).}
We use a two-layer deep Gaussian process~\cite{damianou2013deep} built on GPyTorch's variational deep-GP layers and wrapped with BoTorch's multi-task model API (Mat{\'e}rn~5/2 hidden layer, RBF output layer with linear coregionalization), trained via variational inference~\cite{dutordoir2021deep}.
The DGP stacks GP layers to capture non-linear input transformations that a single-layer GP cannot represent.
However, this additional expressiveness comes at the cost of more parameters, potentially more fragile training, and sensitivity to the input dimensionality.

\subsection{Experimental Protocol}
\label{sec:protocol}

For each dataset, we subsample training sets of size $\ntrain \in \{100, 250, 500\}$ from the training fold, while the test fold remains fixed.
For each $\ntrain$ we draw a fixed training pool of size $\ntrain$ from the dataset and route the remaining rows to a per-$\ntrain$ held-out test set. The five evaluation runs are independent random 80/20 splits inside that training pool, all scored against the same held-out test set for the corresponding $\ntrain$. This controls split variance while preserving a common evaluation set across configurations at a given training budget.
This simulates the low-data regime typical of early-stage materials discovery, where DFT calculations are expensive.

The full factorial design yields $4 \times 3 \times 3 \times 3 = 108$ configurations per dataset, or 324 across the three datasets.
With five random splits and per-property evaluation (8~properties for elastic, 4 for dielectric, 4 for phonon), the total number of model fits is 8{,}640.

We report three complementary metrics:
\begin{itemize}[nosep]
    \item \textbf{\Rtwo{}}: measures the fraction of variance explained, penalizing both bias and scatter.
    \item \textbf{RMSE} (root mean square error): measures absolute prediction error in the units of the target property.
    \item \textbf{Spearman rank correlation} ($\rho$): measures the monotonic relationship between predicted and true values; critical for Bayesian optimization, where candidate \emph{ranking} matters more than exact predictions.
\end{itemize}
All metrics are computed on the held-out test set and averaged across the five random splits.

\subsection{Closed-Loop Bayesian-Optimization Pipeline}
\label{sec:closed_loop_methods}

Beyond the static benchmark, we evaluate the same surrogate stack inside a closed-loop discovery pipeline that couples a pretrained generative diffusion prior to online policy-gradient updates and an uncertainty-aware acquisition gate (Section~\ref{sec:closed_loop_probe}).
The pipeline is a single-component modification of the published MatInvent workflow~\cite{chen2025matinvent}: between the Stable+Unique+Novel (SUN) filter and the property oracle, we insert a Gaussian process trained on the running long-term memory and route only the top-$K$ candidates to the oracle each cycle.

\textbf{Generative backbones.}
We use three architecturally distinct pretrained generators that span the taxonomy of crystal generative models~\cite{generative_review_2025}: MatterGen~\cite{zeni2025mattergen}, a joint diffusion model on atom types, fractional coordinates, and lattice parameters, trained on Materials-Project ground states; CrystalFlow~\cite{luo2025crystalflow}, a graph-equivariant continuous normalizing flow with conditional flow matching; and ADiT~\cite{joshi2025adit}, an all-atom latent diffusion transformer.
Each backbone enters the loop at its publicly released checkpoint, unmodified.

\textbf{Reinforcement-learning update.}
Following MatInvent, we treat each denoising trajectory as a stochastic policy and apply policy-gradient updates with reward-weighted Kullback--Leibler regularization against the pretrained prior: $\mathcal{L} = \mathbb{E}_{x \sim \pi_\theta}[\,r(x) \cdot \log \pi_\theta(x)\,] + \sigma \cdot \text{KL}(\pi_\theta \,\|\, \pi_{\text{prior}})$, with $\sigma = 0.025$.
The update is supplemented by a 100-sample experience-replay buffer and a Tanimoto diversity filter (tolerance 3, buffer 6).
This update rule descends directly from the DDPO~\cite{black2023ddpo} and DPOK~\cite{fan2023dpok} formulations developed for image-domain diffusion models, and is identical in BASE (ungated) and ACC (gated) except for the property calculator.

\textbf{GP-gated acquisition (ACC only).}
Each cycle, the SUN-survivors are featurized with ORB and projected to PCA\,$=$\,50 --- the recommended stack from the static benchmark (Section~\ref{sec:cross_dataset}). A single-task GP with BoTorch's default RBF\,+\,ARD kernel is trained on the long-term memory of oracled candidates accumulated so far.
We score the SUN-survivors by Expected Improvement and select the top $K = 4$ subject to a Determinantal Point Process diversity term over the GP's posterior mean. Only those four reach the oracle; the remaining candidates are assigned a NaN reward so they contribute zero RL gradient.
Cycle 0 is a 16-candidate warm-start: every SUN-survivor goes to the oracle so that the GP has in-loop training data before the gate begins to discriminate.
In addition, the GP's long-term memory is preloaded before cycle 0 with an externally labeled seed pool. For $K_{\text{VRH}}$, this is 500 structures sampled from matminer's \texttt{elastic\_tensor\_2015} dataset with their tabulated $K_{\text{VRH}}$ values. For $C_p$, it is ${\sim}446$ oracle-labeled structures pooled from earlier runs of the same pipeline.
These preloaded points (tagged cycle $-1$) are down-weighted in the GP fit by an age-dependent noise inflation but remain in the training set throughout, so the GP never operates from a 16-point cold start. The per-run oracle budgets reported in Section~\ref{sec:closed_loop_probe} count in-loop oracle calls only.
The same seed pool is the reference set for the novelty analysis (Section~\ref{sec:cl_novelty}).

\textbf{Property oracle.}
The oracle is itself a learned surrogate of DFT.
For heat capacity at room temperature ($C_p$), we relax each candidate with the eSEN-30M-OAM machine-learning interatomic potential~\cite{uma2025} and feed the relaxed structure into phonopy~\cite{togo2015first} for a quasi-harmonic phonon calculation that yields $C_p(T = 300\,\text{K})$ (distinct from the static-benchmark phonon target \texttt{Cv\_300K} of Section~\ref{sec:datasets}, which is a DFPT per-atom heat capacity).
For the Voigt--Reuss--Hill bulk modulus ($K_{\text{VRH}}$), we compute energies under seven uniform strains in the range $\pm 3\%$ with eSEN-30M-OAM and fit a Birch--Murnaghan equation of state~\cite{birch1947finite}.
Beyond this learned oracle, we recompute the bulk modulus of the top candidates with DFT as an oracle-parity check (Section~\ref{sec:dft_methods}); the corresponding heat-capacity validation is substantially more costly and is left to future work.

\textbf{Run protocol.}
Each closed-loop trajectory runs for 20 RL cycles with a per-cycle design budget of up to 16 candidates after SUN filtering; the realized count varies by backbone (MatterGen approaches the budget; CrystalFlow and ADiT typically deliver $\sim$5--10 SUN-survivors per cycle).
We run five random seeds for each $\{$backbone, target, policy$\}$ cell, giving 30 trajectories per target ($C_p$ and $K_{\text{VRH}}$); one ADiT/ACC $C_p$ seed did not complete, leaving 29 on $C_p$ and 30 on $K_{\text{VRH}}$.
Per-seed running-best property values are forward-filled to the full 20-cycle horizon before averaging across seeds, so the cross-seed mean is itself non-decreasing; this is the cross-seed equivalent of the standard convention used by MatInvent's own discovery curves.
Concurrent benchmark frameworks for closed-loop materials discovery~\cite{made_benchmark} and for cross-model evaluation of crystal generators~\cite{lematbench2025} provide complementary protocols; the present pipeline reports a specific gating method run end-to-end across three pretrained priors.

\subsection{DFT Validation of Generated Structures}
\label{sec:dft_methods}

To anchor the closed-loop probe's learned oracle and surrogate to first-principles ground truth, we recompute the bulk modulus of the top generated $K_{\text{VRH}}$ structures with DFT and replay the GP surrogate causally against those values (Section~\ref{sec:cl_dft}).
We validate the bulk-modulus target only: its oracle is a seven-point equation-of-state fit (seven single-point energies per structure), whereas the heat-capacity oracle requires a phonon supercell with tens to over a hundred displacement calculations on 40--580-atom cells, roughly an order of magnitude more expensive; a DFT check of the heat-capacity target is therefore left to future work.

\textbf{Oracle-parity equation of state.}
The DFT bulk moduli are computed to be like-for-like with the eSEN oracle, so that the comparison isolates the oracle's error rather than a generic DFT-versus-MLIP gap. Each generated cell is first refined to its nearest space group with spglib~\cite{togo2018spglib} so that VASP~\cite{kresse1996efficient} can exploit the symmetry. Cells for which spglib finds no symmetry above the tolerance are computed without symmetry (space group P1). The cell and ions are then relaxed under Materials-Project-compatible settings~\cite{pymatgen2013}---the PBE functional~\cite{perdew1996generalized}, projector augmented-wave potentials~\cite{blochl1994projector,kresse1999ultrasoft}, a 680~eV plane-wave cutoff, and spin polarization, with no Hubbard $U$ correction, as none of the validated chemistries falls under the Materials Project $+U$ scheme~\cite{jain2013materials}. From the relaxed cell we apply seven rigid isotropic strains $\varepsilon \in \{-3\%, \dots, +3\%\}$ with frozen fractional coordinates and a single static calculation at each---mirroring the oracle, which likewise does not relax ions at strained volumes---and fit a third-order Birch--Murnaghan equation of state~\cite{birch1947finite} to obtain $K_0$. A single $k$-point grid, fixed from the relaxed reference (reciprocal spacing $0.16$~\AA$^{-1}$), is used for all seven strains, since regenerating the mesh per strained cell perturbs the curvature of the fit; the statics use the tetrahedron method with an energy threshold of $10^{-7}$~eV. Every validated structure yields an in-window fit with a physical pressure derivative $B_0' \in [4.1, 5.1]$.

\textbf{Causal surrogate replay.}
Because the validated structures are members of the closed-loop long-term memory, training the GP on the full memory and then predicting them would measure memorization rather than generalization. We instead replay each ACC run causally: stepping through cycles in order, at cycle $s$ we train the GP on all structures gathered before cycle $s$---including the preloaded warm-start seed pool (Section~\ref{sec:closed_loop_methods})---and predict the freshly generated cycle-$s$ candidates, which the surrogate has not yet seen. The analysis is restricted to the ACC runs, the only ones that employ a GP gate.

\subsection{Synthesizability Scoring of Generated Structures}
\label{sec:synth_methods}

We assess the synthesizability of generated structures with two independent composition-level models and compare them. The first is an ORB-based positive-unlabeled classifier (ORB-PU) trained here; the second is the pretrained composition-graph model of Jang et al.~\cite{jang2024synth} (CGNF).

\textbf{Training data.}
We assemble 109{,}283 Materials Project compositions: 49{,}283 entries whose Materials Project records carry an ICSD cross-reference (experimentally synthesized) as labeled positives, and 60{,}000 theoretical entries with no ICSD cross-reference and energy above hull below 0.5~eV/atom as unlabeled, split into train, validation, and test (76{,}498 / 10{,}928 / 21{,}857) stratified by label. Holding out known positives and treating the unlabeled pool as pseudo-negatives matches the evaluation protocol used by CGNF.

\textbf{Features and classifier.}
Each composition is represented by its mean-pooled ORB embedding reduced with the benchmark PCA (Section~\ref{sec:pca}), concatenated with Magpie descriptors~\cite{ward2016general}; an ORB-energy stability scalar is an optional third block. The classifier is a Mordelet--Vert positive-unlabeled bagging ensemble~\cite{mordelet2014bagging}: each bag keeps all positives and draws a random subsample of the unlabeled pool as pseudo-negatives, sized by a tuned multiple of the positive count and capped at the pool size, and per-bag positive probabilities are averaged.
In the selected configuration (19 bags, negative-sampling ratio 1.29) the cap binds, so every bag trains on the full unlabeled pool and ensemble diversity comes from the base learner's per-bag random seed and column subsampling rather than from the unlabeled draw. An abstention layer marks a composition as out-of-distribution when its standardized nearest-neighbor distance to the positive training set exceeds the 95th percentile of that distance among positives, or when bag disagreement or prediction confidence fall outside fixed thresholds.

\textbf{Hyperparameter tuning.}
Hyperparameters are tuned with Optuna~\cite{akiba2019optuna} (tree-structured Parzen sampler, median pruner, five-fold stratified cross-validation) to maximize the area under the precision--recall curve on the held-out positive-versus-unlabeled split, over feature sets \{Magpie, ORB, ORB+Magpie, ORB+Magpie+stability\} and base learners \{random forest, gradient-boosted trees\}; the bag count and negative-sampling ratio are tuned jointly with each learner's hyperparameters. The selected model is ORB+Magpie with gradient-boosted trees (Supplementary Table~S4).

\textbf{Evaluation.}
On the held-out Materials Project test split the selected ORB-PU model and pretrained CGNF are scored on the same compositions (Supplementary Table~S5); these numbers favor ORB-PU because it is in-distribution whereas CGNF is applied zero-shot, so we do not read them as a method-superiority comparison. The comparison on equal footing is on the generated structures (Section~\ref{sec:cl_synth}), where both models are out-of-distribution.

\section{Results}
\label{sec:results}

We first consider a closed discovery loop, wrapping three pretrained diffusion priors with an online policy-gradient update and a Gaussian-process acquisition gate, and ask whether the gate improves on ungated fine-tuning (Sections~\ref{sec:closed_loop_probe}--\ref{sec:cl_ablation}).
We then test the principle behind the gate's low cost (a Bayesian-optimization surrogate need only \emph{rank} candidates, not predict their values) in a static cross-factorial benchmark of embeddings and surrogates across mechanical, electronic, and vibrational properties (Section~\ref{sec:foundation}).

\subsection{Closed-Loop Probe with Diffusion Backbones}
\label{sec:closed_loop_probe}

The static benchmark indicates that an ORB\,+\,GP surrogate should remain useful in closed-loop Bayesian optimization (BO).
We tested this with three architecturally distinct pretrained diffusion priors, fine-tuned online via the policy-gradient procedure of MatInvent~\cite{chen2025matinvent}, with a Gaussian Process surrogate inserted between the SUN-filter and the property oracle.
The reinforcement-learning side of the procedure follows the lineage of DDPO~\cite{black2023ddpo} and DPOK~\cite{fan2023dpok} from the image domain, ported to crystals by MatInvent and (concurrently, for an autoregressive transformer prior) by CrystalFormer-RL~\cite{cao2025crystalformerrl}.
Existing tutorials and reviews of RL fine-tuning of diffusion models~\cite{uehara2024rldiffusion} catalog several policy-gradient and value-based families, none of which include a Bayesian-optimization gate at acquisition time.
A recent review of generative AI for crystals~\cite{generative_review_2025} likewise surveys the design space without devoting a section to closed-loop, online surrogate-gated refinement, although it notes one offline precedent: Qi et al.'s Latent Conservative Objective Models (LCOM)~\cite{qi2023lcom}, which trains a conservative surrogate over a CDVAE latent space to minimize formation energy.
LCOM differs from the present work in being offline (no online policy update of the generator), single-property (formation energy), and built around a CDVAE rather than a diffusion prior; we view it as a complementary precedent rather than a baseline.
The probe in this section sits in that gap.

We ran two property targets ($C_p$ and $K_{\text{VRH}}$), two policies (BASE and ACC, defined below), three backbones, and five seeds, giving 29 completed $C_p$ runs and 30 $K_{\text{VRH}}$ runs.
The diffusion priors are not retrained from scratch; they are updated only through online REINFORCE-style rollouts.
The oracle is itself a learned surrogate (eSEN-30M-OAM~\cite{uma2025} fed into phonopy~\cite{togo2015first} for $C_p$ and a Birch--Murnaghan fit~\cite{birch1947finite} for $K_{\text{VRH}}$); the bulk-modulus candidates are subsequently validated against DFT (Section~\ref{sec:cl_dft}), while the heat-capacity candidates are not DFT-validated here.
With those caveats noted, the runs are large enough to examine cross-backbone behavior, characterize the GP surrogate as the loop progresses, and report the practical effect of inserting the gate into a published RL pipeline.

\subsubsection{Pipeline and Setup}
\label{sec:cl_pipeline}

The three backbones are MatterGen~\cite{zeni2025mattergen}, CrystalFlow~\cite{luo2025crystalflow}, and ADiT~\cite{joshi2025adit}.
Each runs under two policies.
BASE follows the published MatInvent recipe~\cite{chen2025matinvent} verbatim: policy-gradient updates with reward-weighted KL regularization against the pretrained prior ($\sigma = 0.025$), an experience-replay buffer (size 100, sample size 10, reward cutoff 0.1), and a diversity filter on the long-term memory (tolerance 3, buffer 6).
ACC differs from BASE only in the property calculator: between the SUN filter and the oracle it inserts a Gaussian Process surrogate over ORB\,+\,PCA50 features, scores SUN-survivors by the acquisition rule (Expected Improvement plus a Determinantal Point Process diversity term), and dispatches only the top $K = 4$ to the oracle each cycle; the remaining candidates receive NaN reward, so they do not contribute to the policy-gradient update.
Every other RL hyperparameter (KL coefficient, replay buffer, diversity filter, finetune epochs, batch size $\sim 16$ candidates per cycle) is identical between BASE and ACC.
Each run lasts 20 cycles, and every candidate must pass the Stable+Unique+Novel (SUN) screen before any oracle or surrogate call.

\begin{figure}[H]
    \centering
    \includegraphics[width=\columnwidth]{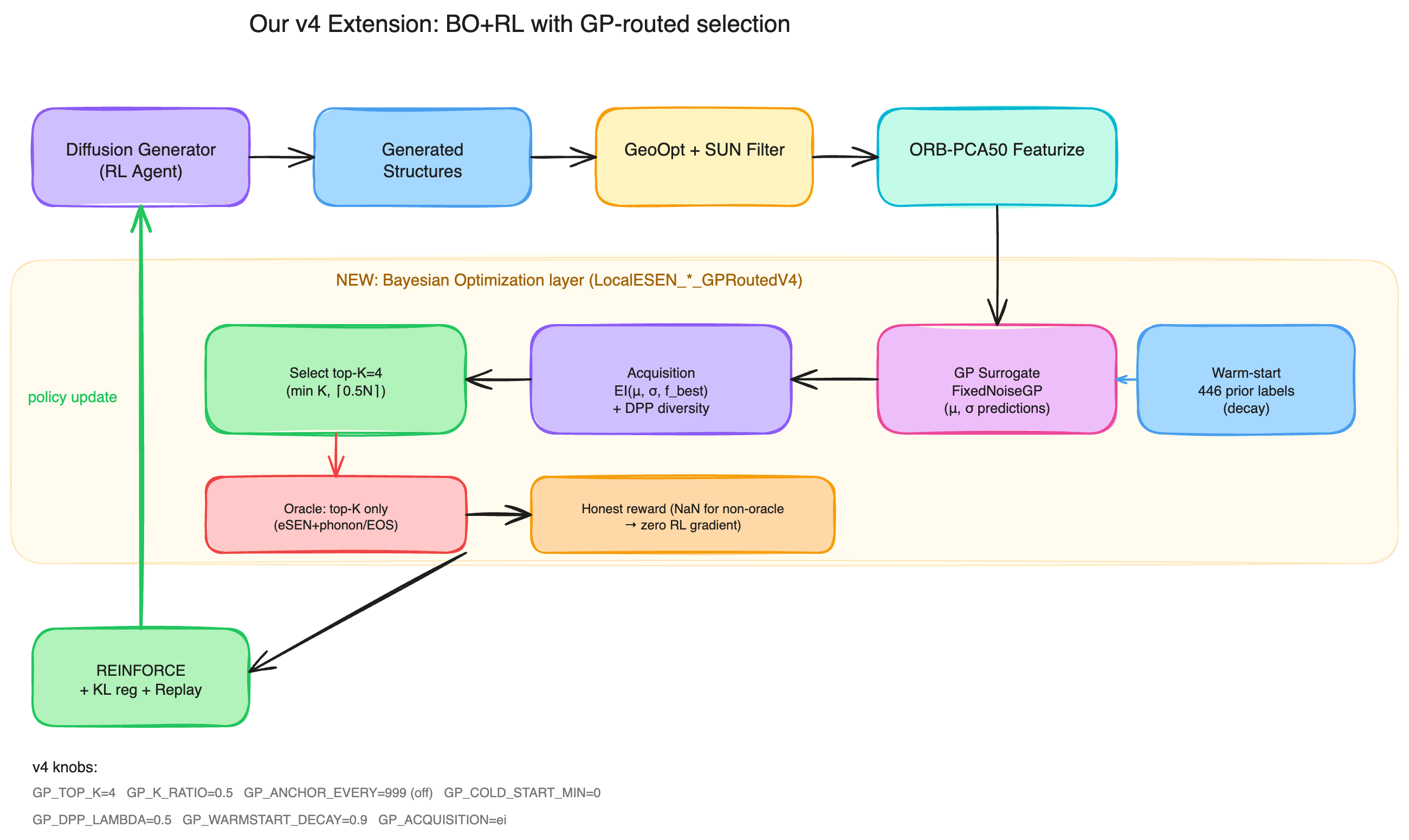}
    \caption{The closed-loop pipeline. The diffusion generator proposes structures; geometry optimization and a SUN filter remove unstable, duplicate, or non-novel candidates; ORB\,+\,PCA50 featurizes the survivors; the GP scores them; an acquisition rule (EI plus a DPP diversity term) selects the top $K=4$; only those reach the oracle (eSEN+phonon for $C_p$, EOS fit for $K_{\text{VRH}}$); REINFORCE then updates the diffusion policy. Non-oracle candidates receive NaN rewards and contribute zero RL gradient, so the surrogate's choice of which $K$ to call is the only training signal.}
    \label{fig:closed_loop_pipeline}
\end{figure}

\FloatBarrier
\subsubsection{Discovery Curves Across Backbones}
\label{sec:cl_curves}

\begin{figure*}[!htb]
    \centering
    \includegraphics[width=\textwidth]{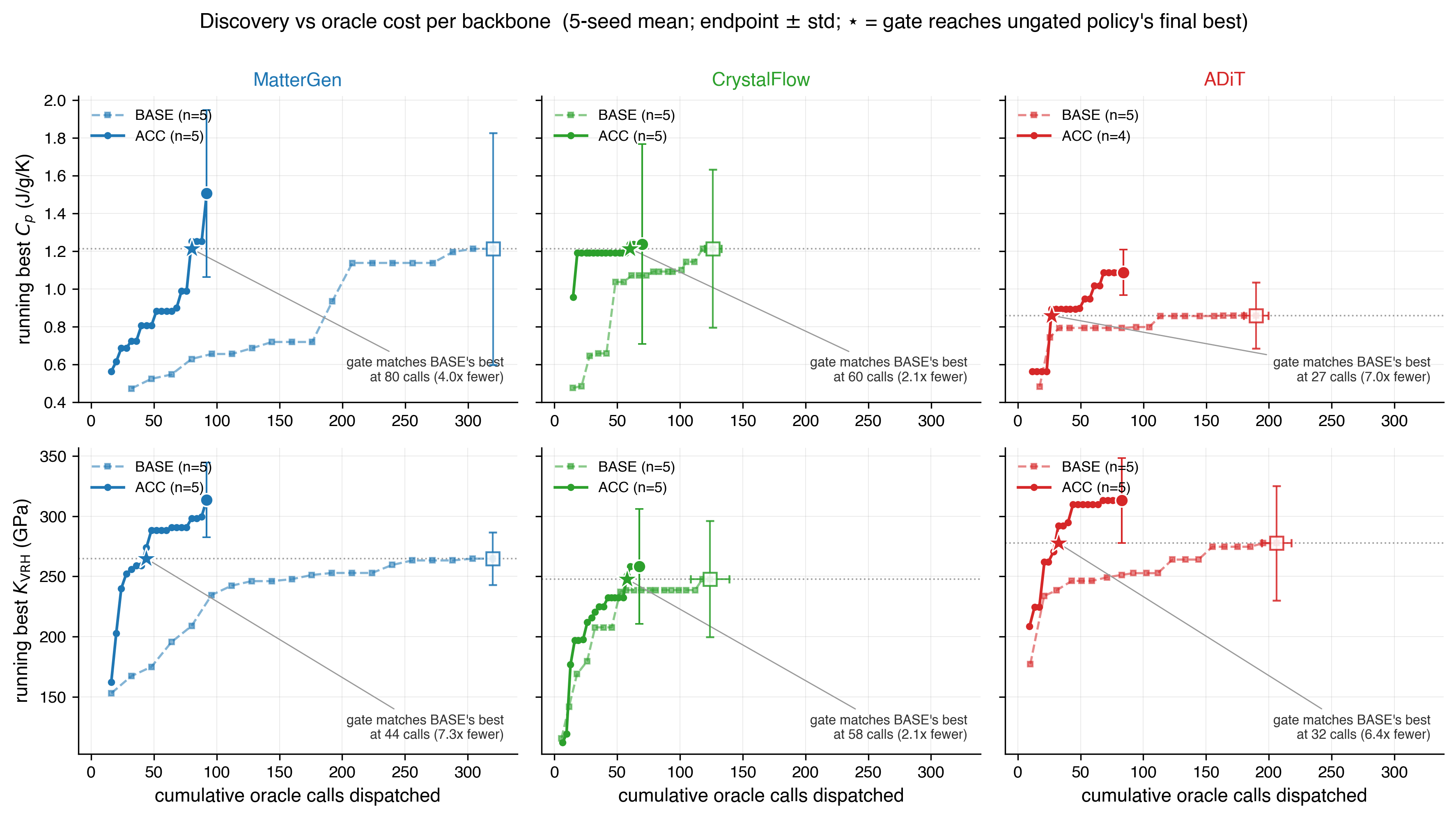}
    \caption{Discovery versus oracle cost, per backbone: running best oracled property (seed-mean) against cumulative oracle calls, for $C_p$ (top row) and $K_{\text{VRH}}$ (bottom row). Solid curves with filled endpoints: GP-gated BO (ACC); dashed curves with open endpoints: vanilla REINFORCE (BASE); endpoint error bars give $\pm$ one between-seed standard deviation on both axes (five seeds per cell except ADiT/ACC-$C_p$, which has four). The dotted guide in each panel marks BASE's final best; the star marks where the gated policy first reaches that value, annotated with the corresponding call count. For every policy the horizontal axis counts each candidate dispatched to the oracle, whether or not the evaluation returns a valid property, so the two policies are placed on one cost basis (every SUN-survivor for BASE, the top-$K$ for ACC); these are the same conventions as the per-cycle schedules in Supplementary Fig.~S2. Per-cycle trajectories against RL cycle, and a combined endpoint summary, are given in Supplementary Figs.~S1--S3.}
    \label{fig:closed_loop_curves}
\end{figure*}

Figure~\ref{fig:closed_loop_curves} plots the running best oracled property against cumulative oracle calls, one panel per backbone; per-seed running-bests are forward-filled to the full cycle horizon before averaging, so the cross-seed mean is itself non-decreasing.
This view pairs discovery with its cost: a curve that rises further up and left finds better materials on a smaller oracle budget.
For $C_p$, the gated policy ends at seed-mean running bests of 1.51 (MatterGen), 1.24 (CrystalFlow), and 1.09~J/g/K (ADiT), above BASE in every backbone.
For $K_{\text{VRH}}$, MatterGen and ADiT plateau near 310~GPa under ACC; the single best structure (375~GPa) is an MoN polymorph in space group $P\,\bar{6}m2$ (\#187) found by ADiT/ACC at seed 7.
The starred markers locate where the gate first matches the ungated policy's end-of-run best: at 2.1--7.3$\times$ fewer oracle calls in all six (backbone, target) cells.
CrystalFlow shows the largest between-seed spread of the three backbones for both targets, so its mean comparison is the noisiest of the panel, but the ACC mean still sits above the BASE mean throughout.
The qualitative pattern is consistent with the static benchmark, which is discussed in detail later in the text and supplementary information: the ORB\,+\,GP surrogate ranks candidates well enough to drive acquisition without breaking the underlying RL update.
Direct numeric comparison with the published MatInvent heat-capacity run (Fig.~2c of \cite{chen2025matinvent}) is limited because the two curves plot different quantities: their 10-seed mean tracks the per-loop \emph{average} $C_p$ of generated structures, reaching ${\approx}0.6$~J/g/K at loop 20 on its way to the 1.5~J/g/K target near loop 60, whereas our curves track the running best. At a matched 20-loop budget the two setups are nonetheless in a similar regime: their loop-20 batch mean sits below our 5-seed BASE running-best mean of $1.2 \pm 0.6$~J/g/K, as a batch-mean versus best-so-far statistic must.

\FloatBarrier
\subsubsection{Surrogate Quality Across Cycles}
\label{sec:cl_surrogate}

\begin{figure*}[!htb]
    \centering
    \includegraphics[width=\textwidth]{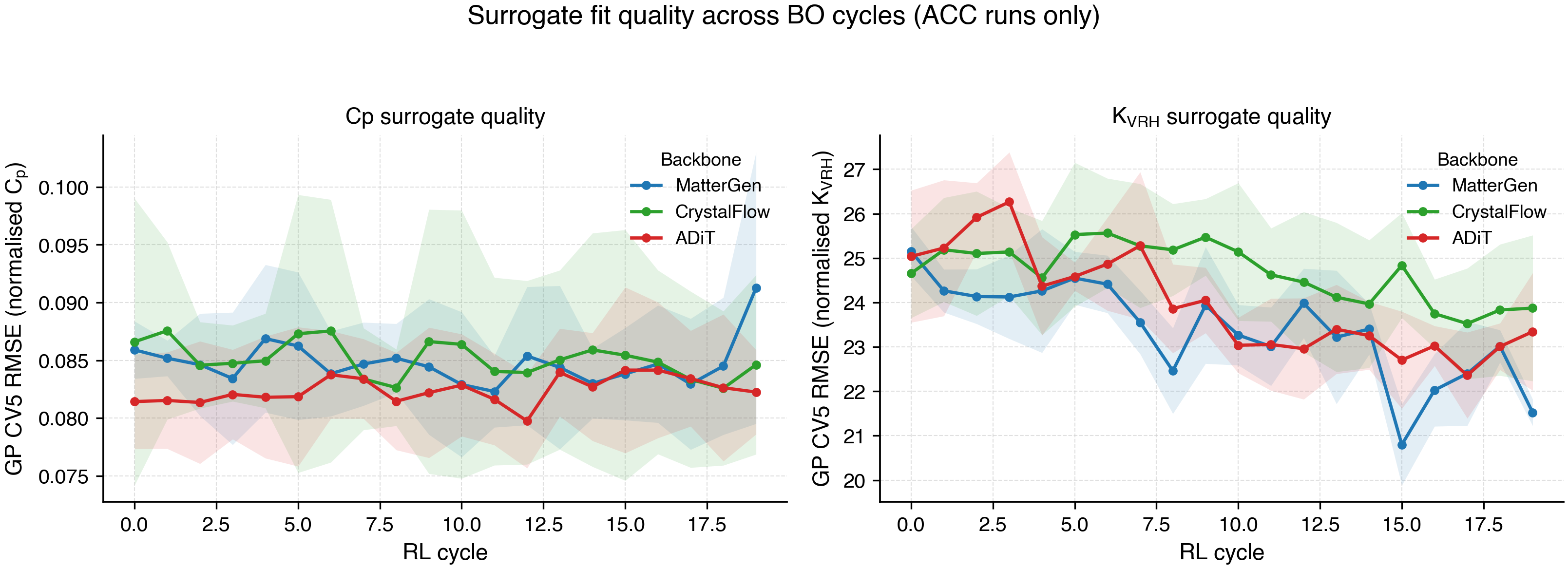}
    \caption{GP CV5 RMSE (normalized property scale) per RL cycle, ACC runs only, seed-mean $\pm$ std (five seeds per cell except ADiT/ACC-$C_p$, which has four). CV5 is computed on the GP's full training set, which includes the pre-loaded labeled seed pool described in Methods plus all in-loop acquisitions (${\sim}450$--$520$ points at cycle~0, growing to ${\sim}510$--$590$ by cycle 19), not on held-out fresh diffusion proposals. The flatness across cycles indicates that the loop does not break the GP's self-consistency as the training pool grows; it does not directly demonstrate generalization to unseen candidates.}
    \label{fig:closed_loop_surrogate}
\end{figure*}

Figure~\ref{fig:closed_loop_surrogate} plots the GP CV5 RMSE per cycle, ACC runs only.
This metric is 5-fold cross-validation on the GP's full training set at that cycle: the pre-loaded labeled seed pool described in Methods, plus the oracled warm-start candidates and all top-$K$ acquisitions up to the current cycle (${\sim}450$ points at cycle~0 growing to ${\sim}510$ by cycle 19 on $C_p$; ${\sim}510$ growing to ${\sim}590$ on $K_{\text{VRH}}$).
It measures how well the GP fits its own oracled labels under leave-one-fold-out resampling; it does \emph{not} measure error on the next batch of fresh diffusion proposals, the candidates the gate actually scores with the GP posterior mean $\mu$.
On $C_p$ the CV5 RMSE sits at 0.080--0.091 (normalized property scale) and is statistically flat across cycles: mean RMSE in cycles 0--5 differs from cycles 15--19 by less than one between-seed standard deviation.
$K_{\text{VRH}}$ shows the same pattern at the 21--26~GPa level, with a mild downward trend in MatterGen and CrystalFlow.
The GP thus remains well-fitted to training data as the training set grows, a necessary but not sufficient condition for surrogate-gated BO health.
A held-out RMSE on the diffusion proposals before they reach the gate is not in our current logs; for the bulk-modulus target, however, the causal surrogate replay of Section~\ref{sec:cl_dft} measures generalization directly as a ranking on those fresh proposals (Spearman $\rho = 0.944$).

\FloatBarrier
\subsubsection{Oracle-Call Savings from the GP Gate}
\label{sec:cl_oracle}

The GP gate's practical benefit is that it caps the number of expensive oracle calls per cycle at $K=4$ regardless of how many candidates pass the SUN filter, replacing the rest with cheap GP-$\mu$ pseudo-evaluations.
Supplementary Fig.~S2 compares the per-cycle schedules of cumulative oracle calls between BASE (every SUN-survivor goes to the oracle) and ACC (top-$K=4$ only) for both targets.
ACC's curve is the same straight line for every backbone: 16 oracle calls for the warm-start at cycle 0, then 4 per cycle, totaling 92 calls per 20-cycle run.
BASE's curve depends on each backbone's SUN-survival rate.

Total oracle calls per run, 5-seed mean: for $C_p$, BASE/ACC is 320/92 in MatterGen ($3.48\times$ saving), 126/70 in CrystalFlow ($1.80\times$), and 190/84 in ADiT ($2.25\times$); for $K_{\text{VRH}}$, BASE/ACC is 320/92 in MatterGen ($3.47\times$), 124/68 in CrystalFlow ($1.83\times$), and 206/83 in ADiT ($2.49\times$).
The pattern is consistent: SUN survival per cycle exceeds $K=4$ in every (backbone, target) cell, with roughly 6--16 candidates dispatched per cycle, so the gate reduces the oracle-call count throughout, by the largest factor for MatterGen, where survival is highest.
Beyond the call count, the gate also changes which candidates are scored: it forces a top-$K$ selection by EI\,+\,DPP rather than letting whatever survives the SUN filter pass through to the oracle, an effect the budget-matched ablation isolates next.
Measured at matched quality rather than matched cycles, the same data give a sharper statement: the gate first reaches BASE's end-of-run best with $2.1$--$7.3\times$ fewer oracle calls in all six (backbone, target) cells (stars in Figure~\ref{fig:closed_loop_curves}).

\FloatBarrier
\subsubsection{Budget-matched ablations: candidate selection versus call count}
\label{sec:cl_ablation}

\begin{figure*}[!htb]
    \centering
    \includegraphics[width=\textwidth]{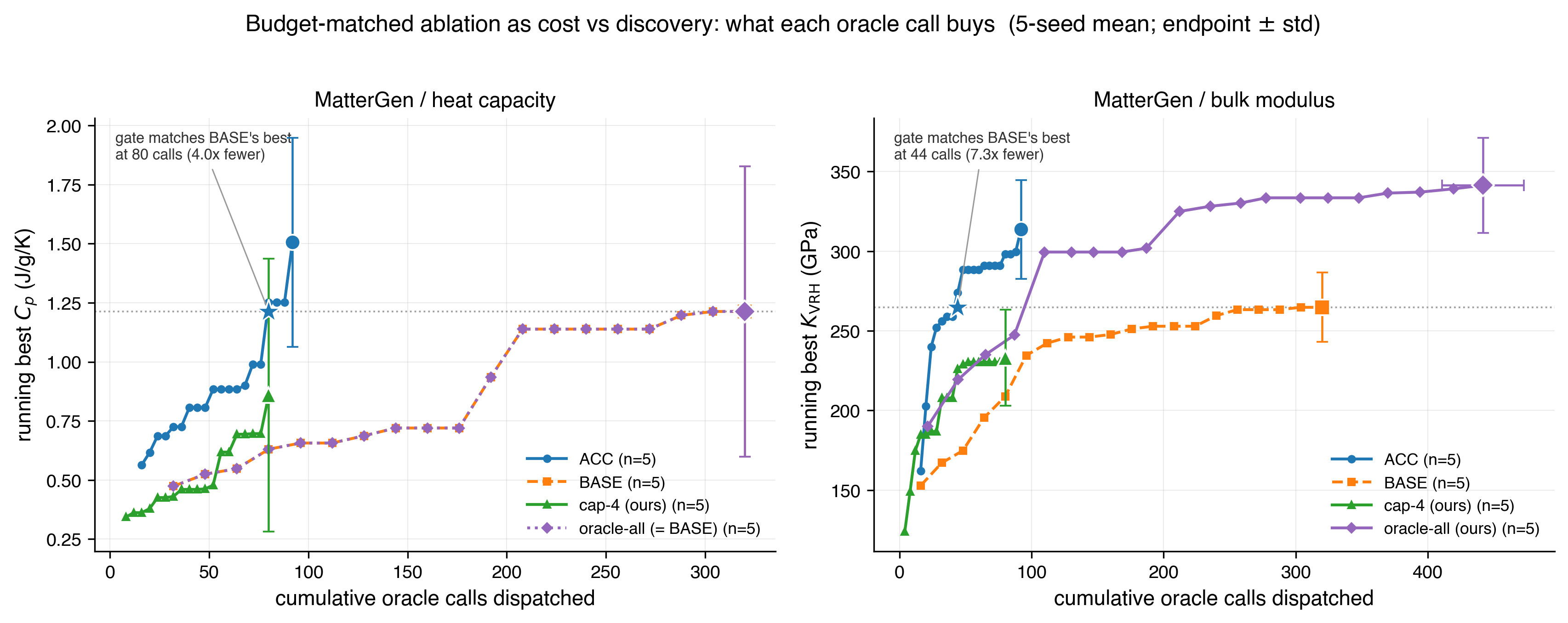}
    \caption{MatterGen oracle-budget ablation as discovery versus cost: running best oracled property (seed-mean) against cumulative oracle calls for $C_p$ (left) and $K_{\text{VRH}}$ (right), with endpoint error bars of $\pm$ one between-seed standard deviation on both axes. ACC (the GP gate, four oracle calls per cycle) ends above both vanilla REINFORCE (BASE) and the budget-matched cap-4 control on both targets; oracle-all (no budget cap, $\sim$25 calls per cycle) marks the brute-force ceiling, reaching $341 \pm 30$~GPa at 442 calls against ACC's $314 \pm 31$~GPa at 92, i.e.\ $4.8\times$ the spend for $\sim$9\% more. The dotted guide marks BASE's final best; the star marks where the gated policy first reaches it (80 calls on $C_p$ and 44 on $K_{\text{VRH}}$, against BASE's 320 on both targets). For every arm the horizontal axis counts each candidate dispatched to the oracle, whether or not the evaluation returns a valid property. Every arm aggregates five seeds; two cap-4 $C_p$ seeds initially terminated on zero-reward cycles and were re-run to completion. Because the SUN cap never binds for MatterGen on $C_p$, oracle-all there coincides with BASE. Per-cycle views of the same arms are given in Supplementary Figs.~S4 and~S5.}
    \label{fig:mg_ablation}
\end{figure*}

The gate changes two things at once relative to vanilla REINFORCE (BASE): it spends fewer oracle calls, and it chooses which candidates to spend them on.
To separate these effects we ran two MatterGen controls that hold the generation front-end fixed (64 proposals per cycle, identical SUN filter) and vary only the back-end selection (Figure~\ref{fig:mg_ablation}).

The first control, \emph{cap-4}, matches ACC's oracle budget exactly (four oracle calls per cycle) but selects the four arbitrarily, taking the first four survivors of the SUN filter rather than ranking them by Expected Improvement.
If the gate's benefit were the reduced call count alone, cap-4 would track ACC.
It does not: by cycle 20 cap-4 reaches only $0.86 \pm 0.58$~J/g/K on $C_p$ and $233 \pm 30$~GPa on $K_{\text{VRH}}$, against ACC's $1.51 \pm 0.44$~J/g/K and $314 \pm 31$~GPa.
At an identical budget, it is the surrogate's \emph{choice} of which four candidates to evaluate, not the reduced call count, that drives discovery.

The second control, \emph{oracle-all}, removes the budget cap and sends every SUN survivor to the oracle (roughly 25 per cycle for MatterGen on $K_{\text{VRH}}$), giving the brute-force ceiling at $\sim$442 oracle calls per run against ACC's 92.
It reaches $341 \pm 30$~GPa on $K_{\text{VRH}}$, about 9\% above ACC at $4.8\times$ the oracle cost.
The gated policy therefore comes within a tenth of exhaustive spending figure of merit while paying roughly a fifth of the budget.
On $C_p$ the SUN cap never binds for MatterGen, so oracle-all coincides with BASE and is plotted as such.

Two of the five cap-4 $C_p$ seeds initially terminated early, on cycles where all four oracle calls returned no valid reward and the policy gradient had nothing to update against; both were re-run to completion, and the re-runs enter the five-seed aggregate.
An overly tight oracle budget can thus starve the loop, a failure mode the gate avoids by spending its four calls on the most promising survivors rather than on whatever the filter happens to emit first.

\FloatBarrier
\subsubsection{Crystal-System Distribution of Generated Top Structures}
\label{sec:cl_crystal_systems}

\begin{figure*}[!htb]
    \centering
    \includegraphics[width=\textwidth]{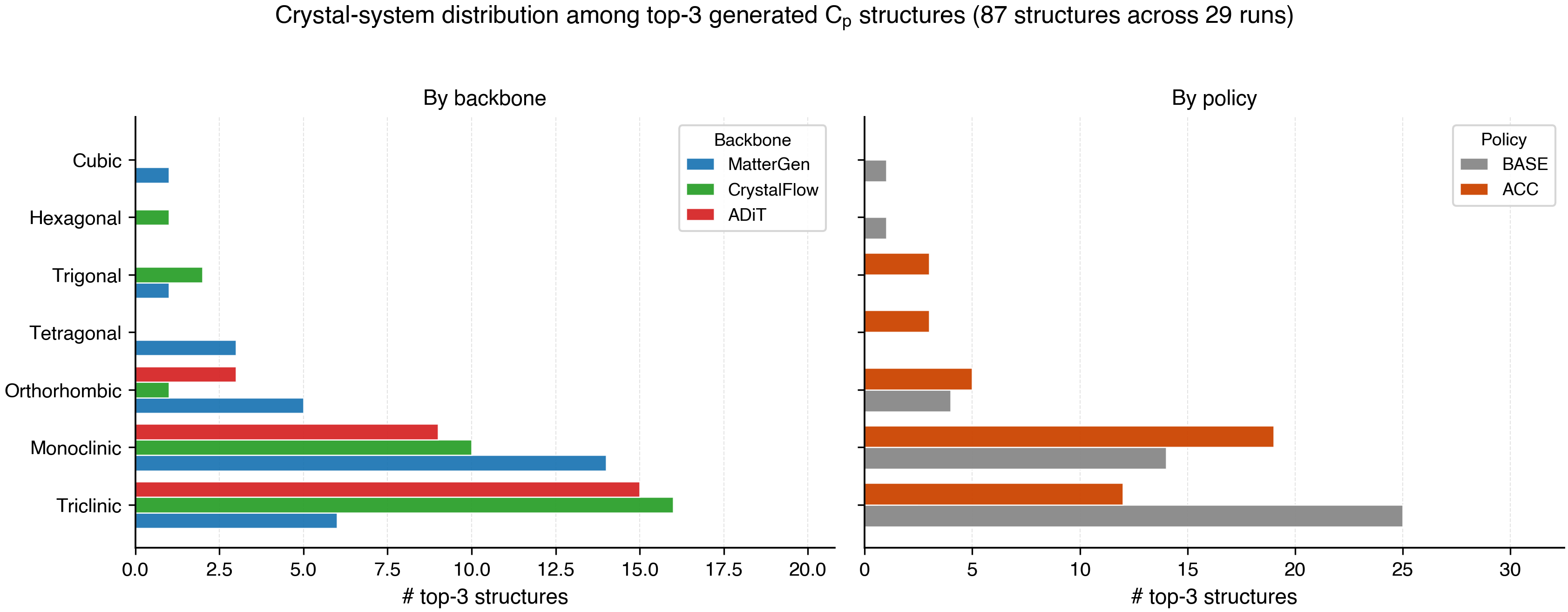}
    \caption{Crystal-system distribution among top-3 generated $C_p$ structures (87 structures across 29 completed runs). Left: by backbone. Right: by policy. CrystalFlow and ADiT skew strongly towards low-symmetry (triclinic, monoclinic) systems; MatterGen is the only backbone producing tetragonal hits; the BASE-vs-ACC split is balanced across systems, so the GP gate is property-driven, not geometry-driven. The matching $K_{\text{VRH}}$ panel is Figure~\ref{fig:closed_loop_spacegroup_bm}.}
    \label{fig:closed_loop_spacegroup}
\end{figure*}

\begin{figure*}[!htb]
    \centering
    \includegraphics[width=\textwidth]{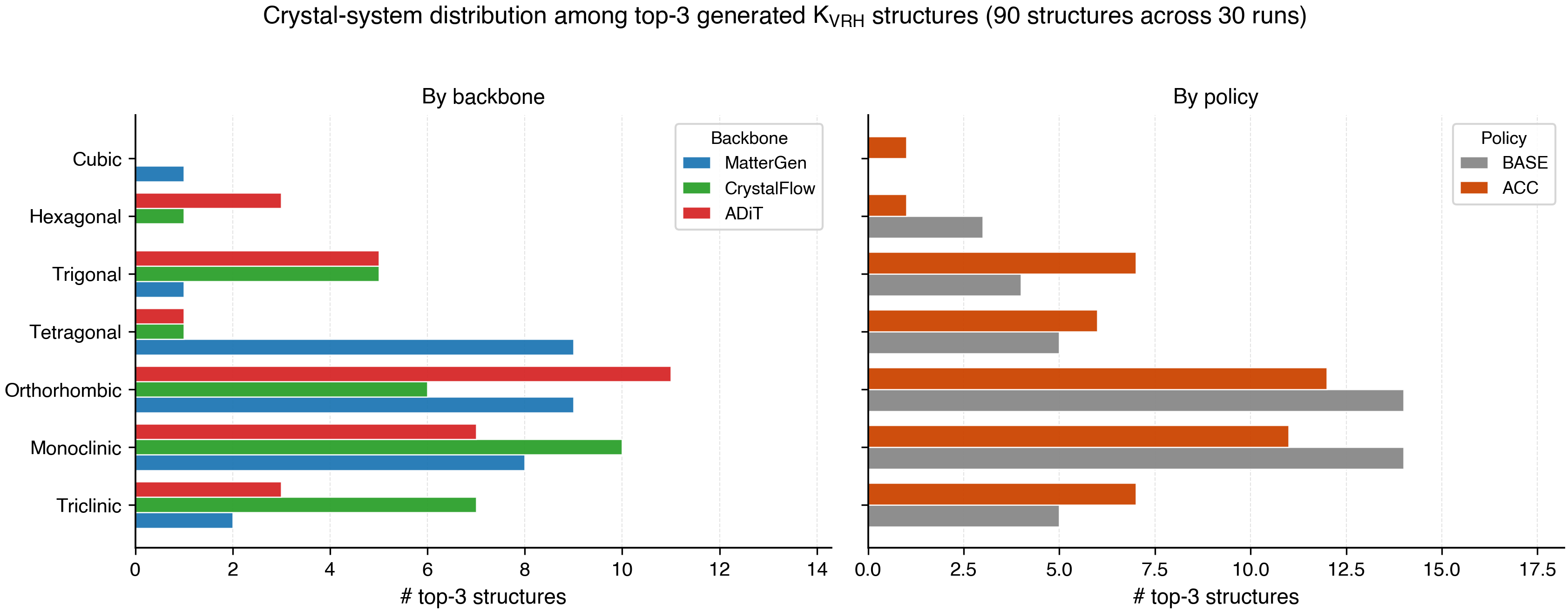}
    \caption{Crystal-system distribution among the top-3 generated $K_{\text{VRH}}$ structures (90 structures across the 30 completed runs). Left: by backbone. Right: by policy. The pattern matches the $C_p$ panel (Figure~\ref{fig:closed_loop_spacegroup}), with an enrichment of cubic and hexagonal systems consistent with high bulk modulus favoring densely packed lattices.}
    \label{fig:closed_loop_spacegroup_bm}
\end{figure*}

The space groups of the top-3 candidates per run reveal a strong asymmetry across backbones (Figure~\ref{fig:closed_loop_spacegroup}).
For $C_p$, CrystalFlow and ADiT both produce more than half of their top structures in triclinic and monoclinic systems, the lowest-symmetry crystal classes, while MatterGen splits more evenly, contributing the bulk of the orthorhombic and tetragonal hits.
All three priors are trained on Materials-Project-derived data (MatterGen on Alex-MP-20; CrystalFlow and ADiT on MP-20), so this spread reflects differences in model architecture and online fine-tuning rather than in training set. We report it as an empirical observation. For $K_{\text{VRH}}$ (Figure~\ref{fig:closed_loop_spacegroup_bm}) the dominance of low-symmetry systems persists, but cubic and hexagonal structures are over-represented relative to $C_p$, reflecting the well-known correlation between high bulk modulus and densely packed cubic/hexagonal lattices.
Splitting by policy (right panel) shows that the GP gate (ACC) does not bias the loop towards any particular crystal system: both BASE and ACC produce the same broad triclinic/monoclinic dominance, so the GP is screening on the property rather than on geometric proxies.

\FloatBarrier
\subsubsection{Per-Backbone Summary}
\label{sec:cl_per_model}

\begin{figure*}[!htb]
    \centering
    \includegraphics[width=0.85\textwidth]{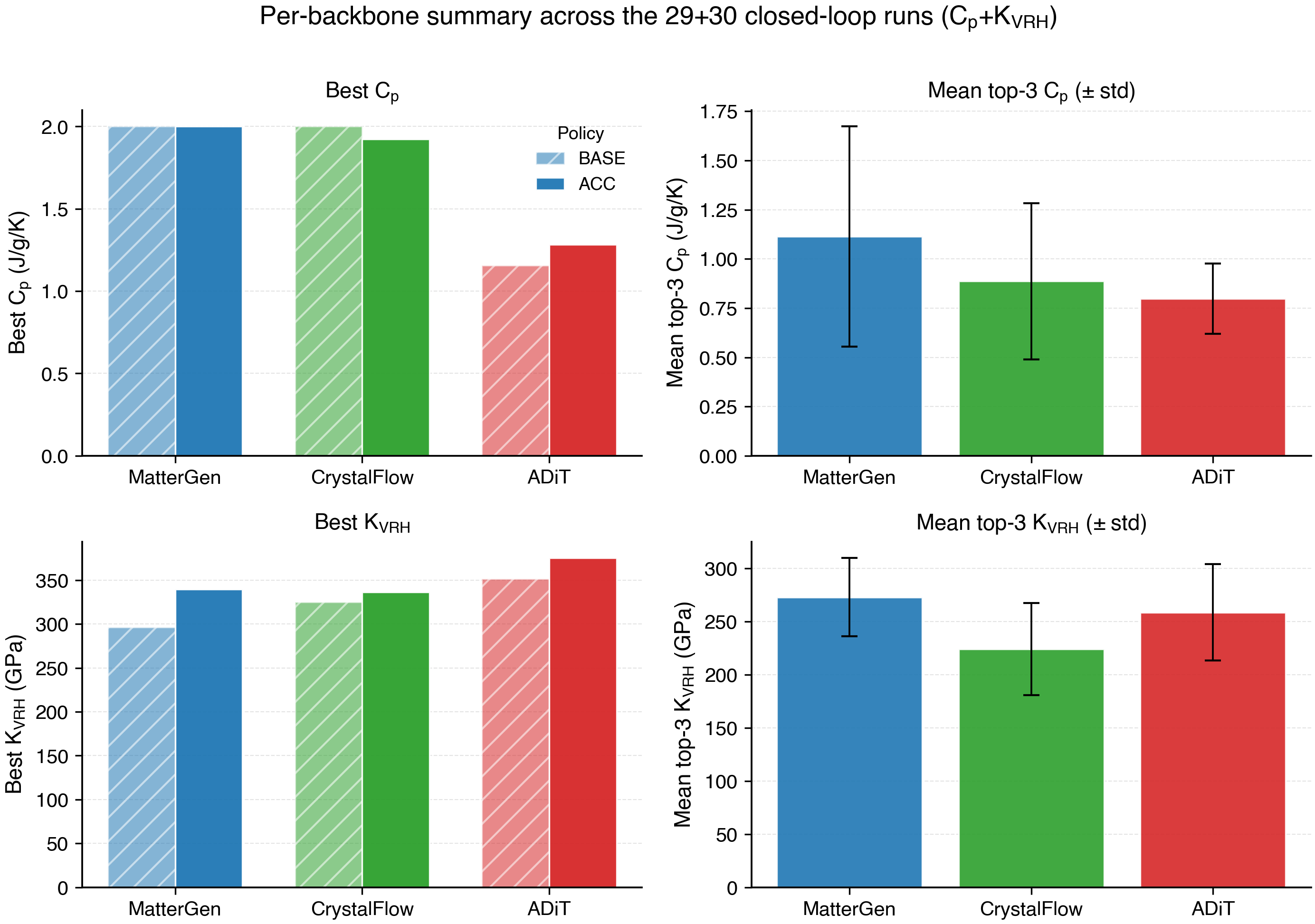}
    \caption{Per-backbone summary across the completed closed-loop runs (29 $C_p$, 30 $K_{\text{VRH}}$). Left column: best property value (hatched = BASE, solid = ACC). Right column: mean top-3 value $\pm$ std across runs. Top row: $C_p$. Bottom row: $K_{\text{VRH}}$. The best-value spread across backbones is small once seed variance is taken into account; ACC's lift over BASE is consistent but modest. Novelty rates against the warm-start seed pool are 100\% ($C_p$) and 96.7\% ($K_{\text{VRH}}$) for all three backbones (Section~\ref{sec:cl_novelty}).}
    \label{fig:closed_loop_per_model}
\end{figure*}

Figure~\ref{fig:closed_loop_per_model} consolidates the summary statistics per backbone across all completed runs (29 $C_p$, 30 $K_{\text{VRH}}$).
Best-value bars (left column) split BASE vs.\ ACC: MatterGen and CrystalFlow both hit the 2.0~J/g/K reward ceiling (oracle-imposed cap, attained on a few seeds) while ADiT plateaus at 1.28~J/g/K; for $K_{\text{VRH}}$, ADiT's 375~GPa hit pulls its best-value bar above MatterGen's 340 and CrystalFlow's 336~GPa.
Mean top-3 (right column) is the more robust comparator, since best-of-run is dominated by favorable seeds: on $C_p$ the three backbones land at 1.11~$\pm$~0.56, 0.89~$\pm$~0.40, and 0.80~$\pm$~0.18~J/g/K, differences smaller than the within-backbone seed spread.
On $K_{\text{VRH}}$ the corresponding means are 273~$\pm$~37, 224~$\pm$~43, and 259~$\pm$~45~GPa, again with overlapping error bars.
Novelty against the warm-start seed pool is uniform across backbones at 100\% for $C_p$ and 96.7\% for $K_{\text{VRH}}$, so we report it verbally rather than as a separate panel; details and the formula-level analysis are in Section~\ref{sec:cl_novelty}.
The ACC-vs-BASE lift on best-of-run is consistent in MatterGen and ADiT (Section~\ref{sec:cl_curves}) but never large enough to outweigh seed variance, so we report it as a trend rather than a measured difference.

\FloatBarrier
\subsubsection{Novelty vs.\ the Warm-Start Seed Pool}
\label{sec:cl_novelty}

A natural concern is that the loop is retrieving the structures we used to warm-start the GP rather than producing new ones.
We checked this with pymatgen's structure matcher~\cite{pymatgen2013} at default tolerances.
For $C_p$, none of the 87 top-3 structures (aggregated across the 29 completed runs) shares a reduced formula with the seed pool, and zero pass strict matching.
For $K_{\text{VRH}}$, 3 of 90 share a reduced formula but again zero pass strict matching.
The GP is acting as a property prior, not as memory.

\begin{figure*}[!htb]
    \centering
    \begin{subfigure}[t]{0.48\textwidth}
        \centering
        \includegraphics[width=\textwidth]{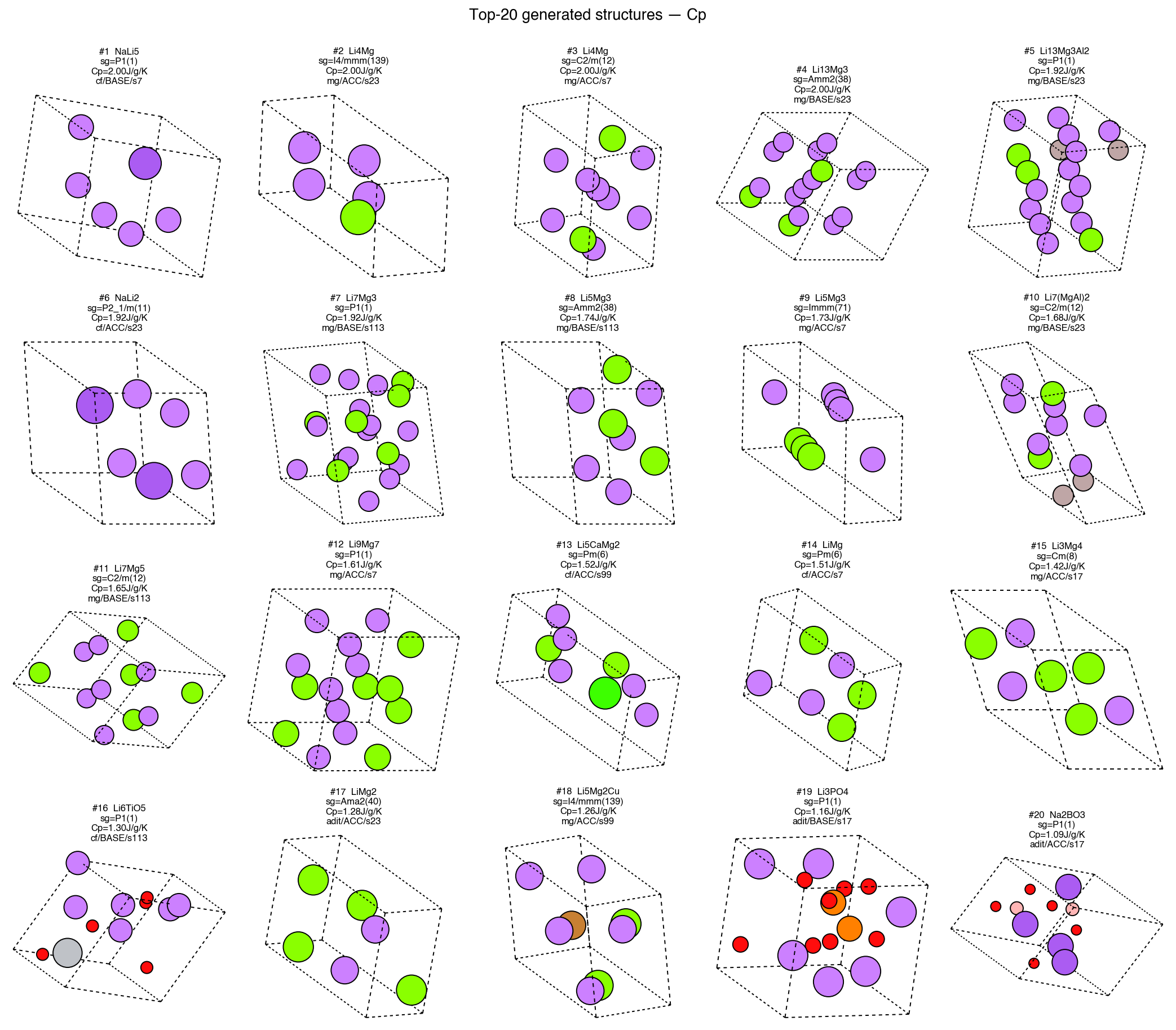}
        \caption{Heat capacity ($C_p$)}
    \end{subfigure}
    \hfill
    \begin{subfigure}[t]{0.48\textwidth}
        \centering
        \includegraphics[width=\textwidth]{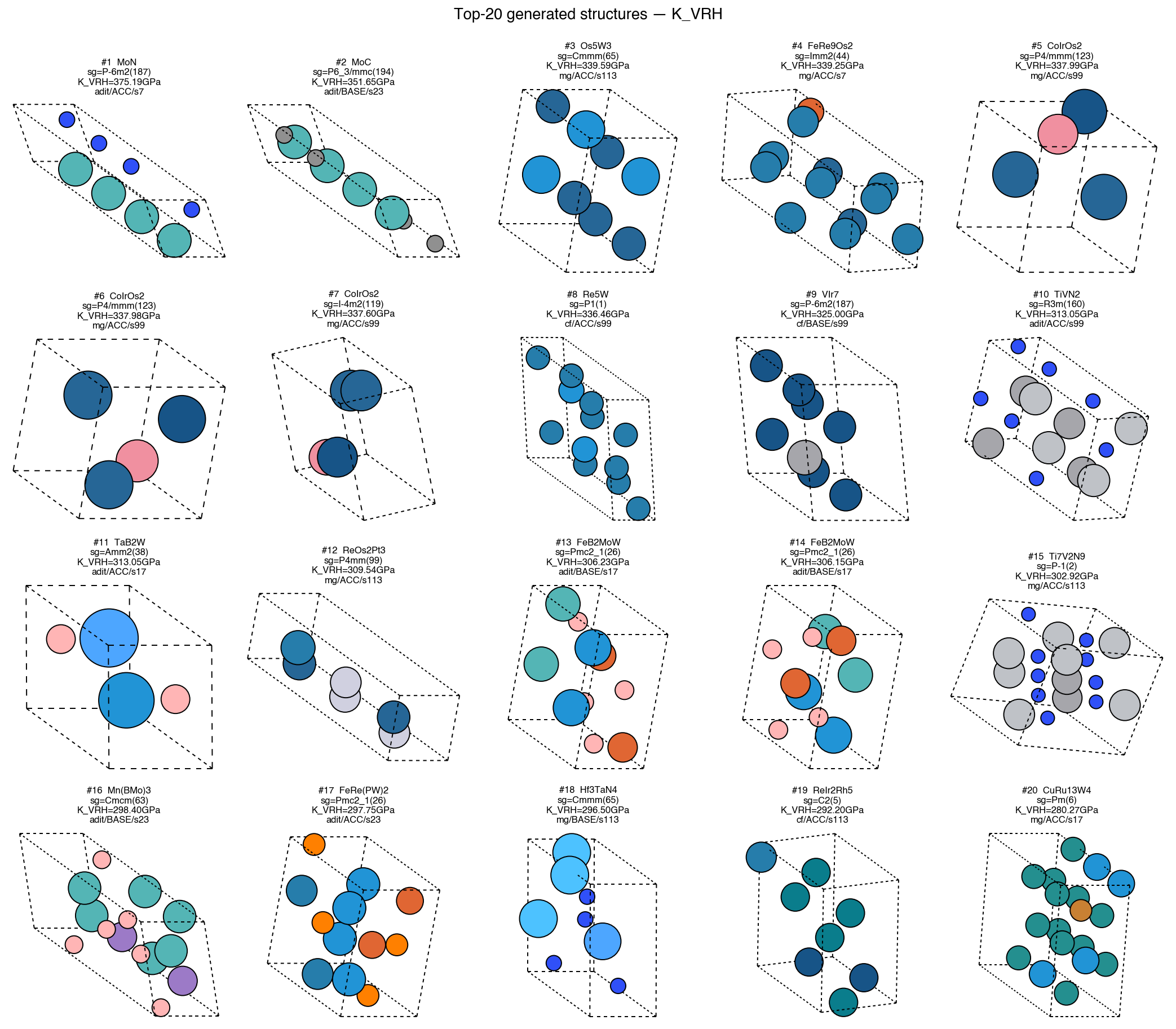}
        \caption{Bulk modulus ($K_{\text{VRH}}$)}
    \end{subfigure}
    \caption{Top-20 generated structures per target (4$\times$5 grid), ranked by the oracle. Border color: blue = MatterGen, green = CrystalFlow, red = ADiT. Each panel reports formula, space group, target value, and the originating (cycle, seed). None pass strict structure matching against the warm-start seed pool.}
    \label{fig:closed_loop_zoo}
\end{figure*}

Figure~\ref{fig:closed_loop_zoo} shows the top-20 structures per target, borders coded by backbone.
The compositions split along physical lines: light-element binaries and ternaries (Li, Mg, Na, O) for $C_p$, transition-metal carbides, nitrides, and borides (Mo, W, Os, Re) for $K_{\text{VRH}}$, roughly what each property's known physical drivers would predict.

\subsubsection{Synthesizability of Generated Structures}
\label{sec:cl_synth}

We score the top generated structures for synthesizability with two independent composition-level models (Section~\ref{sec:synth_methods}): an ORB-based positive-unlabeled classifier (ORB-PU) trained here, and the pretrained CGNF model~\cite{jang2024synth}. Neither model saw these structures in training, so this is a comparison of two scorers on equal, out-of-distribution footing, unlike the in-distribution test-split numbers in Supplementary Section~S6.

\begin{figure}[H]
    \centering
    \includegraphics[width=\columnwidth]{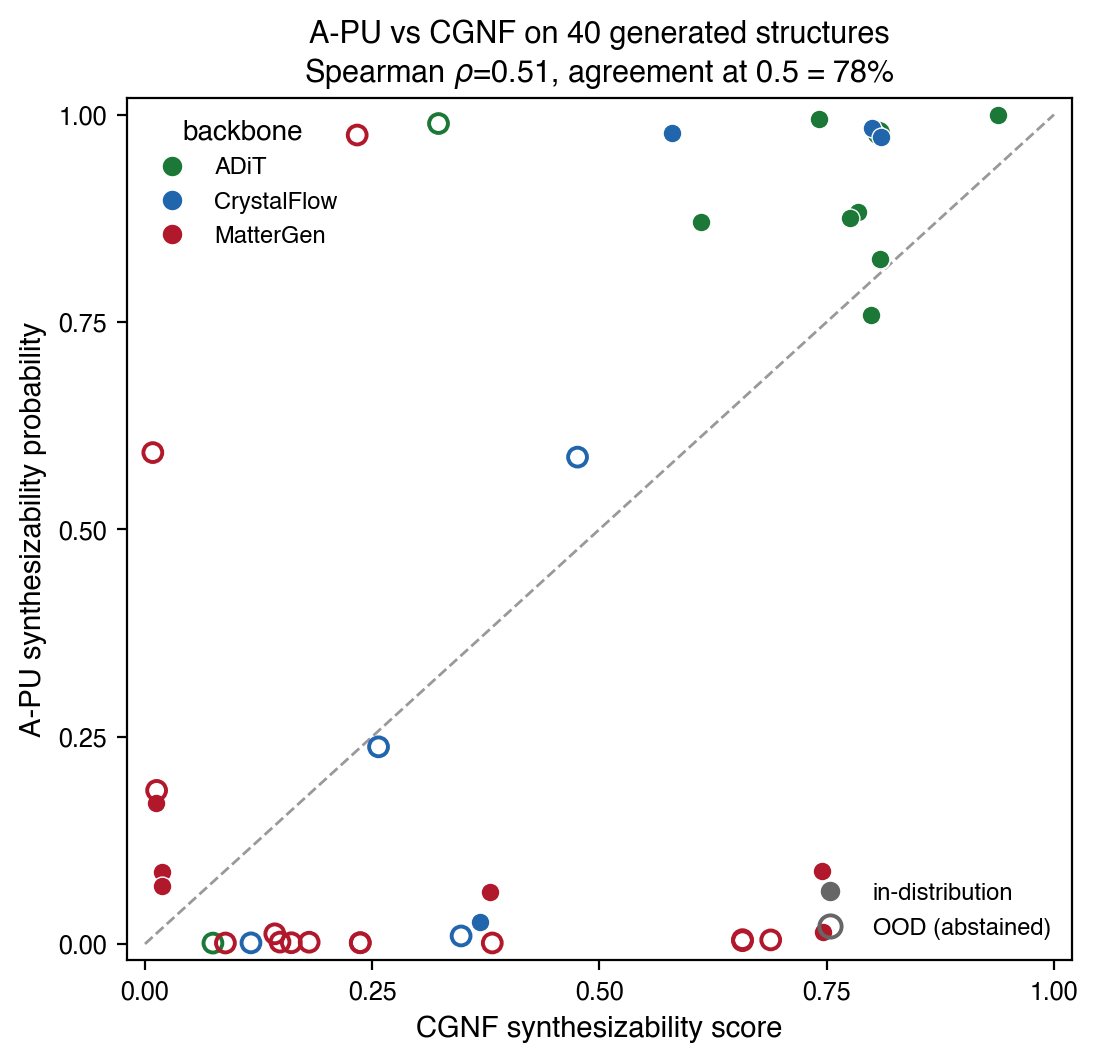}
    \caption{ORB-PU versus CGNF synthesizability scores for the 40 top generated structures, colored by backbone (green = ADiT, blue = CrystalFlow, red = MatterGen). Open markers are structures the ORB-PU model flags as out-of-distribution; the dashed line is parity. Rank agreement is moderate (Spearman $\rho = 0.51$; 78\% agreement at the 0.5 threshold).}
    \label{fig:synth_scatter}
\end{figure}

Across the 40 top structures the two scores agree moderately: Spearman $\rho = 0.51$ and 78\% agreement at the conventional 0.5 decision threshold (Figure~\ref{fig:synth_scatter}). They concur on conventional main-group chemistries, for example Na$_2$BO$_3$, Li$_3$PO$_4$, and TaB$_2$W, and diverge on lithium--magnesium binaries and several heavy intermetallics, where ORB-PU assigns low probabilities and flags out-of-distribution.

\begin{figure*}[!htb]
    \centering
    \includegraphics[width=\textwidth]{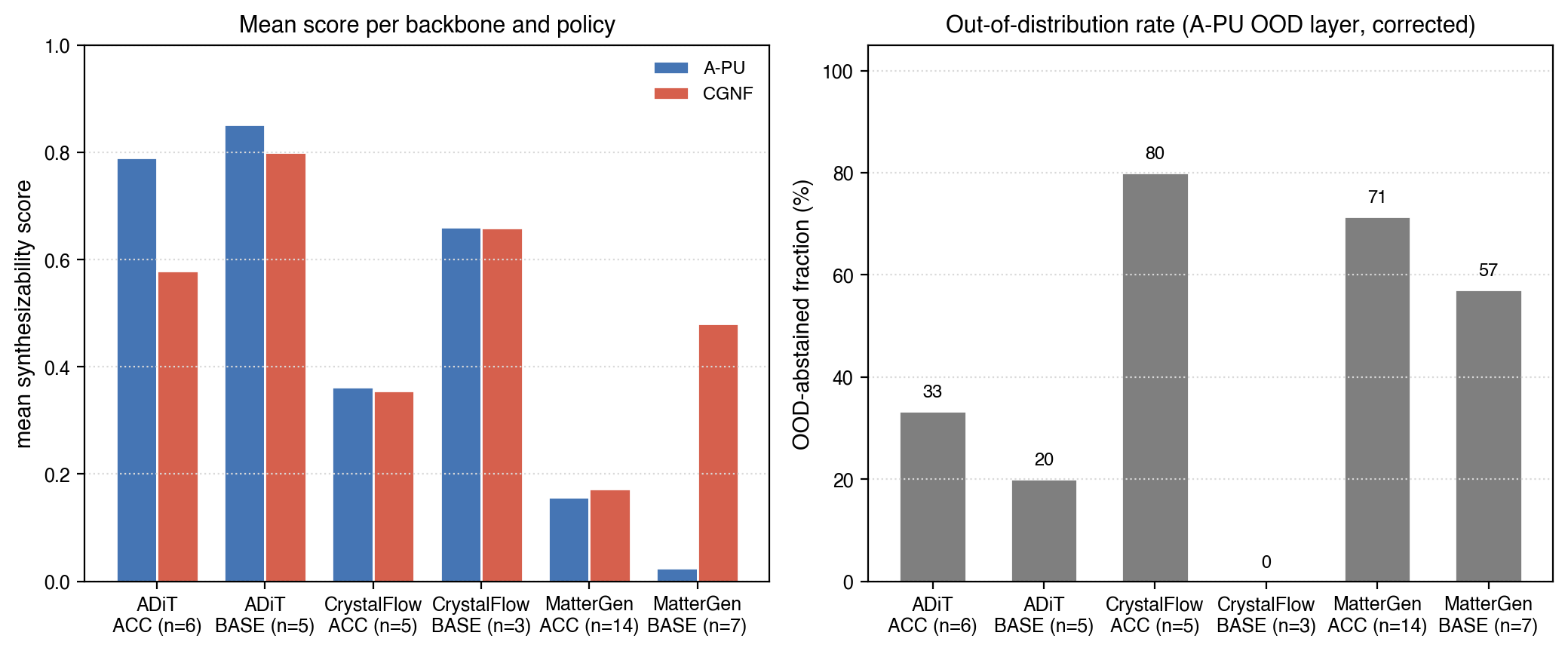}
    \caption{Per-(backbone, policy) synthesizability of the top generated structures ($n = 6, 5, 5, 3, 14, 7$ for ADiT-ACC, ADiT-BASE, CrystalFlow-ACC, CrystalFlow-BASE, MatterGen-ACC, MatterGen-BASE). Left: mean ORB-PU and CGNF scores. Right: fraction flagged out-of-distribution by the ORB-PU criterion. MatterGen's top compositions are predominantly lithium alloys far from the training distribution, giving it among the highest out-of-distribution rates.}
    \label{fig:synth_per_backbone}
\end{figure*}

The ORB-PU out-of-distribution criterion flags 21 of 40 structures (52.5\%), almost all by distance rather than low confidence or bag disagreement. The flagged fraction varies by backbone (Figure~\ref{fig:synth_per_backbone}): it is lowest for the BASE policies of CrystalFlow and ADiT and highest for the ACC runs of CrystalFlow and MatterGen, the latter's top compositions being predominantly lithium alloys that sit far from the Materials-Project training distribution. Because roughly half the set is out-of-distribution for both models, the absolute scores on generated structures are extrapolative and should be interpreted jointly with the per-backbone OOD rate, which itself measures how far each generator's outputs sit from known compositions. Synthesizability here is a composition-level, post-hoc check rather than an in-loop gate.

\FloatBarrier
\subsubsection{DFT Validation of the Bulk-Modulus Discoveries}
\label{sec:cl_dft}

The closed-loop oracle and surrogate are both learned models, so two questions sit underneath every $K_{\text{VRH}}$ result and neither is answered by the loop's own logs: is the eSEN oracle a faithful stand-in for DFT on the structures it selects, and does the GP predict the property of \emph{freshly generated} structures rather than reproducing values it has already seen? We answer both for the bulk-modulus target with an oracle-parity DFT recomputation of the validated top structures and a causal replay of the surrogate (Section~\ref{sec:dft_methods}).

\begin{figure*}[!htb]
    \centering
    \includegraphics[width=\textwidth]{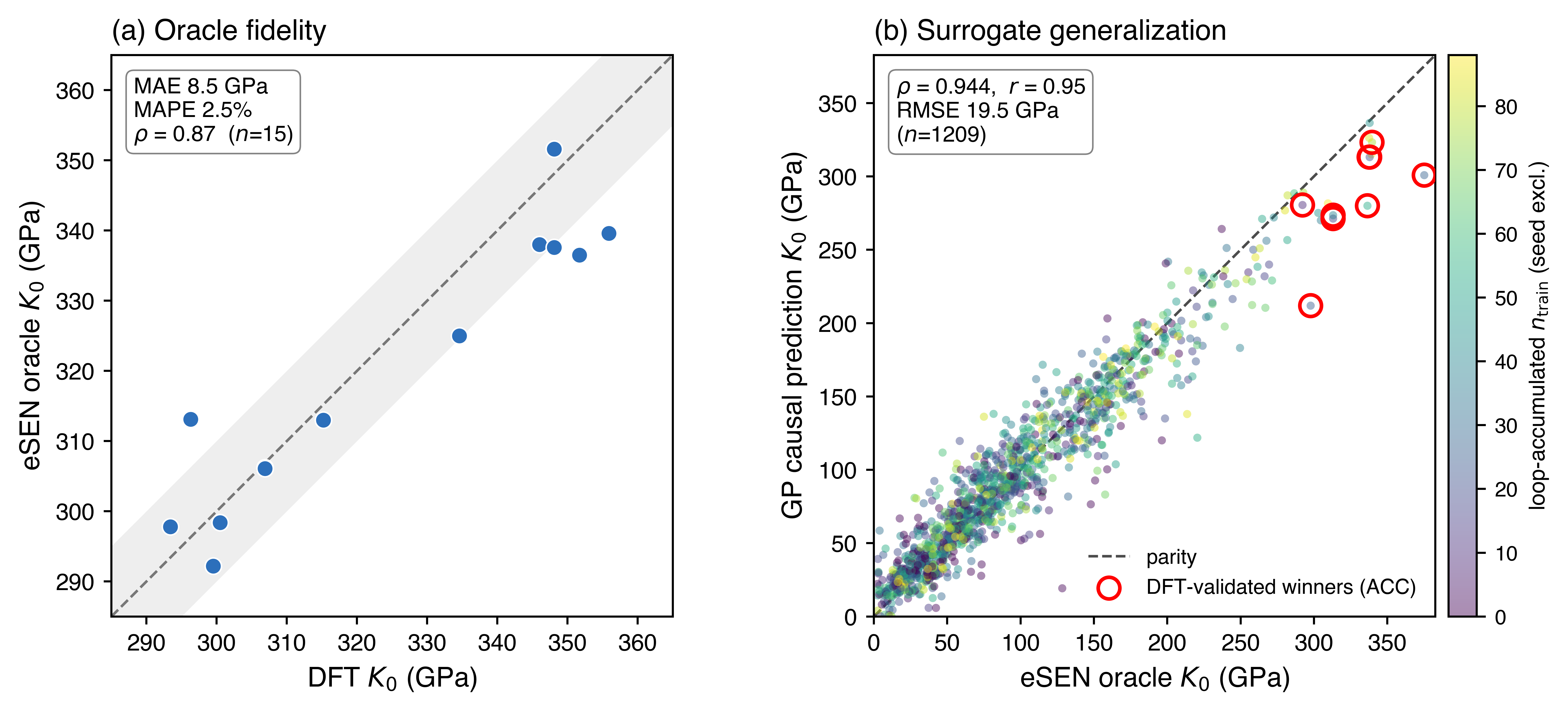}
    \caption{DFT validation of the bulk-modulus discoveries. (a)~Oracle fidelity: the eSEN oracle $K_0$ against ground-truth oracle-parity DFT $K_0$ for the 15 validated structures (MAE 8.5~GPa, MAPE 2.5\%, Spearman $\rho = 0.87$); the shaded band is $\pm 10$~GPa. (b)~Surrogate generalization: causal step-ahead GP predictions against the eSEN oracle over all 1{,}209 freshly generated structures, colored by the loop-accumulated training count (warm-start seed excluded), with the DFT-validated winners circled in red. The GP ranks the unseen candidates with $\rho = 0.944$ while modestly underpredicting the hardest extremes.}
    \label{fig:dft_bm}
\end{figure*}

\textbf{The oracle is a faithful DFT proxy.}
Across the 15 bulk-modulus structures with completed DFT, the eSEN oracle tracks ground-truth DFT closely (Figure~\ref{fig:dft_bm}a): mean absolute error 8.5~GPa, mean absolute percentage error 2.5\%, Spearman $\rho = 0.87$, and a small bias of $-2.3$~GPa, with every point within or near a $\pm 10$~GPa band of parity (per-structure values in Supplementary Table~S6). The largest single disagreements are MoN ($+21.6$~GPa) and TiVN$_2$ ($+16.8$~GPa); the rest agree to within roughly 10~GPa. Because the validated structures are all hard refractory phases spanning a narrow 293--356~GPa range, the correlation is restricted-range and the small percentage error is the more meaningful statistic. This licenses using eSEN as the loop's reward and DFT as a post-hoc check rather than an in-loop calculator.

\textbf{The surrogate generalizes as a ranker.}
Replaying the GP causally---at each cycle training only on earlier structures plus the warm-start seed and predicting the freshly generated candidates---the surrogate orders the unseen structures well, with a pooled Spearman $\rho = 0.944$ over 1{,}209 step-ahead predictions (Figure~\ref{fig:dft_bm}b). This is a direct measurement of the generalization that the in-loop cross-validated diagnostic of Section~\ref{sec:cl_surrogate} could not provide: it is computed on each cycle's fresh proposals before they reach the gate, not on the accumulated training set. The ranking is already strong in the first cycles and rises only modestly as loop-gathered data accumulate, so most of the surrogate's discriminative power comes from the warm start rather than from in-loop labels.

\textbf{The surrogate underpredicts the rare extremes.}
On the DFT-validated winners the GP modestly underpredicts the absolute bulk modulus of the hardest structures (Table~\ref{tab:dft_bm_winners}), pulling the top extremes toward the training bulk by 42.5~GPa on average, with its posterior standard deviation ($\sigma \approx 13$--$32$~GPa) flagging the shortfall. The effect is real but bounded: the surrogate's contribution is uncertainty-aware \emph{ordering}, not absolute prediction at the extremes---the rank-not-regress regime in which the gate operates.

\begin{table}[H]
    \centering
    \caption{DFT-validated bulk-modulus winners: ground-truth DFT $K_0$, the eSEN oracle value, and the causal GP prediction with its posterior standard deviation (all in GPa). The oracle tracks DFT to a few percent; the GP ranks the structures well (Figure~\ref{fig:dft_bm}b) but underpredicts the hardest by tens of GPa.}
    \label{tab:dft_bm_winners}
    \small
    \begin{tabular}{@{}lccc@{}}
        \toprule
        \textbf{Structure} & \textbf{DFT $K_0$} & \textbf{eSEN $K_0$} & \textbf{GP $K_0$ ($\pm\sigma$)} \\
        \midrule
        MoN            & 354 & 375 & 301 ($\pm$26) \\
        Os$_5$W$_3$    & 356 & 340 & 323 ($\pm$21) \\
        CoIrOs$_2$     & 347 & 338 & 313 ($\pm$16) \\
        Re$_5$W        & 352 & 336 & 280 ($\pm$32) \\
        TiVN$_2$       & 296 & 313 & 274 ($\pm$17) \\
        TaB$_2$W       & 315 & 313 & 271 ($\pm$18) \\
        FeRe(PW)$_2$   & 293 & 298 & 212 ($\pm$13) \\
        ReIr$_2$Rh$_5$ & 299 & 292 & 280 ($\pm$13) \\
        \bottomrule
    \end{tabular}
\end{table}

\FloatBarrier
\subsection{Static benchmark: foundation-model embeddings as surrogate features}
\label{sec:foundation}

The static benchmark evaluates four featurizers (SOAP, MACE, ORB, UMA/eSEN) $\times$ three probabilistic surrogates (GP, MTGP, DGP) $\times$ three PCA dimensionalities $\times$ three training-set sizes across three property datasets, 8{,}640 cross-validated fits in total.
It identifies the default stack used in the loop (ORB\,+\,GP\,+\,PCA\,=\,50) and reports both regression accuracy (\Rtwo{}, RMSE) and ranking quality (Spearman $\rho$), the latter because acquisition in the loop selects candidates by their relative order.
Unless stated otherwise, benchmark numbers refer to $\ntrain = 500$.

\subsection{Featurizer ranking across datasets}
\label{sec:featurizer_ranking}

\begin{figure*}[!htb]
    \centering
    \begin{subfigure}[t]{0.32\textwidth}
        \centering
        \includegraphics[width=\textwidth]{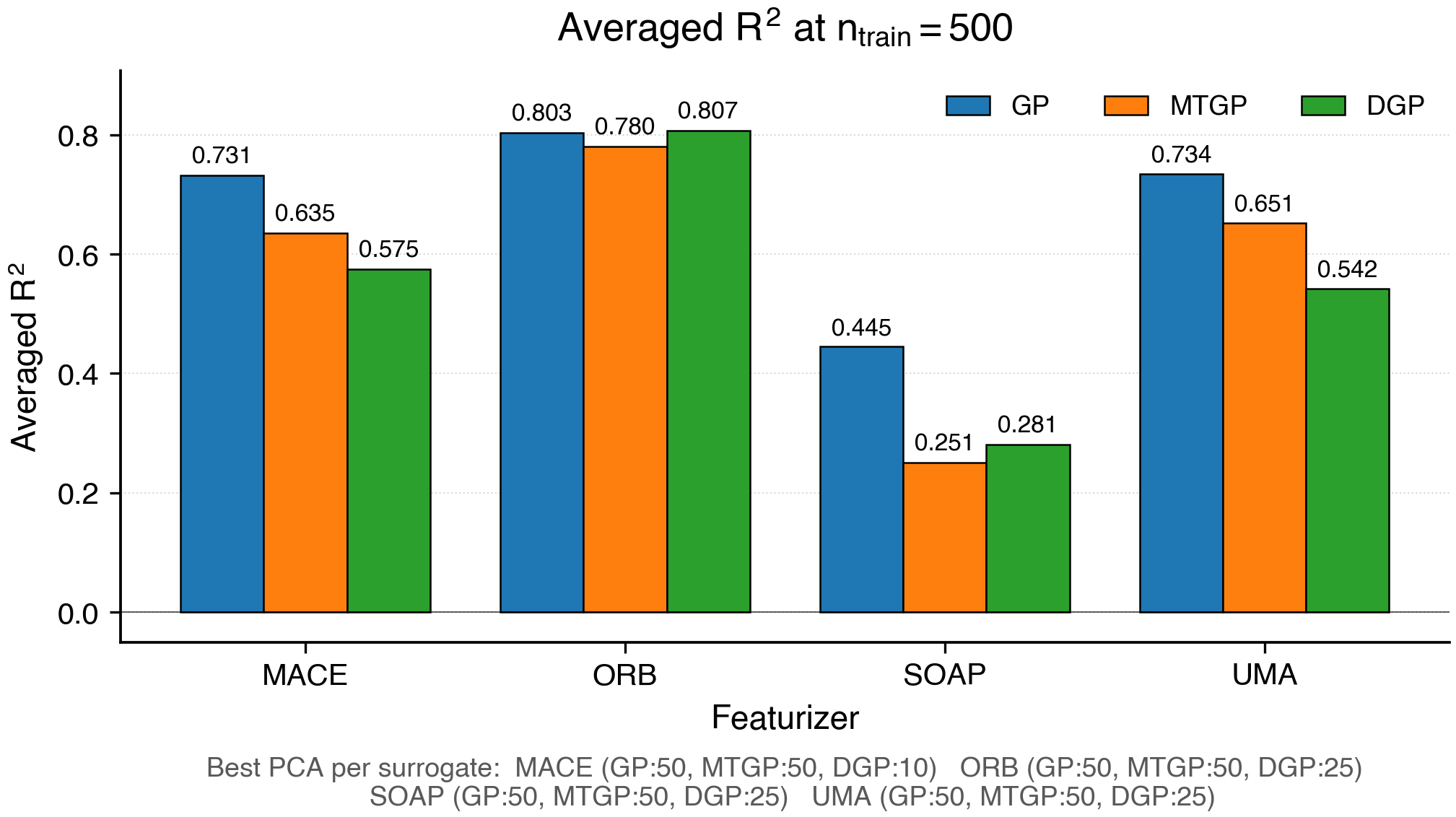}
        \caption{Elastic tensor (8 properties)}
        \label{fig:bar_elastic}
    \end{subfigure}
    \hfill
    \begin{subfigure}[t]{0.32\textwidth}
        \centering
        \includegraphics[width=\textwidth]{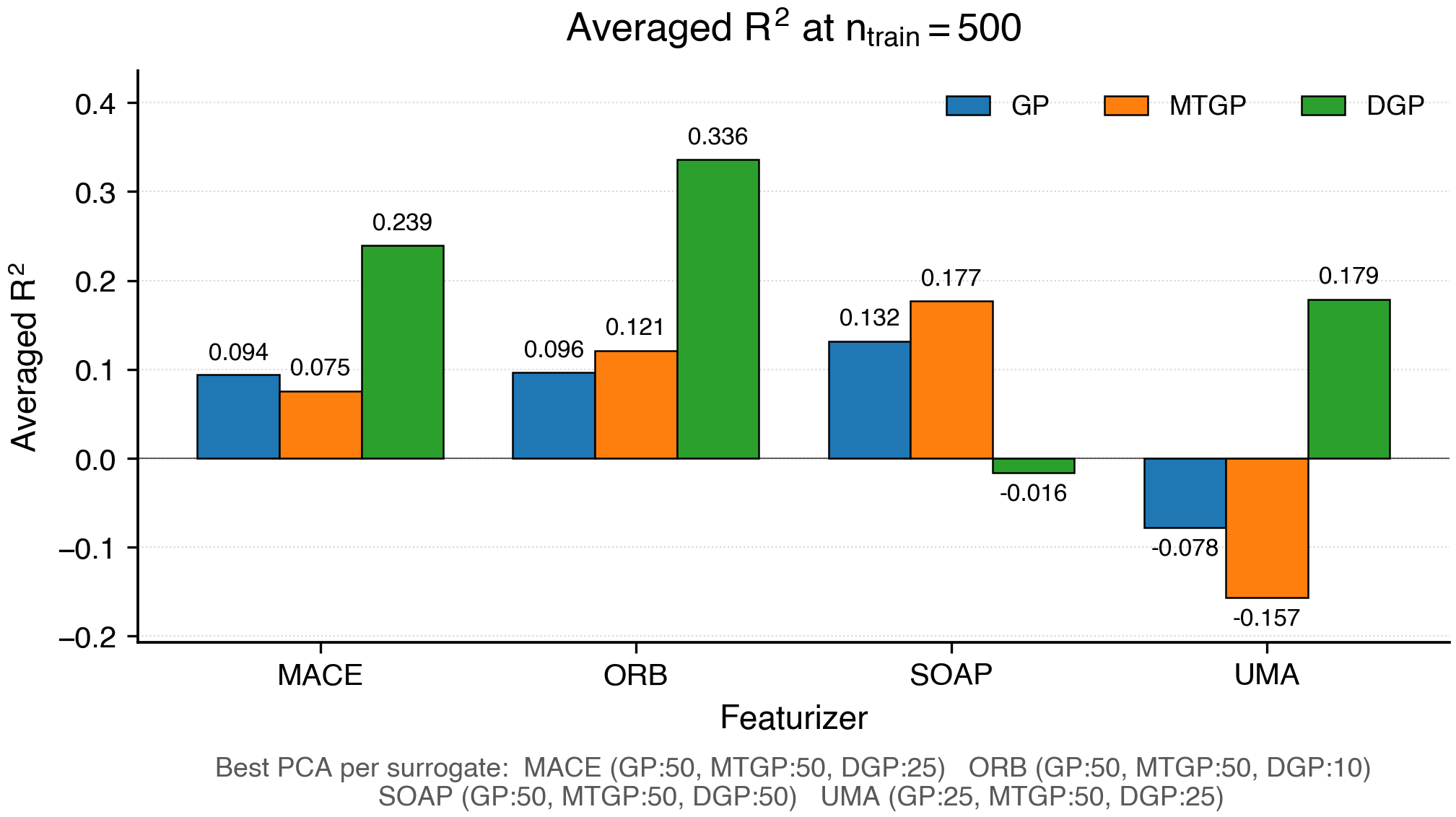}
        \caption{Dielectric constant (4 properties)}
        \label{fig:bar_dielectric}
    \end{subfigure}
    \hfill
    \begin{subfigure}[t]{0.32\textwidth}
        \centering
        \includegraphics[width=\textwidth]{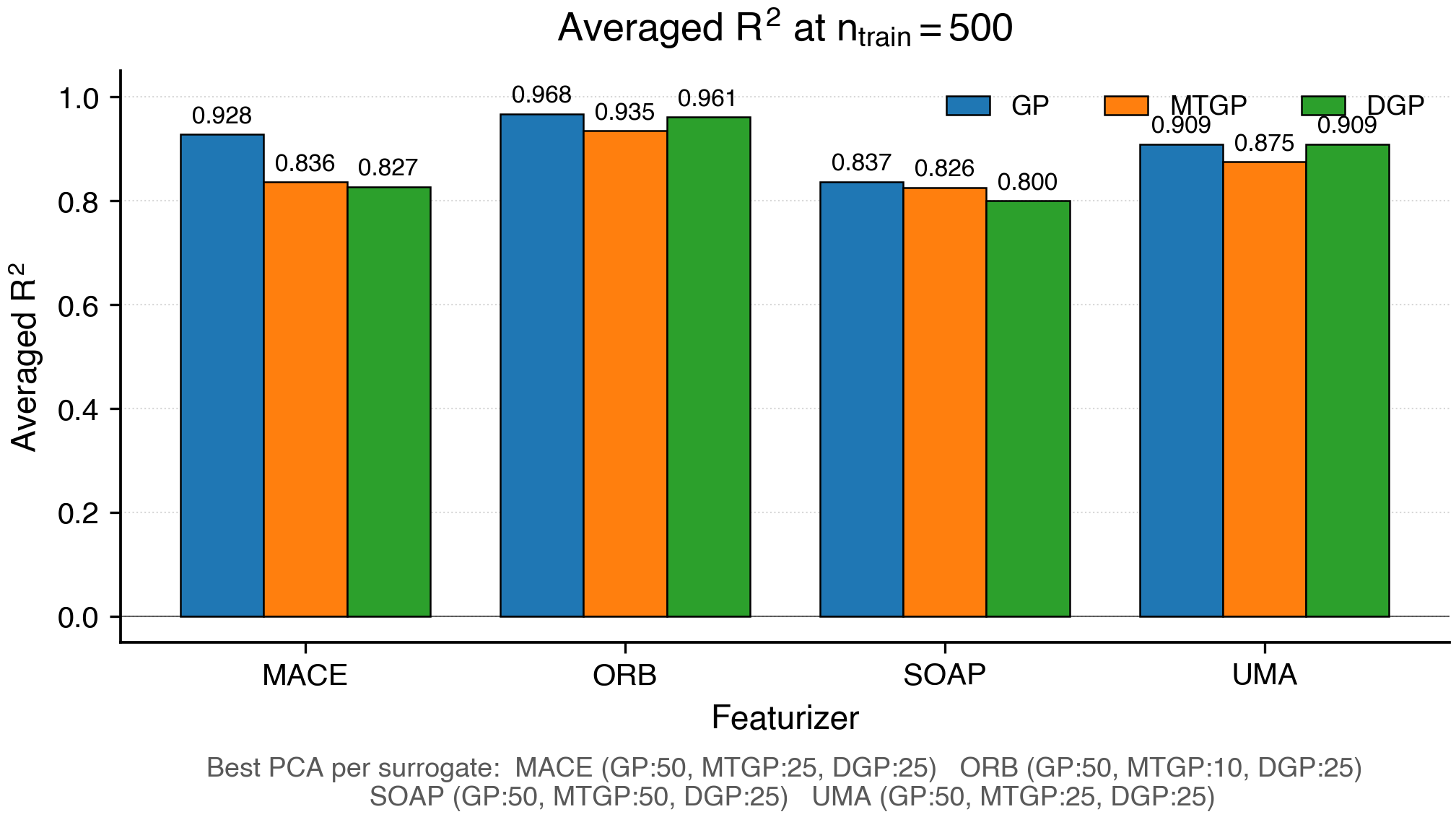}
        \caption{Phonon thermodynamics (4 properties)}
        \label{fig:bar_phonon}
    \end{subfigure}
    \caption{Averaged \Rtwo{} across all target properties at $\ntrain = 500$, grouped by featurizer and surrogate model, for each of the three benchmark datasets. ORB embeddings rank first in most featurizer--surrogate cells; SOAP edges ORB on the dielectric dataset under GP and MTGP. Note the different $y$-axis scales, which reflect the different \Rtwo{} ranges across datasets.}
    \label{fig:bar_r2_all}
\end{figure*}

Figure~\ref{fig:bar_r2_all} summarizes averaged \Rtwo{} for every featurizer--surrogate combination at $\ntrain = 500$ on the three datasets.
Pre-trained ORB ranks first on nearly every (dataset, surrogate) pair, with the ordering ORB $>$ UMA $\approx$ MACE $\gg$ SOAP recurring across surrogates.
Tables~\ref{tab:best_configs_elastic}, \ref{tab:best_configs_dielectric}, and~\ref{tab:best_configs_phonon} report the best configuration per pair on the elastic, dielectric, and phonon datasets.
On the elastic dataset the best configuration is ORB\,+\,DGP at PCA\,=\,25 ($\bar{R}^2 = 0.807$), with ORB\,+\,GP at PCA\,=\,50 a close and far more stable second ($0.803$).
Accuracy drops on the electronic dielectric dataset ($\bar{R}^2 \leq 0.34$), where SOAP becomes competitive under GP and MTGP, which suggests its explicit local-environment encoding carries dielectric-relevant information that the energy-trained embeddings emphasize less.

\begin{table*}[!htb]
    \centering
    \caption{Best configuration per featurizer--surrogate pair on the elastic tensor dataset ($\ntrain = 500$). For each combination, we report the PCA dimensionality that maximizes averaged \Rtwo{} across all 8~properties, together with the corresponding averaged RMSE and Spearman~$\rho$.}
    \label{tab:best_configs_elastic}
    \small
    \begin{tabular}{@{}llcccc@{}}
        \toprule
        \textbf{Surrogate} & \textbf{Featurizer} & \textbf{Best PCA} & \textbf{Avg.\ \Rtwo{}} & \textbf{Avg.\ RMSE} & \textbf{Avg.\ Spearman} \\
        \midrule
        \multirow{4}{*}{GP}
        & ORB     & 50 & \textbf{0.803} & \textbf{17.37} & \textbf{0.842} \\
        & UMA     & 50 & 0.734 & 21.42 & 0.811 \\
        & MACE    & 50 & 0.731 & 20.46 & 0.820 \\
        & SOAP    & 50 & 0.445 & 33.63 & 0.688 \\
        \midrule
        \multirow{4}{*}{MTGP}
        & ORB     & 50 & \textbf{0.780} & \textbf{19.30} & \textbf{0.843} \\
        & UMA     & 50 & 0.651 & 25.80 & 0.780 \\
        & MACE    & 50 & 0.635 & 25.76 & 0.758 \\
        & SOAP    & 50 & 0.251 & 41.34 & 0.529 \\
        \midrule
        \multirow{4}{*}{DGP}
        & ORB     & 25 & \textbf{0.807} & \textbf{17.84} & \textbf{0.826} \\
        & MACE    & 10 & 0.575 & 25.53 & 0.659 \\
        & UMA     & 25 & 0.542 & 26.80 & 0.696 \\
        & SOAP    & 25 & 0.281 & 38.78 & 0.521 \\
        \bottomrule
    \end{tabular}
\end{table*}

\begin{table}[H]
    \centering
    \caption{Best configuration per featurizer--surrogate pair on the dielectric constant dataset ($\ntrain = 500$).}
    \label{tab:best_configs_dielectric}
    \footnotesize
    \begin{tabular}{@{}llcccc@{}}
        \toprule
        \textbf{Surr.} & \textbf{Feat.} & \textbf{PCA} & \textbf{\Rtwo{}} & \textbf{RMSE} & \textbf{$\rho$} \\
        \midrule
        \multirow{4}{*}{GP}
        & SOAP    & 50 & 0.132 & 7.60 & 0.549 \\
        & ORB     & 50 & 0.096 & 7.84 & \textbf{0.746} \\
        & MACE    & 50 & 0.094 & 7.94 & 0.695 \\
        & UMA     & 25 & $-$0.078 & 8.14 & 0.609 \\
        \midrule
        \multirow{4}{*}{MTGP}
        & SOAP    & 50 & \textbf{0.177} & \textbf{7.27} & 0.583 \\
        & ORB     & 50 & 0.121 & 7.85 & \textbf{0.773} \\
        & MACE    & 50 & 0.075 & 7.90 & 0.715 \\
        & UMA     & 50 & $-$0.157 & 8.49 & 0.656 \\
        \midrule
        \multirow{4}{*}{DGP}
        & ORB     & 10 & \textbf{0.336} & \textbf{6.88} & \textbf{0.778} \\
        & MACE    & 25 & 0.239 & 7.31 & 0.754 \\
        & UMA     & 25 & 0.179 & 7.35 & 0.742 \\
        & SOAP    & 50 & $-$0.016 & 7.46 & 0.019 \\
        \bottomrule
    \end{tabular}
\end{table}

\begin{table}[H]
    \centering
    \caption{Best configuration per featurizer--surrogate pair on the phonon thermodynamics dataset ($\ntrain = 500$). For each combination, we report the PCA dimensionality that maximizes averaged \Rtwo{} across all 4~properties, together with the corresponding averaged RMSE and Spearman~$\rho$. The averaged RMSE is dominated by the free-energy term (J\,mol-atom$^{-1}$), as the four targets carry differing units.}
    \label{tab:best_configs_phonon}
    \footnotesize
    \begin{tabular}{@{}llcccc@{}}
        \toprule
        \textbf{Surr.} & \textbf{Feat.} & \textbf{PCA} & \textbf{\Rtwo{}} & \textbf{RMSE} & \textbf{$\rho$} \\
        \midrule
        \multirow{4}{*}{GP}
        & ORB     & 50 & \textbf{0.968} & \textbf{159.0} & \textbf{0.986} \\
        & MACE    & 50 & 0.928 & 257.6 & 0.969 \\
        & UMA     & 50 & 0.909 & 348.9 & 0.952 \\
        & SOAP    & 50 & 0.837 & 405.0 & 0.954 \\
        \midrule
        \multirow{4}{*}{MTGP}
        & ORB     & 10 & \textbf{0.935} & \textbf{238.7} & \textbf{0.967} \\
        & UMA     & 25 & 0.875 & 375.2 & 0.938 \\
        & MACE    & 25 & 0.836 & 411.6 & 0.932 \\
        & SOAP    & 50 & 0.826 & 439.3 & 0.948 \\
        \midrule
        \multirow{4}{*}{DGP}
        & ORB     & 25 & \textbf{0.961} & \textbf{197.3} & \textbf{0.982} \\
        & UMA     & 25 & 0.909 & 337.5 & 0.949 \\
        & MACE    & 25 & 0.827 & 397.6 & 0.934 \\
        & SOAP    & 25 & 0.800 & 473.4 & 0.921 \\
        \bottomrule
    \end{tabular}
\end{table}

Per-configuration heatmaps and PCA-sensitivity curves for all three datasets are in the Supplementary Information (Supplementary Figs.~S6 and~S7).
Two patterns from those panels matter for the loop: the GP is the most stable surrogate at PCA\,=\,50, and the DGP, although it reaches the best \Rtwo{} at PCA\,$\leq$\,25, collapses at PCA\,=\,50 on all three datasets, which is why we do not adopt it as a default. The collapse of DGP at high-dimensional PCA regime can be attributed to the fact that, the learnable parameter count for DGP explodes with higher feature dimension. Hence, DGP performs poorly in small training data regime.

\FloatBarrier
\subsection{Data efficiency and property difficulty}
\label{sec:learning_curves}

\begin{figure}[H]
    \centering
    \includegraphics[width=\columnwidth]{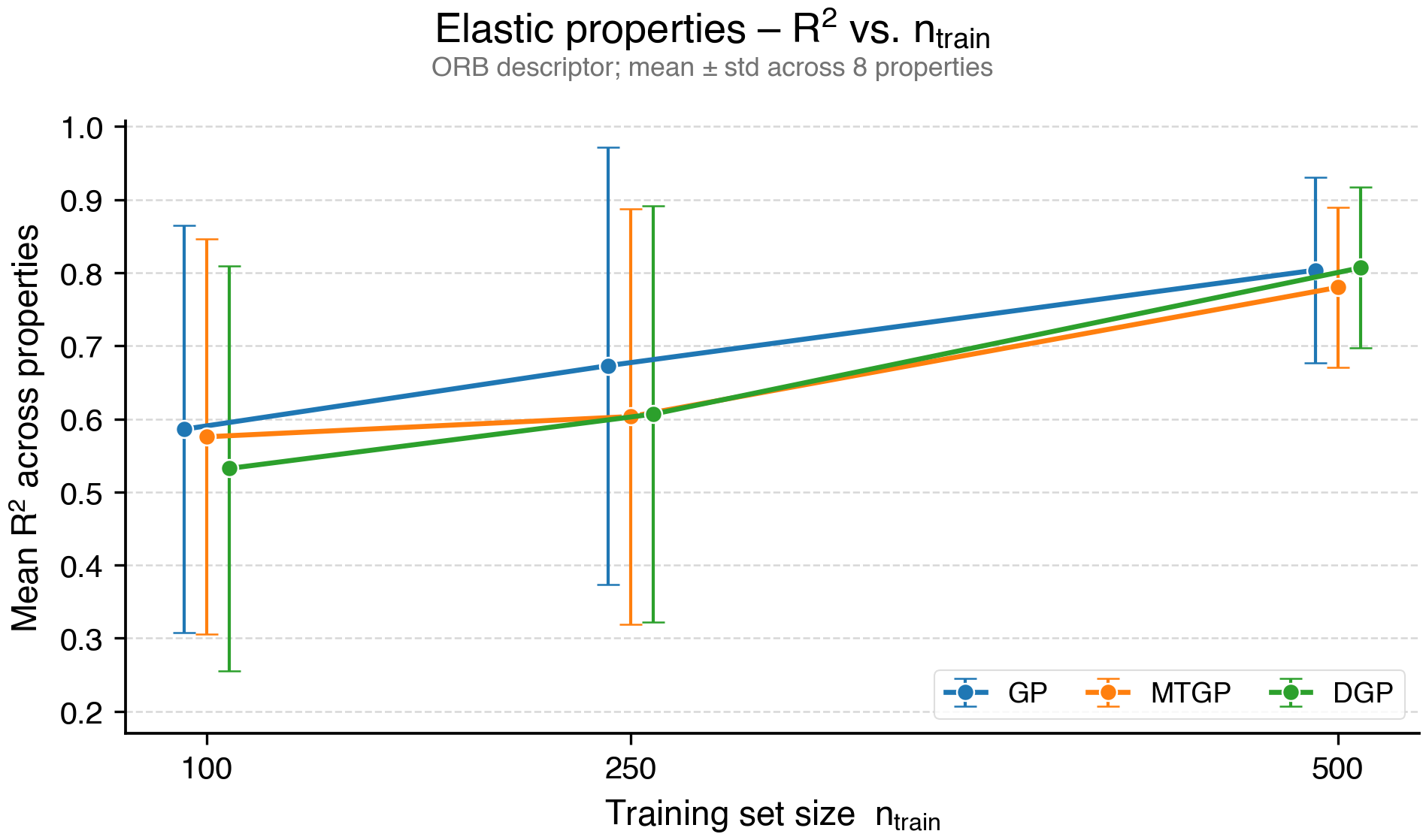}
    \caption{Learning curves showing averaged \Rtwo{} as a function of training set size for the GP, MTGP, and DGP surrogates on ORB embeddings, elastic tensor dataset. ORB\,+\,GP at $\ntrain = 100$ approaches the performance of SOAP\,+\,GP at $\ntrain = 500$ (Table~\ref{tab:best_configs_elastic}).}
    \label{fig:learning_curves}
\end{figure}

ORB embeddings are data-efficient.
On the elastic dataset, ORB\,+\,GP at $\ntrain = 100$ already reaches $\bar{R}^2 \approx 0.59$, above SOAP\,+\,GP at $\ntrain = 500$ ($0.445$), so the pre-trained representation is worth roughly 400 additional training samples (Figure~\ref{fig:learning_curves}, Table~\ref{tab:learning_curves}).
The DGP shows the largest relative gain with added data but starts lowest and needs about 500 samples to match the GP, so in the low-data regime at the start of a campaign the GP is the safer choice.

\begin{table}[H]
    \centering
    \caption{Learning curve summary for ORB embeddings across all three surrogates on the elastic tensor dataset. \Rtwo{} values are averaged across all 8~elastic properties at the best PCA setting for each configuration.}
    \label{tab:learning_curves}
    \small
    \begin{tabular}{@{}lcccc@{}}
        \toprule
        \textbf{Surrogate} & $\ntrain{=}100$ & $\ntrain{=}250$ & $\ntrain{=}500$ & \textbf{Gain} \\
        \midrule
        GP  & 0.586 & 0.673 & 0.803 & +37\% \\
        MTGP & 0.576 & 0.603 & 0.780 & +36\% \\
        DGP  & 0.532 & 0.607 & 0.807 & +52\% \\
        \bottomrule
    \end{tabular}
\end{table}

\begin{figure*}[!htb]
    \centering
    \begin{subfigure}[t]{0.32\textwidth}
        \centering
        \includegraphics[width=\textwidth]{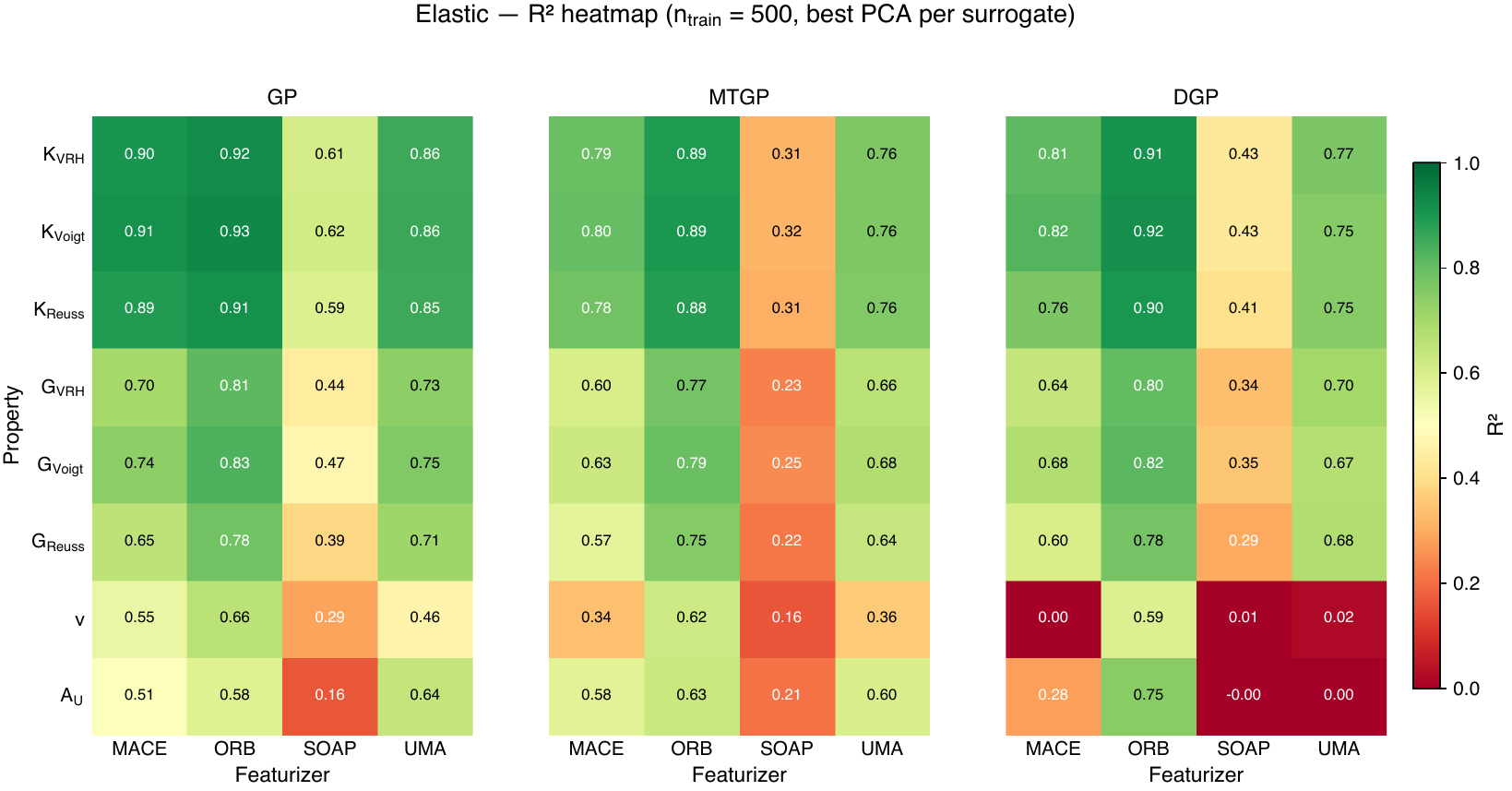}
        \caption{Elastic tensor}
        \label{fig:diff_elastic}
    \end{subfigure}
    \hfill
    \begin{subfigure}[t]{0.32\textwidth}
        \centering
        \includegraphics[width=\textwidth]{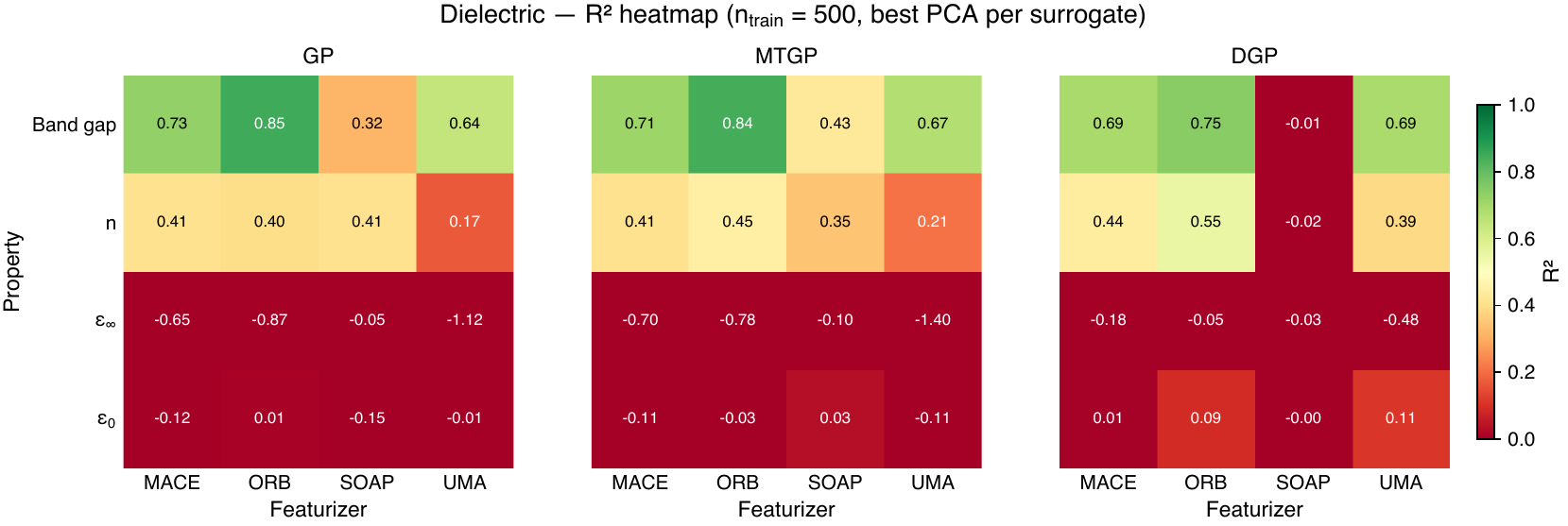}
        \caption{Dielectric constant}
        \label{fig:diff_dielectric}
    \end{subfigure}
    \hfill
    \begin{subfigure}[t]{0.32\textwidth}
        \centering
        \includegraphics[width=\textwidth]{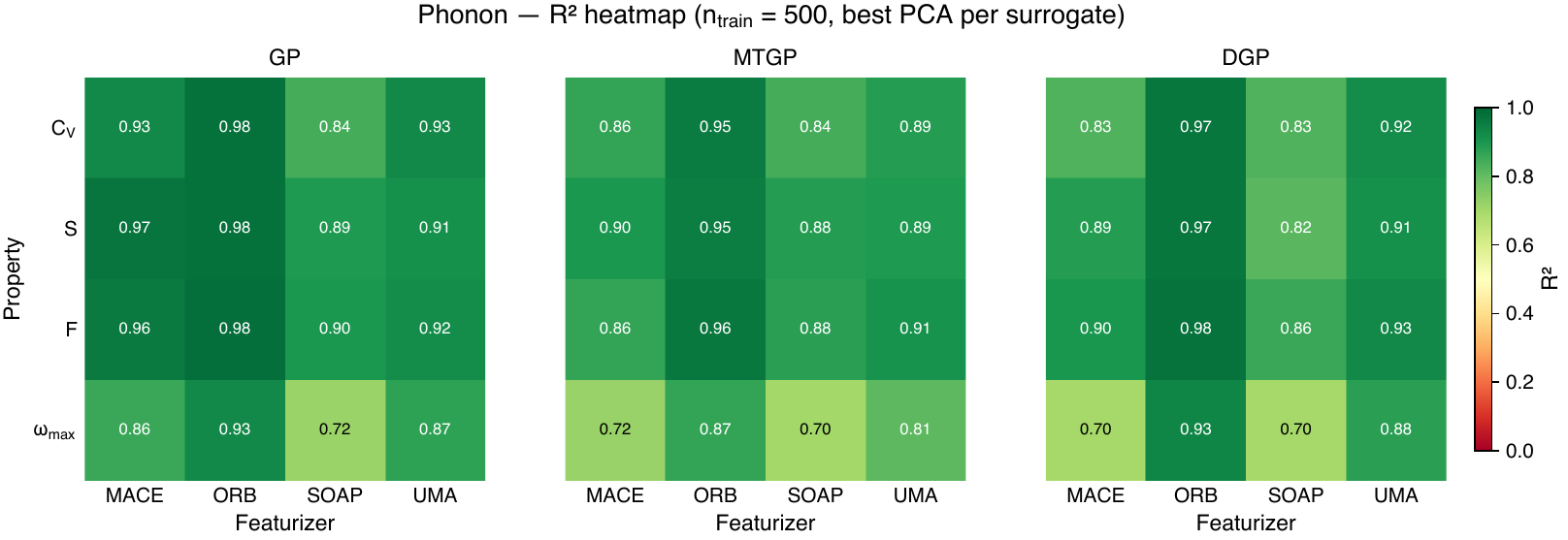}
        \caption{Phonon thermodynamics}
        \label{fig:diff_phonon}
    \end{subfigure}
    \caption{Property difficulty matrices showing the best-achievable \Rtwo{} per property across all featurizer--PCA combinations at $\ntrain = 500$, broken down by surrogate model. (a)~Elastic: bulk modulus variants are easiest; anisotropy and Poisson's ratio are hardest. (b)~Dielectric: band gap is easiest; polycrystalline dielectric constants are hardest. (c)~Phonon thermodynamics: heat capacity, entropy, free energy, and the maximum phonon frequency at 300\,K are all well predicted.}
    \label{fig:difficulty_all}
\end{figure*}

The three datasets form a difficulty gradient (Figure~\ref{fig:difficulty_all}).
Elastic moduli are the easiest: the three bulk-modulus variants reach $R^2 > 0.91$, the shear moduli $0.78$--$0.83$, with elastic anisotropy and Poisson's ratio hardest.
On the dielectric dataset only band gap is predicted well ($R^2 = 0.852$); the polycrystalline constants fall below zero.
On the phonon-thermodynamics dataset all four targets (heat capacity $C_v$, entropy $S$, free energy $F$, and the maximum phonon frequency at 300\,K) are well predicted ($R^2 = 0.93$--$0.98$ at the best configuration), with ORB the strongest featurizer.
Per-property difficulty rankings for each dataset, the per-dataset radar profiles, and the dielectric and phonon learning curves are in the Supplementary Information (Supplementary Tables~S1--S3, Supplementary Figs.~S8--S10).

\FloatBarrier
\subsection{Cross-dataset synthesis}
\label{sec:cross_dataset}

Table~\ref{tab:cross_dataset} collects the best configuration and the easiest and hardest property for each dataset.
Three patterns recur.
ORB gives the best Spearman correlation in every dataset--surrogate cell, and the best \Rtwo{} in all but the GP and MTGP rows of the dielectric dataset, where SOAP is higher (Table~\ref{tab:best_configs_dielectric}); its \Rtwo{} margin is largest on the elastic data ($\Delta R^2 \approx 0.07$ over UMA and MACE) and smallest on the phonon data ($\approx 0.04$).
The DGP collapses at PCA\,=\,50 on all three datasets, a consequence of its two-layer variational parameters outnumbering the data at $\ntrain = 500$, so it is accurate but fragile.
The stack the closed-loop probe uses, ORB\,+\,GP\,+\,PCA\,=\,50, follows from these patterns: ORB for accuracy and the GP for stability.

\begin{table*}[!htb]
    \centering
    \caption{Cross-dataset comparison at $\ntrain = 500$. For each dataset we report the configuration that maximizes averaged \Rtwo{} (with its averaged $\bar{R}^{2}$ and $\bar{\rho}$), and the easiest and hardest properties identified by their best \Rtwo{} across all featurizer--surrogate--PCA combinations (which need not coincide with the configuration in column 2).}
    \label{tab:cross_dataset}
    \footnotesize
    \begin{tabular}{@{}llccll@{}}
        \toprule
        \textbf{Dataset} & \textbf{Best Config} & \textbf{Avg.\ \Rtwo{}} & \textbf{Avg.\ $\rho$} & \textbf{Easiest Property} & \textbf{Hardest Property} \\
        \midrule
        Elastic tensor & ORB+DGP+PCA25 & 0.807 & 0.826 & $K_{\text{Voigt}}$ (0.931) & Poisson (0.658) \\
        Dielectric     & DGP+ORB+PCA10 & 0.336 & 0.778 & band\_gap (0.852) & poly\_elec ($-$0.033) \\
        Phonon thermo. & GP+ORB+PCA50  & 0.968 & 0.986 & F\_300K (0.984) & max\_phonon\_freq (0.934) \\
        \bottomrule
    \end{tabular}
\end{table*}

Representative parity plots and a per-descriptor cross-dataset comparison are in the Supplementary Information (Supplementary Figs.~S11--S12).

\FloatBarrier

\section{Discussion}
\label{sec:discussion}

\subsection{Reading the closed-loop probe}
\label{sec:diffusion_lessons}

The closed-loop runs in Section~\ref{sec:closed_loop_probe} insert a Gaussian-process surrogate into the published MatInvent~\cite{chen2025matinvent} workflow without otherwise changing it.
This follows two directions: adding uncertainty-aware property predictors to provide informative gradients, and moving the workflow toward closed-loop discovery.
A Gaussian process is one such predictor, and the Expected-Improvement-plus-diversity gate is one minimal closed-loop step.
Reviews of reinforcement-learning based fine-tuning for diffusion models~\cite{uehara2024rldiffusion} catalog policy-gradient and value-based families but, to our reading, do not include a Bayesian-optimization acquisition step. Recent surveys of generative models for crystals~\cite{generative_review_2025} cover the design space without a section on online surrogate gating, while noting the offline latent-space surrogate of Qi et al.~\cite{qi2023lcom} as a related precedent.
Concurrent benchmarks of closed-loop discovery~\cite{made_benchmark} and cross-model evaluation of crystal generators~\cite{lematbench2025} are complementary. They supply environments and metrics, whereas we report one gating method run end to end across three pretrained priors.

The runs support several observations. The static-benchmark prediction carries into the loop.
Both targets are well-ranked by the ORB\,+\,GP surrogate the benchmark recommends, and the gated policy (ACC) matches or exceeds ungated fine-tuning (BASE) on MatterGen and ADiT for both $C_p$ and $K_{\text{VRH}}$.
The surrogate's five-fold cross-validated RMSE on its accumulated training set stays flat across cycles (Figure~\ref{fig:closed_loop_surrogate}), so the loop does not collapse the surrogate's self-consistency.
We do not log a per-cycle held-out regression error on each cycle's fresh proposals; for the bulk-modulus target, however, a causal replay of the surrogate (Section~\ref{sec:cl_dft}) measures its generalization directly as a ranking on unseen proposals ($\rho = 0.944$), and the gain is in any case consistent with the static-benchmark prediction and is not accompanied by surrogate breakdown on the training points.

Seed distributions are more informative than best-of-run values.
Best-value summaries in Figure~\ref{fig:closed_loop_per_model} make MatterGen and CrystalFlow look interchangeable on $C_p$, and place ADiT far ahead on $K_{\text{VRH}}$ on the strength of a single 375~GPa polymorph.
The mean of the top three candidates narrows these gaps to within overlapping error bars on both targets ($C_p$: $1.11 \pm 0.56$, $0.89 \pm 0.40$, $0.80 \pm 0.18$~J/g/K; $K_{\text{VRH}}$: $273 \pm 37$, $224 \pm 43$, $259 \pm 45$~GPa).
At a limited compute budget, reporting maxima per backbone overstates the difference between priors, and seed distributions are the safer comparator.

The priors differ more in the geometry of their top structures than in the property values those structures reach.
The crystal-system histograms in Figure~\ref{fig:closed_loop_spacegroup} show CrystalFlow and ADiT drawing most top hits from triclinic and monoclinic systems, while MatterGen produces the only tetragonal hits and a flatter distribution.
This tracks each prior's training: MatterGen learns Materials Project ground states, which are symmetry-rich, whereas the learned manifolds of CrystalFlow and ADiT contain many more low-symmetry than high-symmetry states.
When the target property favors high symmetry, a symmetric prior helps; when it does not, the bias is neutral.

Across backbones, the gain from the gate is consistent but modest, with its size set by the prior.
The discovery curves separate ACC from BASE by 0.05--0.23~J/g/K on $C_p$ and 10--50~GPa on $K_{\text{VRH}}$, largest where the prior already concentrates in the target region (MatterGen, ADiT) and smallest where seed-to-seed variance is high (CrystalFlow).
The gate cannot rescue runs whose diffusion samples never enter the relevant region.

The gate also saves oracle calls, by a margin that depends on the backbone.
The gate caps oracle calls at $K = 4$ per cycle, so it saves calls whenever the rate of structures surviving the stability filter exceeds four per cycle.
Every (backbone, target) cell clears that rate, with roughly 6--16 SUN-survivors dispatched per cycle, so the gated policy reduces oracle calls in all of them, by $1.8$--$3.5\times$ (Section~\ref{sec:cl_oracle}; Supplementary Fig.~S2).
The gate therefore lowers oracle cost in every cell we ran, while in general bounding it to the fixed per-cycle budget.

Pretrained priors alone do not break through their own ceilings.
The $\sim$1.2~J/g/K plateau in $C_p$ and the $\sim$310~GPa plateau in mean $K_{\text{VRH}}$ reflect what each backbone can reach under policy-gradient rollouts without property-conditional updates.
Moving past these ceilings will require fine-tuning the priors toward the target property, or a more exploratory acquisition rule than Expected Improvement with diversity.
The probe supports the surrogate-gated diffusion loop as a working unit; it does not single out any one backbone or oracle as the recommended choice for a discovery campaign.

\FloatBarrier
\subsection{Contribution of the static benchmark}
\label{sec:benchmark_lessons}

The benchmark explains why the loop's stack (ORB\,+\,GP\,+\,PCA\,=\,50) is a reasonable default rather than the only choice.
Among the four embeddings, pre-trained ORB was the most reliable across the three datasets and most surrogates, which we attribute to the scale and diversity of its pre-training (OMat24, over $10^{8}$ configurations) and to a compact embedding whose leading principal components already carry property-relevant variation: ORB at 10 components matches or exceeds SOAP at 50 on the elastic and phonon datasets.
The exception is the electronic dielectric dataset under GP and MTGP, where SOAP's explicit local-environment encoding gives a higher \Rtwo{} than ORB; ORB recovers under the DGP.
UMA/eSEN and MACE track each other closely, suggesting they sit at a similar representational level that only larger pre-training corpora would surpass.

Surrogate choice follows a simple rule.
The GP at PCA\,=\,50 is a safe default that trains in seconds and needs no tuning beyond marginal-likelihood optimization.
The DGP gives the best \Rtwo{} on the elastic and dielectric datasets when its input is kept to PCA\,$\leq$\,25, but at PCA\,=\,50 with $\ntrain = 500$ its variational parameters outnumber the data and the fit degrades, so PCA dimension should be treated as a tuned hyperparameter when using it (Supplementary Fig.~S13).
The multi-task GP helps only when a few ($T \leq 4$) physically related properties are predicted together, especially with a weaker featurizer; with eight largely independent elastic targets it shows negative transfer~\cite{yamada2019predicting}, and the independent GP is preferable.
The dielectric learning curves (Supplementary Fig.~S9) show Spearman correlations holding with training size even where \Rtwo{} falls.

The datasets span a gradient of difficulty set by the distance between crystal geometry and the target property.
Elastic and phonon-thermodynamic targets, governed by bond stiffness, atomic mass, and lattice vibrations, are predicted well; the electronic dielectric responses, which depend on many-body polarization and electronic structure, are not ($\bar{R}^2 \lesssim 0.34$).
For geometry-dominated targets the pipeline applies directly; for the electronic dielectric targets it can still guide a ranked search ($\bar{\rho}$ up to $\approx 0.78$ even where $\bar{R}^2 \lesssim 0.34$), though absolute prediction would need electronic-structure-aware embeddings.
A practical consequence for Bayesian optimization is data efficiency: ORB embeddings reach SOAP\,+\,GP's $\ntrain = 500$ accuracy at $\ntrain = 100$, which shortens the warm-up phase of a campaign where each evaluation is an expensive calculation.
Surrogate accuracy is necessary but not sufficient, however, since acquisition also depends on uncertainty calibration, which we do not assess here.

\section{Conclusion}
\label{sec:conclusion}

A surrogate gate turns a generative diffusion loop from an oracle-bound search into a budgeted one.
Placing a Gaussian process with an Expected Improvement along with diversity rule between the structure filter and the property oracle of RL-steered generative workflow, we found that the gate matches or exceeds ungated fine-tuning on MatterGen and ADiT for both heat capacity and bulk modulus while holding oracle calls to a fixed per-cycle budget.
Budget-matched ablations attribute the gain to the surrogate's ranking-based selection rather than the reduced number of oracle calls: at four oracle calls per cycle the gate outperforms arbitrary selection, and it comes within roughly nine percent of exhaustive oracle search at about a fifth of the cost in the most competitive instance.
None of the top structures it retrieves match the warm-start pool, so the gate steers the generator rather than replaying it.
In our benchmark of 8{,}640 surrogate fits,
among the embeddings we tested, pre-trained ORB with a Gaussian process was the most reliable choice across mechanical, electronic, and vibrational properties, and we adopt it as the default in the loop.

There are some limitations that can be addressed in future works. The oracle is a learned surrogate, DFT-validated here for the bulk-modulus target but not for heat capacity, the in-loop ablations probe one backbone in depth, and we report seed distributions rather than best-of-run maxima.
We close part of this gap here, validating the bulk-modulus discoveries against DFT and measuring the surrogate's generalization as a causal held-out ranking; the heat-capacity DFT validation, property-conditional fine-tuning of the priors, and a per-cycle held-out regression signal for the gate remain the natural next steps.
The complete pipeline, aggregated results, and analysis scripts are released as open-source software.

\section*{Code and Data Availability}
The elastic and dielectric datasets are publicly available through the matminer library~\cite{ward2018matminer}; the phonon dataset is drawn from the Materials Project DFPT phonon database~\cite{petretto2018high}. The complete static-benchmark pipeline (featurization, surrogate training, and evaluation scripts), together with the aggregated holdout CSVs that drive every figure in the paper, is available at \url{https://github.com/sheikhahnaf/matgen-BO/tree/main}. The same repository contains the closed-loop probe of Section~\ref{sec:closed_loop_probe}, including our modified MatInvent codebase with diffusion adapters for MatterGen, CrystalFlow, and ADiT, the GP-routed acquisition gate (Expected-Improvement plus Determinantal-Point-Process diversity), the heat-capacity and bulk-modulus oracles, all SLURM batches, and the curated per-trajectory results used to make the closed-loop figures, together with the ORB-PU synthesizability classifier of Section~\ref{sec:synth_methods}; its eight Optuna-tuned checkpoints along with the checkpoints of the generative models from the RL cycles are archived at \url{https://huggingface.co/SheikhAhnaf/apu-synthesizability-checkpoints}.

\section*{Conflicts of Interest}
There are no conflicts of interest to declare.

\section*{Author Contributions}
\textbf{Sk Md Ahnaf Akif Alvi:} Conceptualization, Methodology, Software, Validation, Investigation, Data curation, Visualization, Writing -- original draft.
\textbf{Jan Janssen:} Methodology, Resources, Writing -- review \& editing.
\textbf{Danny Perez:} Conceptualization, Methodology, Supervision, Writing -- review \& editing.
\textbf{Douglas Allaire:} Methodology, Supervision, Writing -- review \& editing.
\textbf{Raymundo Arr\'{o}yave:} Conceptualization, Funding acquisition, Project administration, Supervision, Writing -- review \& editing.

\section*{Acknowledgments}
This material is based on work supported by the Texas A\&M University System National Laboratories Office of the Texas A\&M University System and Los Alamos National Laboratory as part of the Joint Research Collaboration Program. The authors acknowledge the support from the U.S. Department of Energy (DOE) ARPA-E CHADWICK Program through Project DE-AR0001988. RA acknowledges the support from the WoodNext Foundation for their support of the Aggie AI Foundry. The authors acknowledge computational resources provided by the Texas A\&M High Performance Research Computing (HPRC) facility. Sk Md Ahnaf Akif Alvi acknowledges an NSF ACCESS Explore allocation (project MAT250131, ``Advanced Surrogates-based Bayesian Optimization for Materials Discovery in High Entropy Alloy Spaces''); this work used the ACES accelerator testbed at Texas A\&M University through allocation MAT250131 from the Advanced Cyberinfrastructure Coordination Ecosystem: Services \& Support (ACCESS) program, which is supported by U.S.\ National Science Foundation grants \#2138259, \#2138286, \#2138307, \#2137603, and \#2138296.

\bibliographystyle{unsrtnat}
\bibliography{refs}

\end{document}


\maketitle

This Supplementary Information collects the closed-loop per-cycle views and the static-benchmark deep-dive figures and per-property tables referenced from the main text. All static-benchmark values are at $\ntrain = 500$ unless stated otherwise; figure and table numbers are prefixed with S and are cited from the main text as ``Supplementary Fig.\ S$n$'' and ``Supplementary Table~S$n$''.

\section{Closed-loop trajectories and oracle-call schedules}
\label{si:closedloop_views}

\begin{figure}[H]
    \centering
    \includegraphics[width=\textwidth]{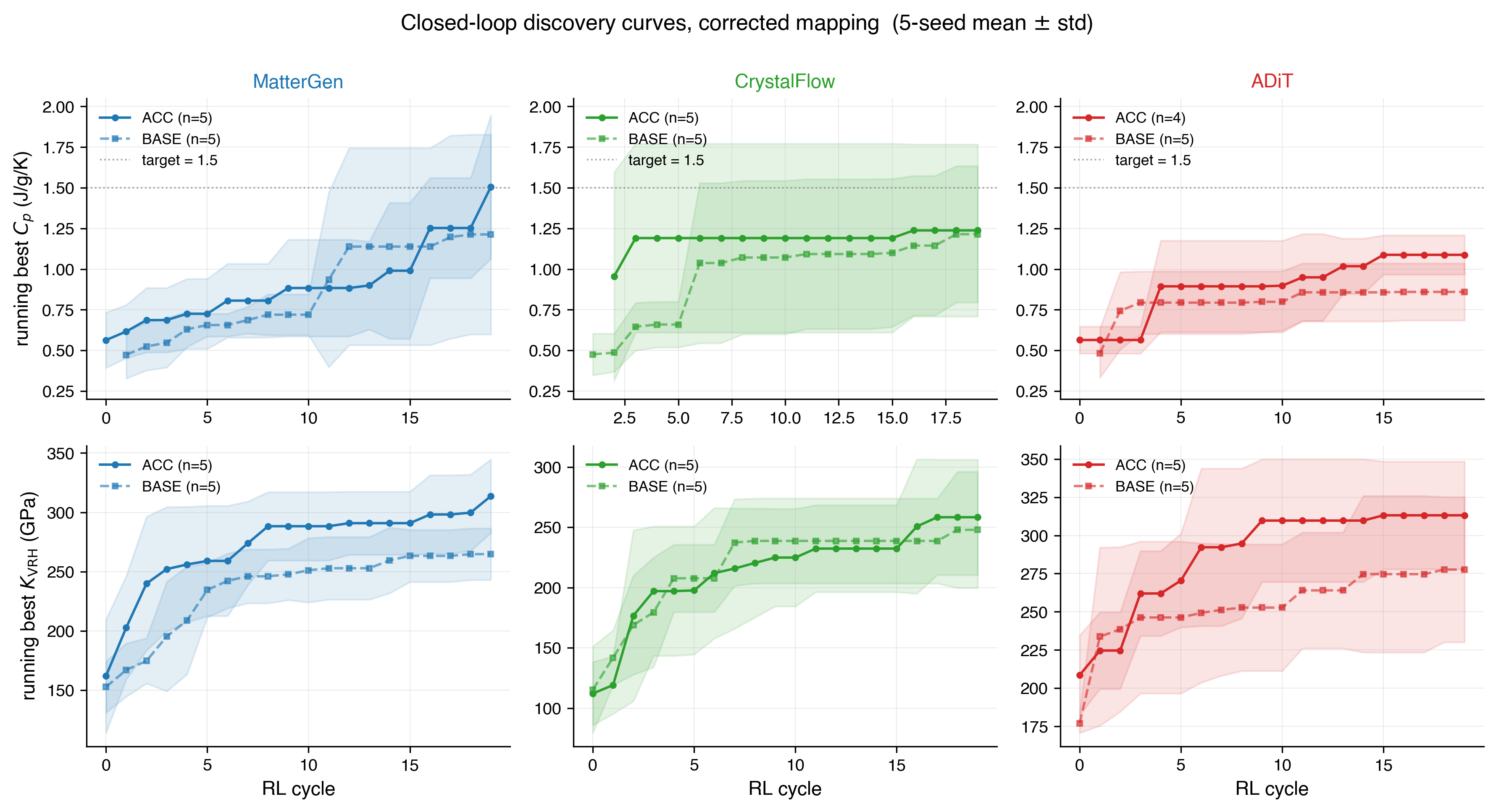}
    \caption{Running best property value over 20 RL cycles, seed-mean $\pm$ std (five seeds per cell except ADiT/ACC-$C_p$, which has four). Top row: $C_p$ (target $\to$ 1.5~J/g/K, dotted line). Bottom row: $K_{\text{VRH}}$ (max). Solid lines and circles: GP-gated BO (ACC); dashed lines and squares: vanilla REINFORCE (BASE). MatterGen and ADiT show a consistent ACC-over-BASE lift on both targets; CrystalFlow's seed variance is too large to separate the two policies.}
    \label{si:fig:cl_curves}
\end{figure}

\begin{figure}[H]
    \centering
    \includegraphics[width=\textwidth]{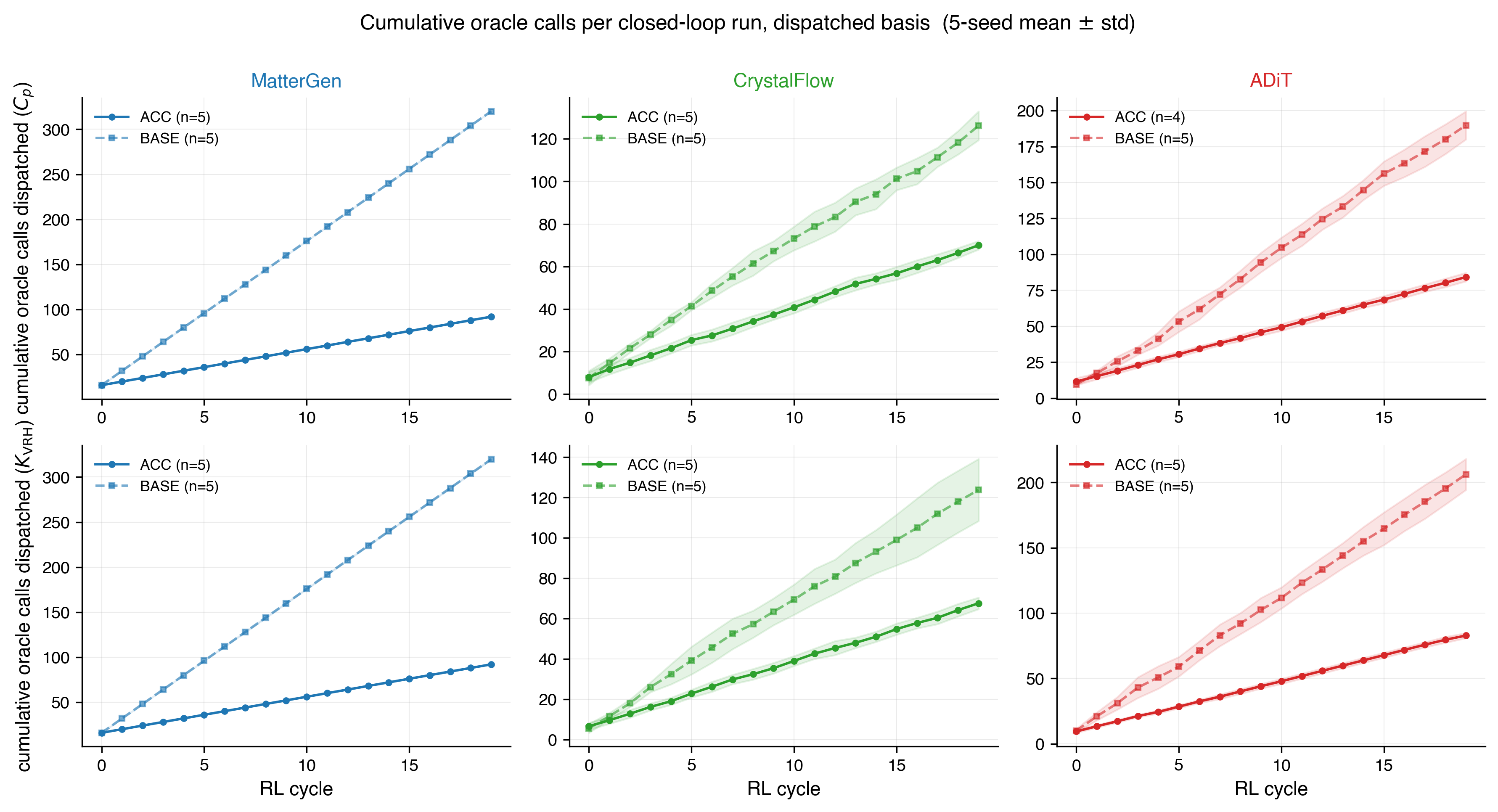}
    \caption{Cumulative oracle calls per closed-loop run on a dispatched basis (every SUN-survivor counts as one call for BASE; the top-$K$ for ACC), seed-mean $\pm$ std (five seeds per cell except ADiT/ACC-$C_p$, which has four). ACC's straight line at +4 per cycle (after a warm-start of 16) reflects the fixed top-$K=4$ gate. BASE rises with each backbone's SUN-survival rate. Because SUN survival exceeds four per cycle in every cell, the gate reduces the oracle-call count throughout ($1.8$--$3.5\times$), by the largest factor for MatterGen.}
    \label{si:fig:cl_oracle}
\end{figure}

\begin{figure}[H]
    \centering
    \includegraphics[width=\textwidth]{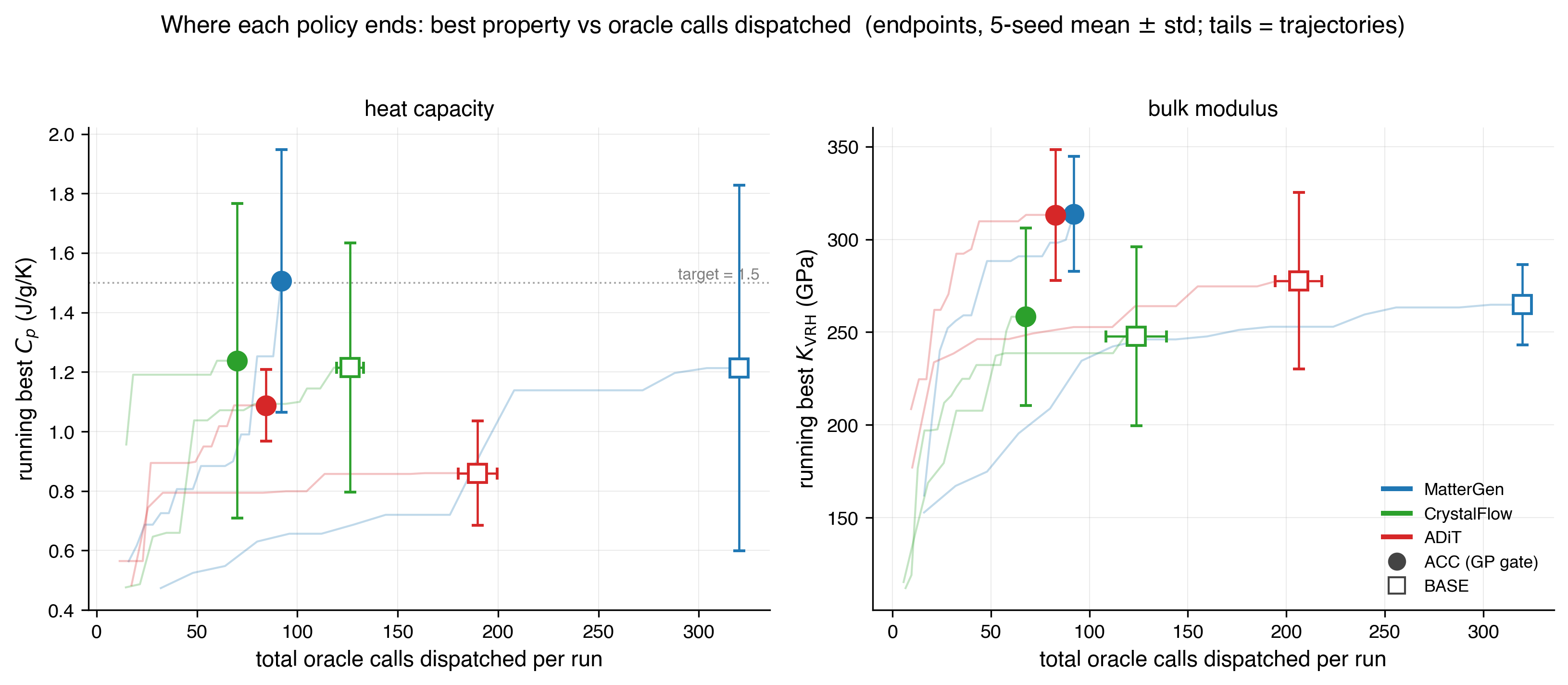}
    \caption{Endpoint summary of discovery versus oracle cost across backbones: best oracled property at the end of each run against total oracle calls dispatched, 5-seed mean $\pm$ std on both axes (filled circles: ACC; open squares: BASE; faint lines: the underlying trajectories). This combined view compresses main-text Fig.~3 to its endpoints; on the dispatched basis the gated runs concentrate in the upper-left (higher value, fewer calls) on both targets.}
    \label{si:fig:cl_comet}
\end{figure}

\begin{figure}[H]
    \centering
    \includegraphics[width=\textwidth]{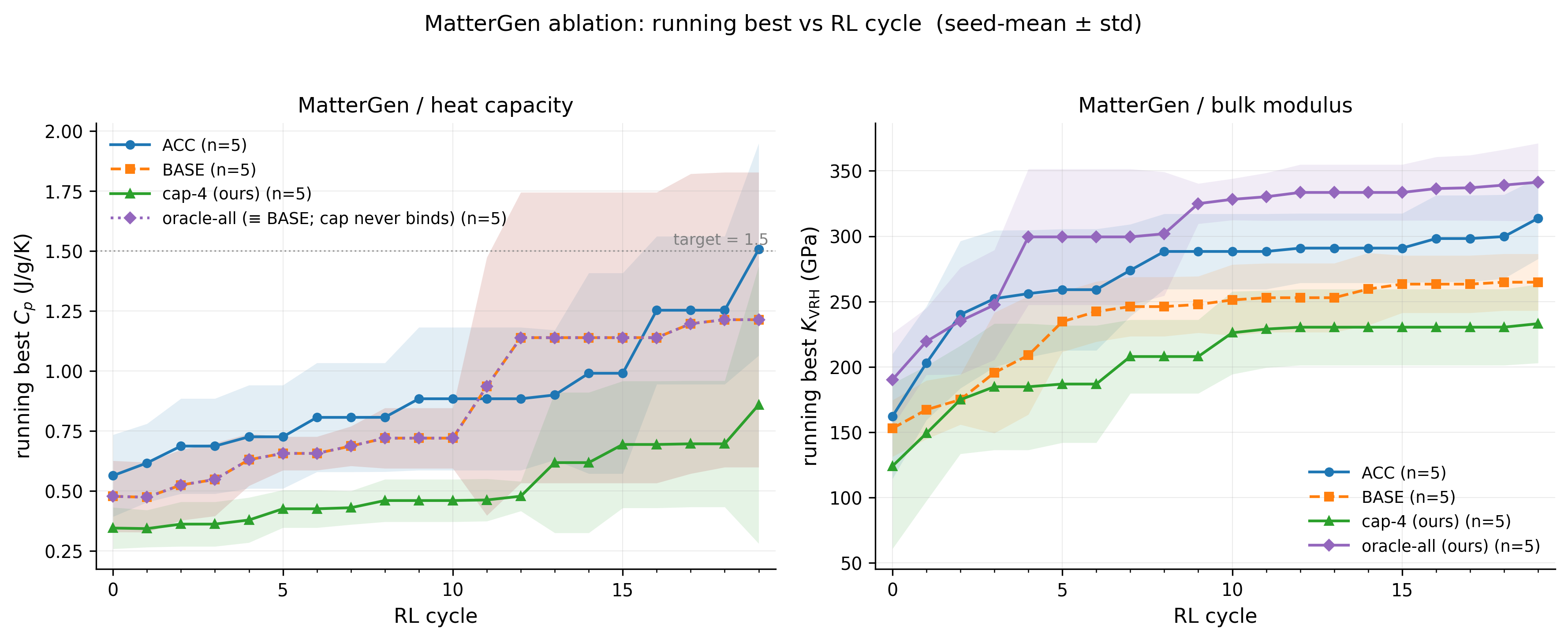}
    \caption{Per-cycle view of the MatterGen oracle-budget ablation: running-best $C_p$ (left) and $K_{\text{VRH}}$ (right) over 20 RL cycles, seed-mean $\pm$ std. ACC (the GP gate, four oracle calls per cycle) leads both vanilla REINFORCE (BASE) and the budget-matched cap-4 control on both targets; oracle-all (no budget cap, $\sim$25 calls per cycle) marks the brute-force ceiling. Every arm aggregates five seeds; two cap-4 $C_p$ seeds initially terminated on zero-reward cycles and were re-run to completion. Because the SUN cap never binds for MatterGen on $C_p$, oracle-all there coincides with BASE.}
    \label{si:fig:abl_curves}
\end{figure}

\begin{figure}[H]
    \centering
    \includegraphics[width=\textwidth]{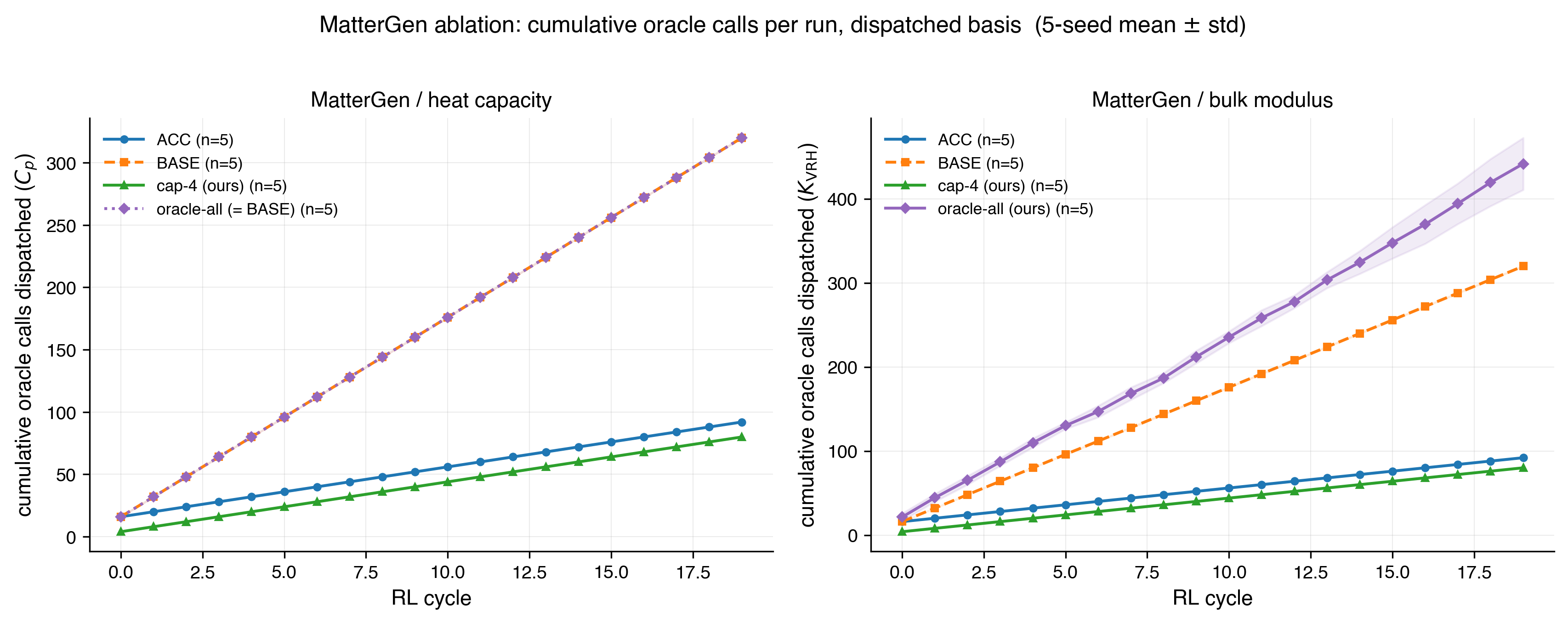}
    \caption{Cumulative oracle calls per run for the four ablation arms, seed-mean $\pm$ std. ACC and cap-4 hold fixed budgets (at most 92 and 80 calls per 20-cycle run, respectively); BASE and oracle-all scale with each cycle's SUN-survival count. ACC reaches within $\sim$9\% of oracle-all's $K_{\text{VRH}}$ discovery while making $4.8\times$ fewer oracle calls (main-text Fig.~5).}
    \label{si:fig:abl_calls}
\end{figure}

The main text summarizes the closed-loop runs as discovery versus oracle cost (main-text Figs.~3 and~5). The figures in this section give the underlying per-cycle views: running best property value against RL cycle (Fig.~\ref{si:fig:cl_curves}), the cumulative oracle-call schedules that the cost axis integrates (Fig.~\ref{si:fig:cl_oracle}), a combined endpoint summary across backbones (Fig.~\ref{si:fig:cl_comet}), and the per-cycle views of the MatterGen oracle-budget ablation (Figs.~\ref{si:fig:abl_curves} and~\ref{si:fig:abl_calls}).

\textbf{Candidate funnel: SUN survival and oracle failure.}
Each cycle the generator proposes 64 candidates, and only those passing the Stable+Unique+Novel (SUN) screen are eligible for the oracle, up to a per-cycle cap of 16. The fraction clearing this screen is set by the backbone rather than by the policy: averaged over the 20 cycles and five seeds, it is $39\%$ for MatterGen, $18\%$ for ADiT, and $13\%$ for CrystalFlow on $C_p$, with the same ordering and near-identical values on $K_{\text{VRH}}$ ($39\%$, $18\%$, $13\%$). The gated and ungated policies agree to within one percentage point in every cell (BASE vs.\ ACC: $39$ vs.\ $39$, $18$ vs.\ $18$, $13$ vs.\ $14$ on $C_p$), so the gate changes only which survivors are scored, not how many survive; all arms therefore draw from a survivor pool of the same size and composition, which is what makes the budget-matched ablation a back-end-only comparison. The roughly threefold spread across backbones reflects different bottlenecks: CrystalFlow is limited by stability, with only $28\%$ of its relaxed structures stable, whereas ADiT is stable at a rate close to MatterGen's ($70\%$ vs.\ $77\%$) but is reduced by the uniqueness and novelty requirements. The practical consequence is the per-cycle oracle load: MatterGen yields about 20 survivors per cycle and is clipped to the 16-call cap, while CrystalFlow ($\sim$6) and ADiT ($\sim$10) fall below it, so for those two backbones the oracle budget is set by generation throughput rather than by the cap.

Clearing the SUN screen does not guarantee a usable label. For $C_p$ the oracle (an eSEN relaxation followed by a phonopy calculation) returns no finite heat capacity for a large fraction of the dispatched candidates: averaged over cycles and seeds, the phonon calculation fails on $47\%$ of MatterGen's dispatched structures, $61\%$ of CrystalFlow's, and $63\%$ of ADiT's under the ungated policy. This loss is specific to the vibrational oracle; the bulk-modulus oracle (a Birch--Murnaghan equation-of-state fit) essentially never fails ($\approx 0\%$ across backbones), which is why the two targets diverge so sharply downstream of the same generator. The gate lowers the failure rate by several points on every backbone (from $47\%$ to $39\%$ for MatterGen, $61\%$ to $56\%$ for CrystalFlow, and $63\%$ to $53\%$ for ADiT): because Expected Improvement ranks toward higher-$C_p$, more stable structures, a somewhat larger share of the calls it spends come back valid. For $C_p$ the two losses compound, so the realized yield of valid heat-capacity labels is roughly 8--9 per cycle for MatterGen but only 2--4 for CrystalFlow and ADiT; for $K_{\text{VRH}}$ only the SUN-survival loss applies.

\section{Surrogate behavior across configurations}
\label{si:surrogate}

\begin{figure}[H]
    \centering
    \begin{subfigure}[t]{0.32\textwidth}
        \centering
        \includegraphics[width=\textwidth]{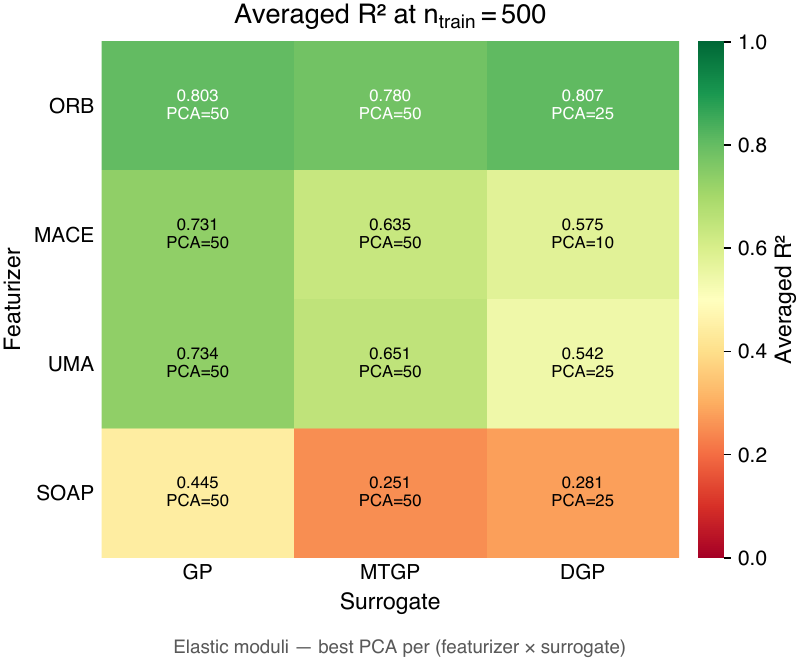}
        \caption{Elastic tensor}
        \label{si:fig:heatmap_elastic}
    \end{subfigure}
    \hfill
    \begin{subfigure}[t]{0.32\textwidth}
        \centering
        \includegraphics[width=\textwidth]{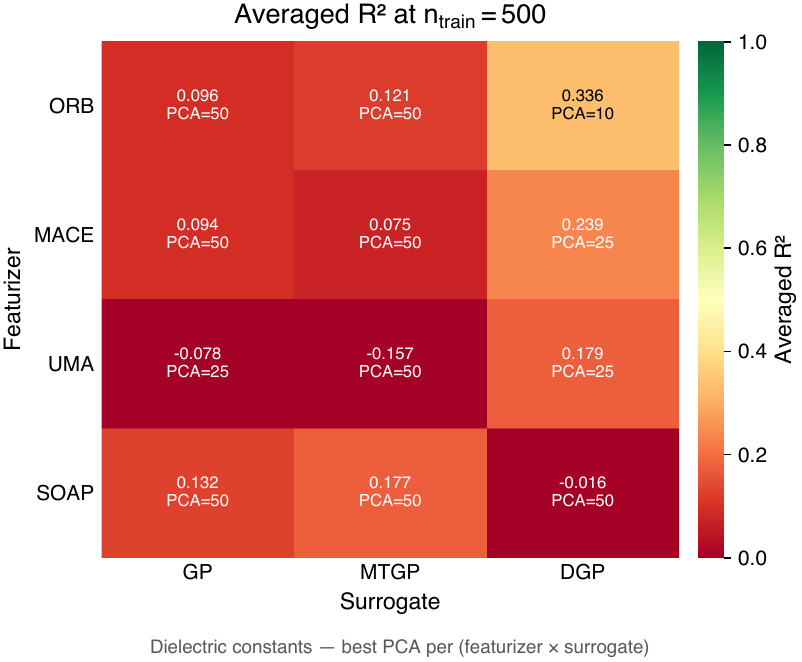}
        \caption{Dielectric constant}
        \label{si:fig:heatmap_dielectric}
    \end{subfigure}
    \hfill
    \begin{subfigure}[t]{0.32\textwidth}
        \centering
        \includegraphics[width=\textwidth]{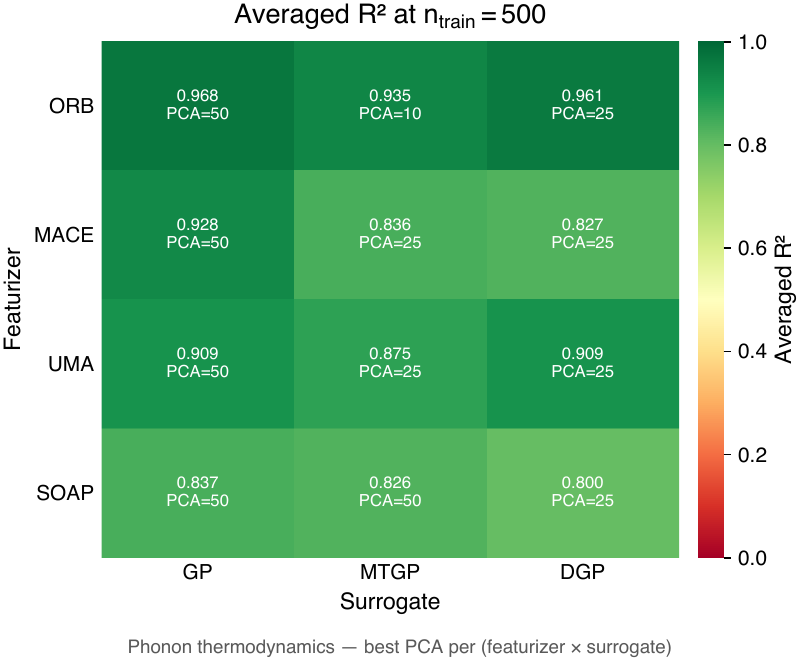}
        \caption{Phonon thermodynamics}
        \label{si:fig:heatmap_phonon}
    \end{subfigure}
    \caption{Heatmaps of averaged \Rtwo{} for each featurizer--surrogate combination at $\ntrain = 500$ across all three datasets. Rows are descriptor--PCA combinations; columns are surrogate models. The DGP column is bimodal across all datasets: strong at low PCA, collapsing at PCA\,=\,50.}
    \label{si:fig:heatmap_all}
\end{figure}

\begin{figure}[H]
    \centering
    \begin{subfigure}[t]{0.32\textwidth}
        \centering
        \includegraphics[width=\textwidth]{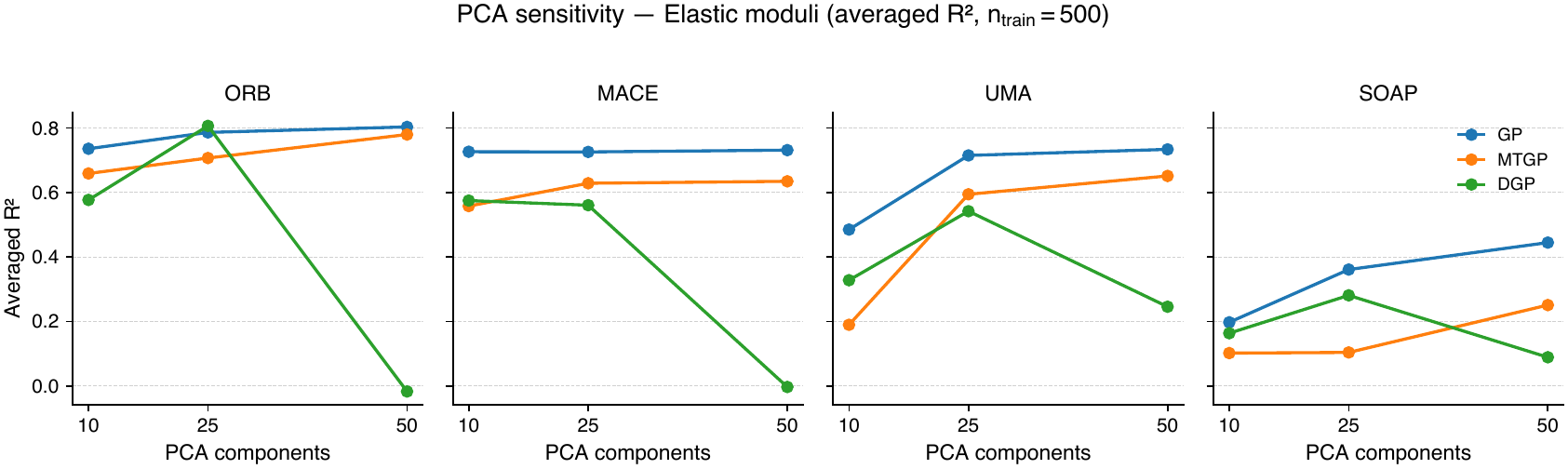}
        \caption{Elastic tensor}
        \label{si:fig:pca_elastic}
    \end{subfigure}
    \hfill
    \begin{subfigure}[t]{0.32\textwidth}
        \centering
        \includegraphics[width=\textwidth]{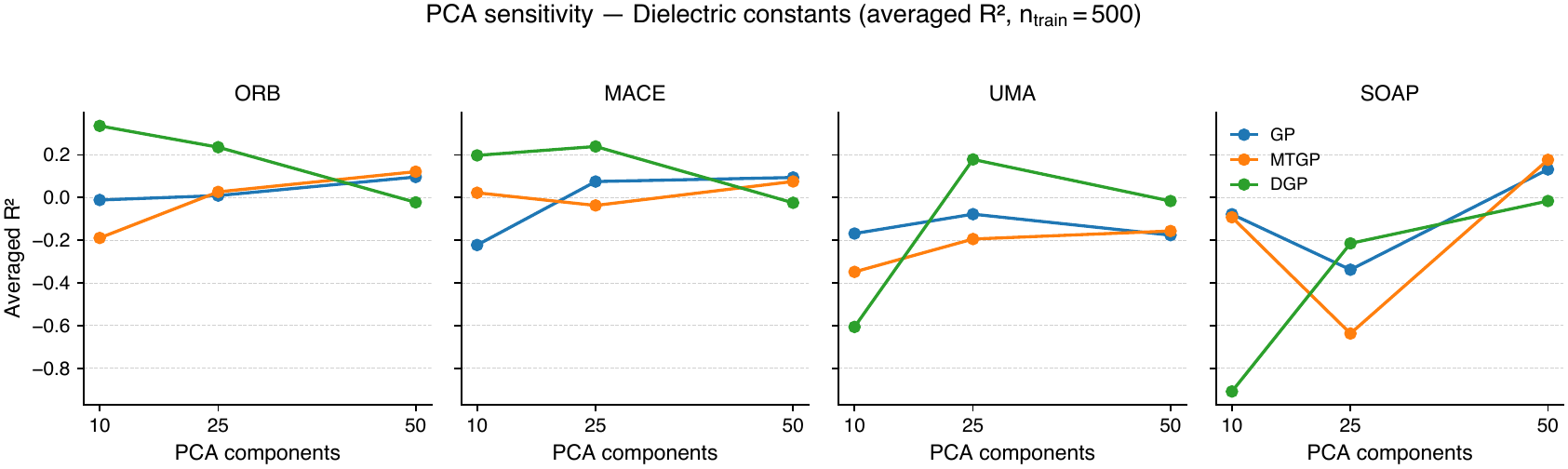}
        \caption{Dielectric constant}
        \label{si:fig:pca_dielectric}
    \end{subfigure}
    \hfill
    \begin{subfigure}[t]{0.32\textwidth}
        \centering
        \includegraphics[width=\textwidth]{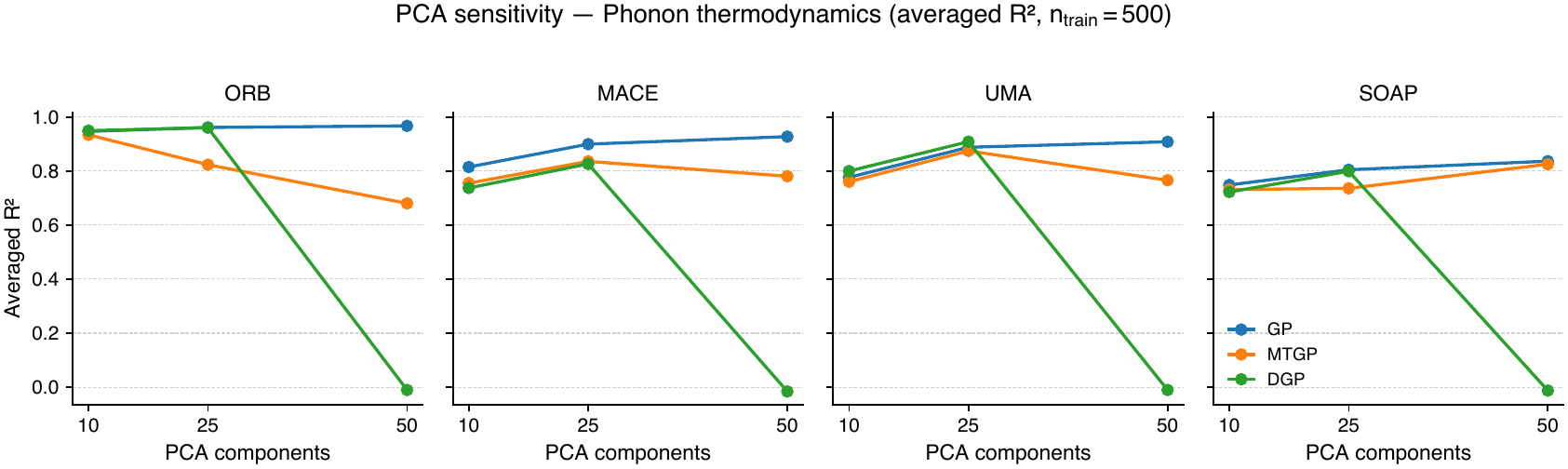}
        \caption{Phonon thermodynamics}
        \label{si:fig:pca_phonon}
    \end{subfigure}
    \caption{Effect of PCA dimensionality on averaged \Rtwo{} at $\ntrain = 500$ across all three datasets. GP and MTGP show monotonic improvement or robustness as PCA increases. DGP (dashed lines) collapses at PCA\,=\,50 across all datasets.}
    \label{si:fig:pca_all}
\end{figure}

Figure~\ref{si:fig:heatmap_all} shows averaged \Rtwo{} for every featurizer--surrogate--PCA combination, and Figure~\ref{si:fig:pca_all} shows the same data as a function of PCA dimensionality. The GP and MTGP improve or plateau as PCA increases to 50, while the DGP reaches its best \Rtwo{} at PCA\,$\leq$\,25 and collapses at PCA\,=\,50 on all three datasets. This collapse is consistent across property domains and reflects the two-layer variational DGP being over-parameterized relative to the data at this sample size, rather than a property-specific artifact.

\FloatBarrier
\section{Per-property difficulty and per-dataset profiles}
\label{si:difficulty}

\begin{figure}[H]
    \centering
    \begin{subfigure}[t]{0.32\textwidth}
        \centering
        \includegraphics[width=\textwidth]{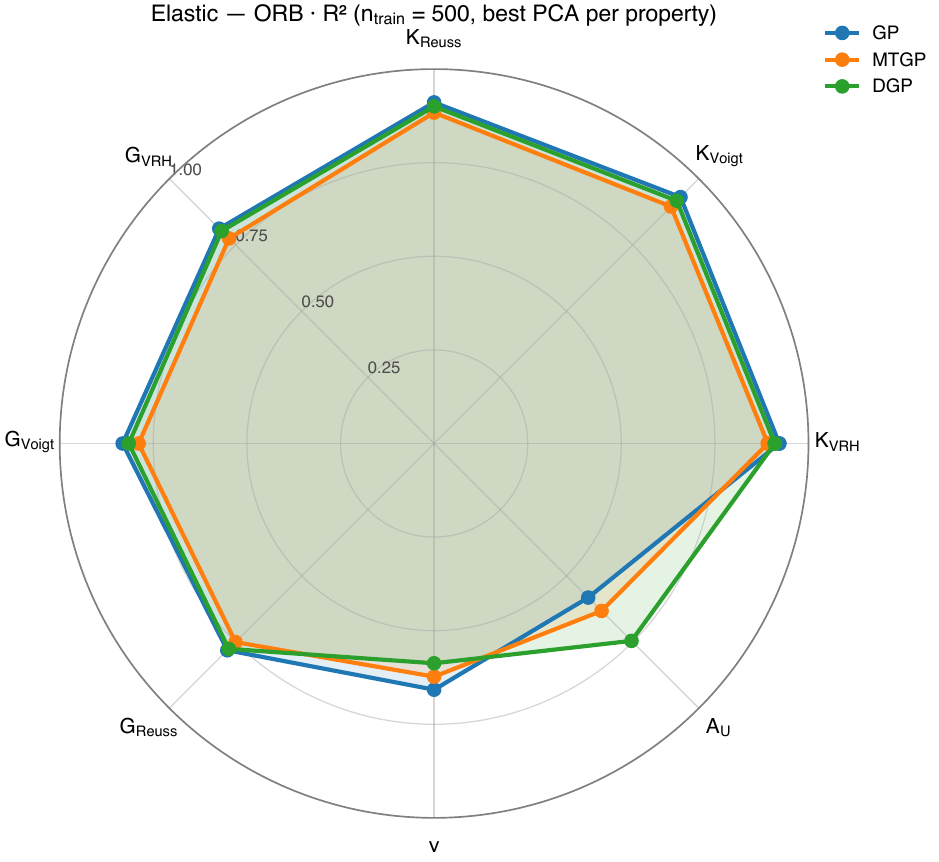}
        \caption{Elastic tensor}
        \label{si:fig:radar_elastic}
    \end{subfigure}
    \hfill
    \begin{subfigure}[t]{0.32\textwidth}
        \centering
        \includegraphics[width=\textwidth]{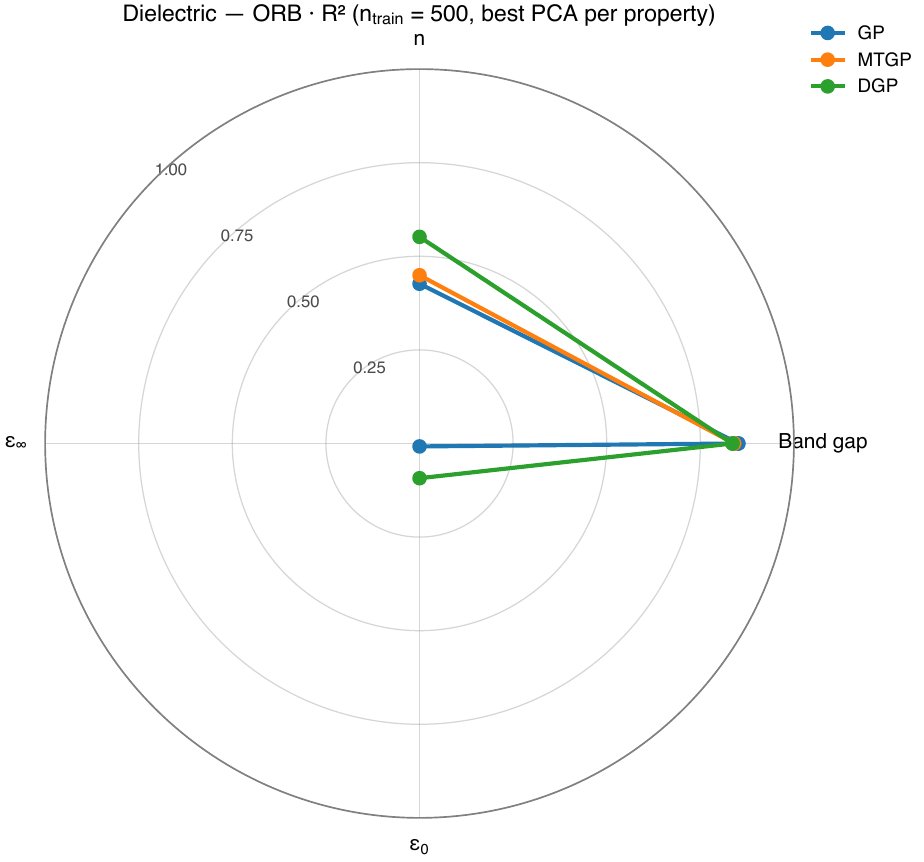}
        \caption{Dielectric constant}
        \label{si:fig:radar_dielectric}
    \end{subfigure}
    \hfill
    \begin{subfigure}[t]{0.32\textwidth}
        \centering
        \includegraphics[width=\textwidth]{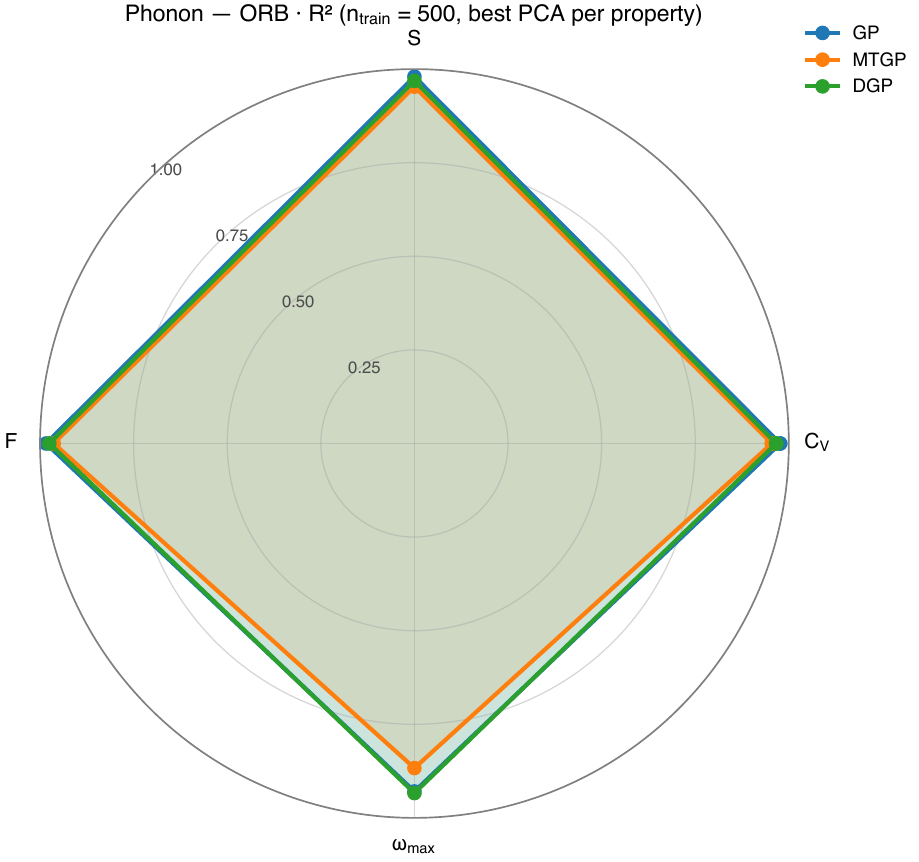}
        \caption{Phonon thermodynamics}
        \label{si:fig:radar_phonon}
    \end{subfigure}
    \caption{Radar charts showing ORB's per-property \Rtwo{} across GP, MTGP, and DGP at $\ntrain = 500$ for each dataset. The elastic dataset shows the most uniform performance across properties; the dielectric dataset shows strong asymmetry between easy and hard properties.}
    \label{si:fig:radar_all}
\end{figure}

The radar charts in Figure~\ref{si:fig:radar_all} give ORB's per-property \Rtwo{} profile per dataset, and Tables~\ref{si:tab:property_difficulty_elastic}--\ref{si:tab:property_difficulty_phonon} rank each property by best-achievable \Rtwo{}. The elastic dataset is the most uniform across properties; the dielectric dataset shows a sharp split between geometry-accessible properties (band gap) and electronically governed ones (polycrystalline dielectric constants), while the phonon-thermodynamics targets are uniformly well predicted.

\begin{table}[H]
    \centering
    \caption{Property difficulty ranking on the elastic tensor dataset ($\ntrain = 500$). Each row reports the best \Rtwo{} achieved across all featurizer and PCA combinations.}
    \label{si:tab:property_difficulty_elastic}
    \small
    \begin{tabular}{@{}lllcc@{}}
        \toprule
        \textbf{Property} & \textbf{Best Feat.} & \textbf{Surr.} & \textbf{PCA} & \textbf{\Rtwo{}} \\
        \midrule
        $K_{\text{Voigt}}$   & ORB & GP  & 50 & 0.931 \\
        $K_{\text{VRH}}$     & ORB & GP  & 50 & 0.922 \\
        $K_{\text{Reuss}}$   & ORB & GP  & 50 & 0.911 \\
        $G_{\text{Voigt}}$   & ORB & GP  & 50 & 0.832 \\
        $G_{\text{VRH}}$     & ORB & GP  & 50 & 0.812 \\
        $G_{\text{Reuss}}$   & ORB & GP  & 50 & 0.781 \\
        Elastic aniso.       & ORB & DGP & 25 & 0.746 \\
        Poisson's ratio      & ORB & GP  & 50 & 0.658 \\
        \bottomrule
    \end{tabular}
\end{table}

\begin{table}[H]
    \centering
    \caption{Property difficulty ranking on the dielectric constant dataset ($\ntrain = 500$).}
    \label{si:tab:property_difficulty_dielectric}
    \small
    \begin{tabular}{@{}lllcc@{}}
        \toprule
        \textbf{Property} & \textbf{Best Feat.} & \textbf{Surr.} & \textbf{PCA} & \textbf{\Rtwo{}} \\
        \midrule
        band\_gap        & ORB  & GP   & 50 & 0.852 \\
        $n$              & ORB  & DGP  & 10 & 0.552 \\
        poly\_total      & UMA  & DGP  & 25 & 0.111 \\
        poly\_electronic & SOAP & DGP  & 50 & $-$0.033 \\
        \bottomrule
    \end{tabular}
\end{table}

\begin{table}[H]
    \centering
    \caption{Property difficulty ranking on the phonon thermodynamics dataset ($\ntrain = 500$). All four vibrational-thermodynamic targets are well predicted.}
    \label{si:tab:property_difficulty_phonon}
    \small
    \begin{tabular}{@{}lllccc@{}}
        \toprule
        \textbf{Property} & \textbf{Best Feat.} & \textbf{Surr.} & \textbf{PCA} & \textbf{\Rtwo{}} & \textbf{$\rho$} \\
        \midrule
        F\_300K           & ORB  & GP   & 50 & 0.984 & 0.994 \\
        S\_300K           & ORB  & GP   & 50 & 0.980 & 0.991 \\
        Cv\_300K          & ORB  & GP   & 50 & 0.977 & 0.992 \\
        max\_phonon\_freq & ORB  & DGP  & 25 & 0.934 & 0.969 \\
        \bottomrule
    \end{tabular}
\end{table}

\FloatBarrier
\section{Learning curves for the dielectric and phonon datasets}
\label{si:learning_curves}

\begin{figure}[H]
    \centering
    \begin{subfigure}[t]{0.48\textwidth}
        \centering
        \includegraphics[width=\textwidth]{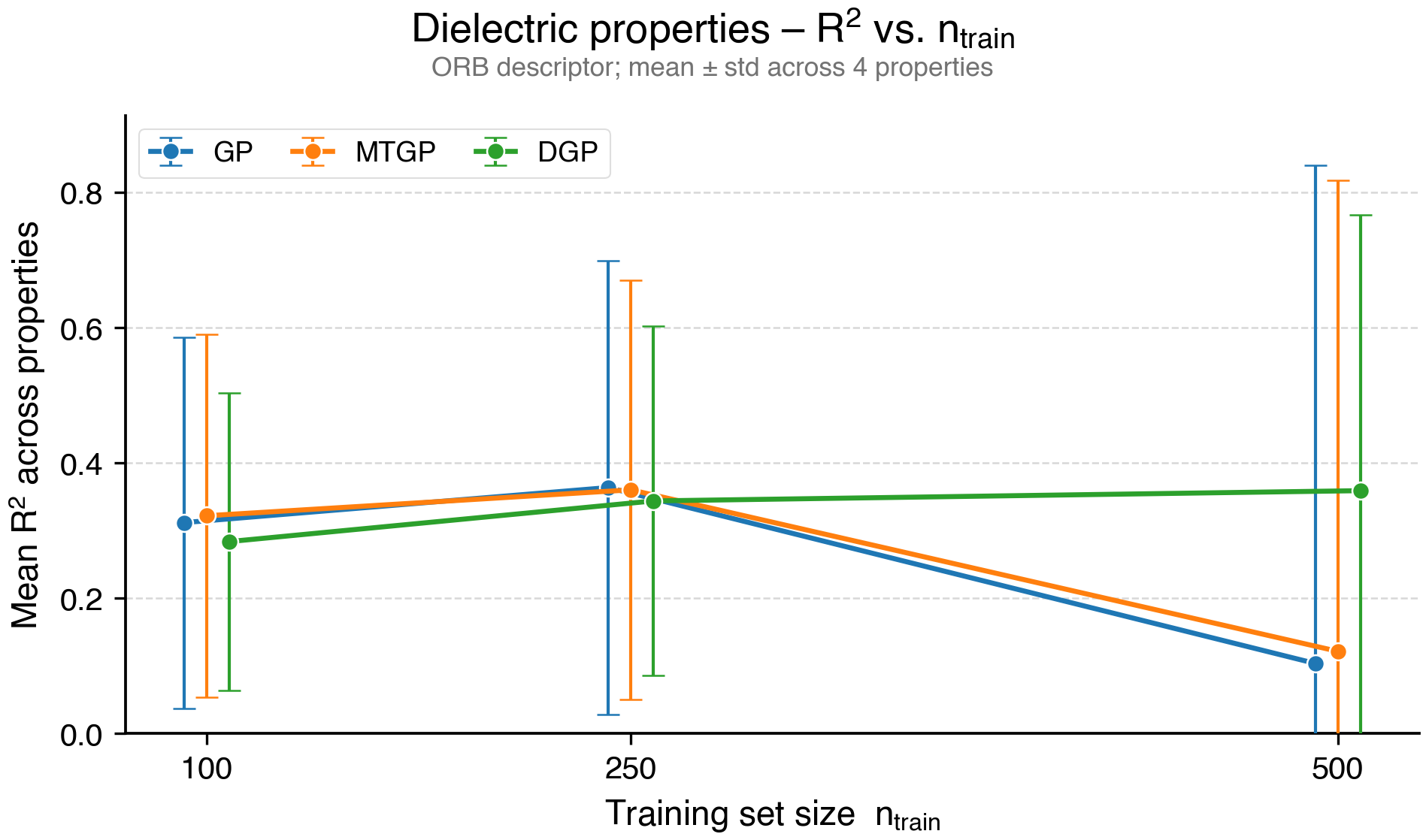}
        \caption{Averaged \Rtwo{}}
        \label{si:fig:lc_dielectric_r2}
    \end{subfigure}
    \hfill
    \begin{subfigure}[t]{0.48\textwidth}
        \centering
        \includegraphics[width=\textwidth]{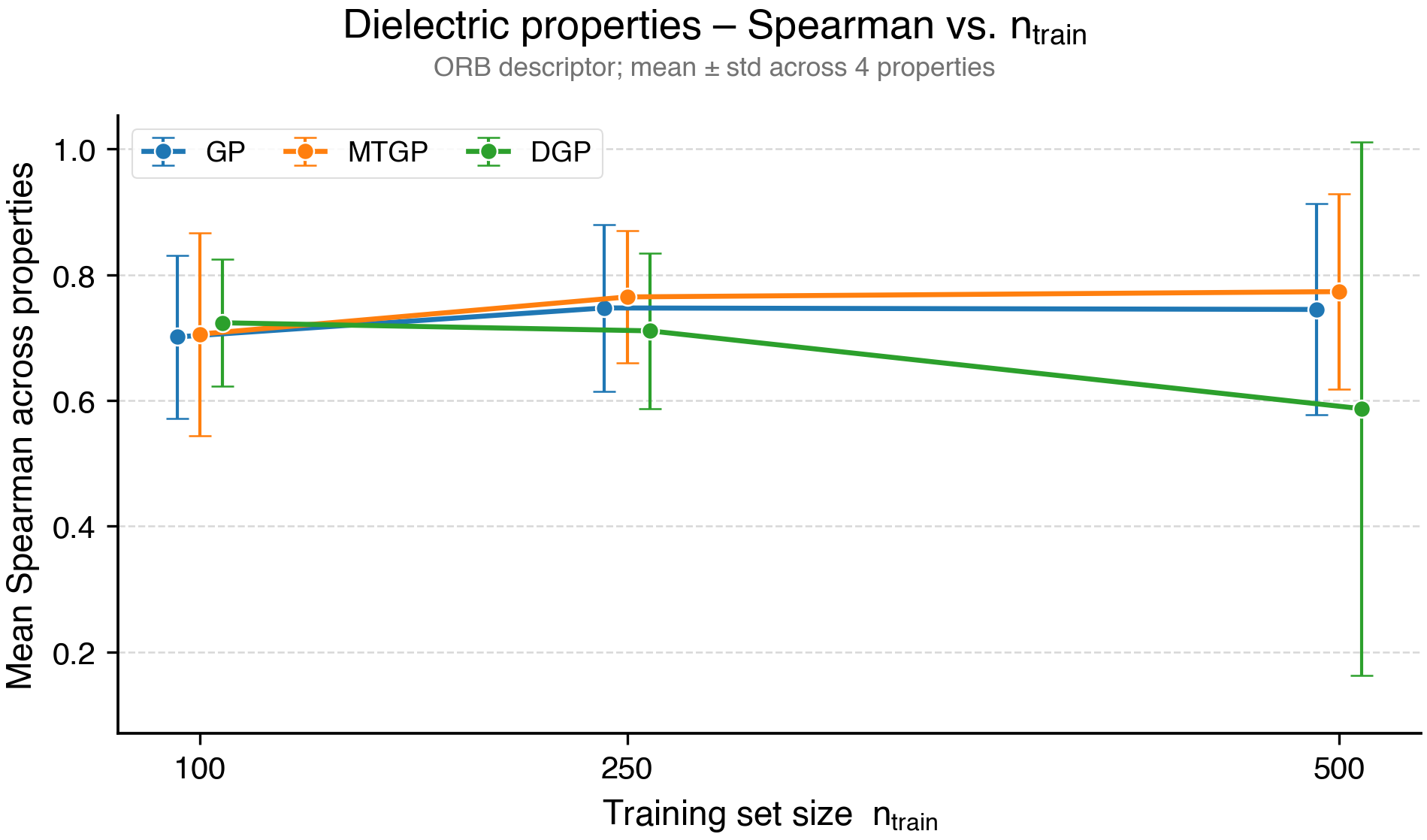}
        \caption{Averaged Spearman $\rho$}
        \label{si:fig:lc_dielectric_spearman}
    \end{subfigure}
    \caption{Learning curves for the dielectric constant dataset (ORB descriptor, best PCA per surrogate). GP and MTGP \Rtwo{} values decrease with training size due to heavy-tailed property distributions, while Spearman correlations remain stable, reinforcing that ranking accuracy is more robust than regression accuracy for electronically complex properties.}
    \label{si:fig:learning_curves_dielectric}
\end{figure}

\begin{figure}[H]
    \centering
    \begin{subfigure}[t]{0.48\textwidth}
        \centering
        \includegraphics[width=\textwidth]{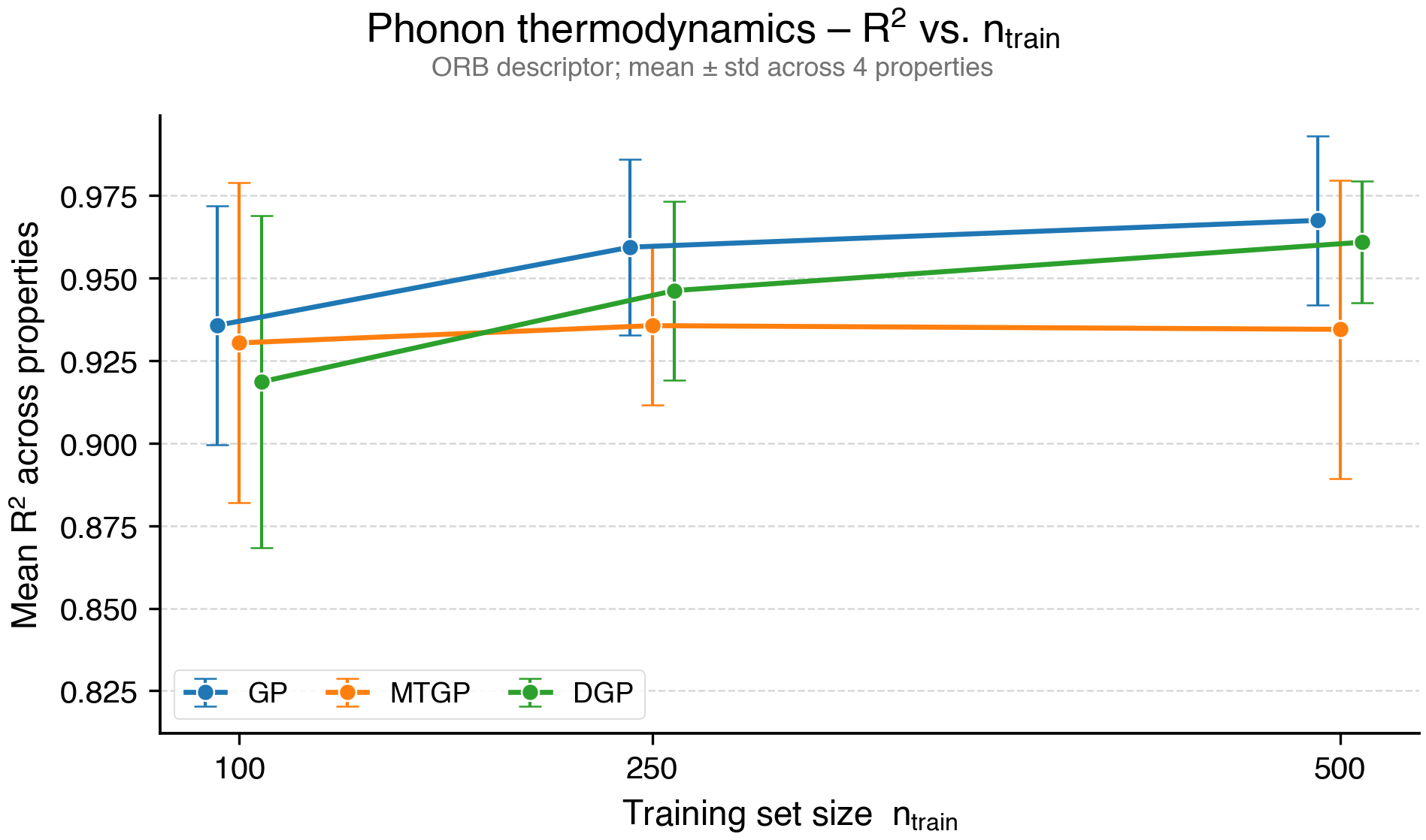}
        \caption{Averaged \Rtwo{}}
        \label{si:fig:lc_phonon_r2}
    \end{subfigure}
    \hfill
    \begin{subfigure}[t]{0.48\textwidth}
        \centering
        \includegraphics[width=\textwidth]{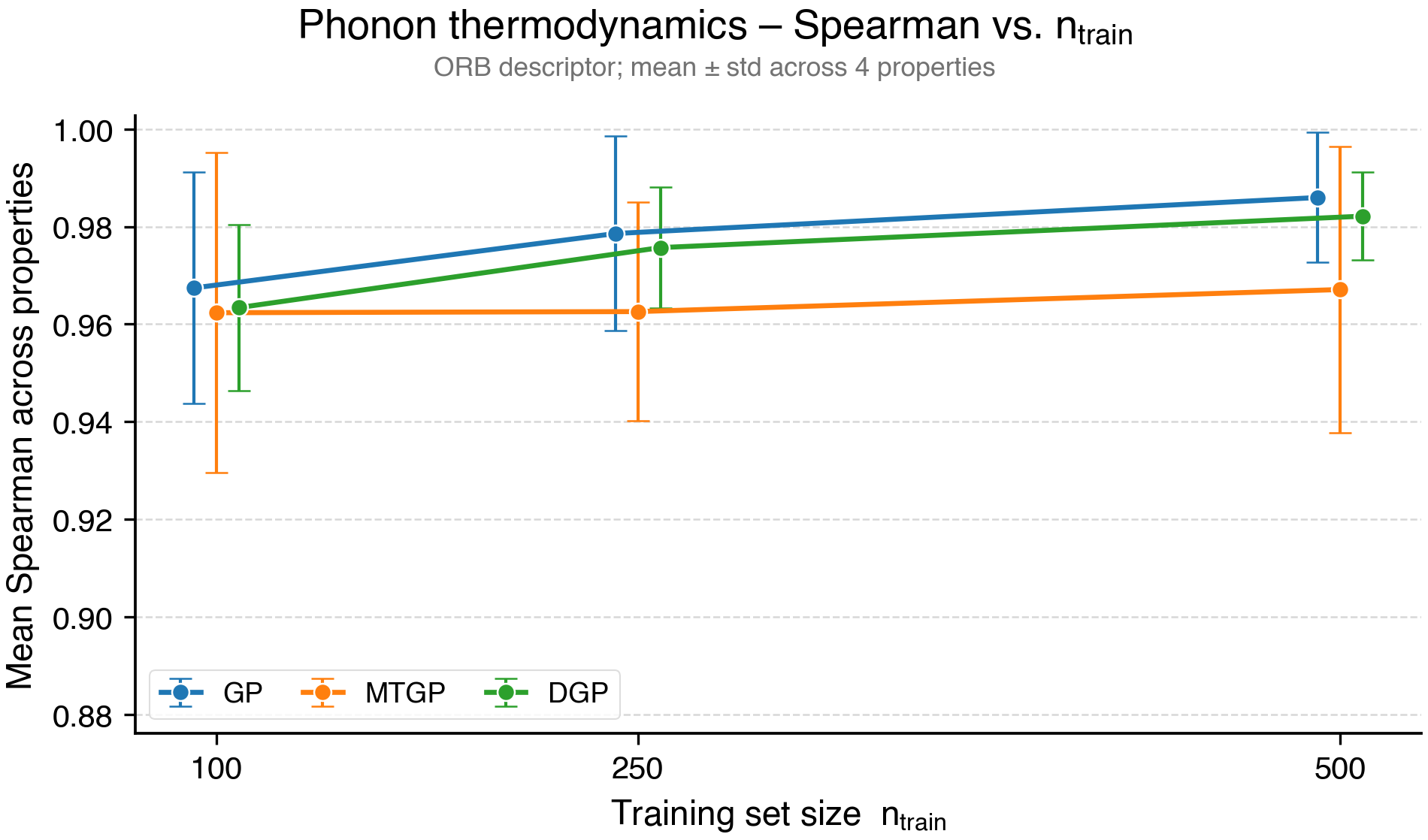}
        \caption{Averaged Spearman $\rho$}
        \label{si:fig:lc_phonon_spearman}
    \end{subfigure}
    \caption{Learning curves for the phonon thermodynamics dataset (ORB descriptor, best PCA per surrogate). All surrogates predict the four targets well across training sizes, with ORB\,+\,GP the strongest ($\bar{R}^2 = 0.968$, $\bar{\rho} = 0.986$ at $\ntrain = 500$).}
    \label{si:fig:learning_curves_phonon}
\end{figure}

Figures~\ref{si:fig:learning_curves_dielectric} and~\ref{si:fig:learning_curves_phonon} show learning curves for the dielectric and phonon datasets. On the dielectric dataset, GP and MTGP \Rtwo{} values can fall as the training set grows, because heavy-tailed property distributions add more extreme outliers that dominate the squared-error metric, while the Spearman correlations remain stable or improve; the phonon-thermodynamic targets, by contrast, are well predicted and improve steadily with training size.

\FloatBarrier
\section{Parity and cross-descriptor comparison}
\label{si:metrics}

\begin{figure}[H]
    \centering
    \begin{subfigure}[t]{0.32\textwidth}
        \centering
        \includegraphics[width=\textwidth]{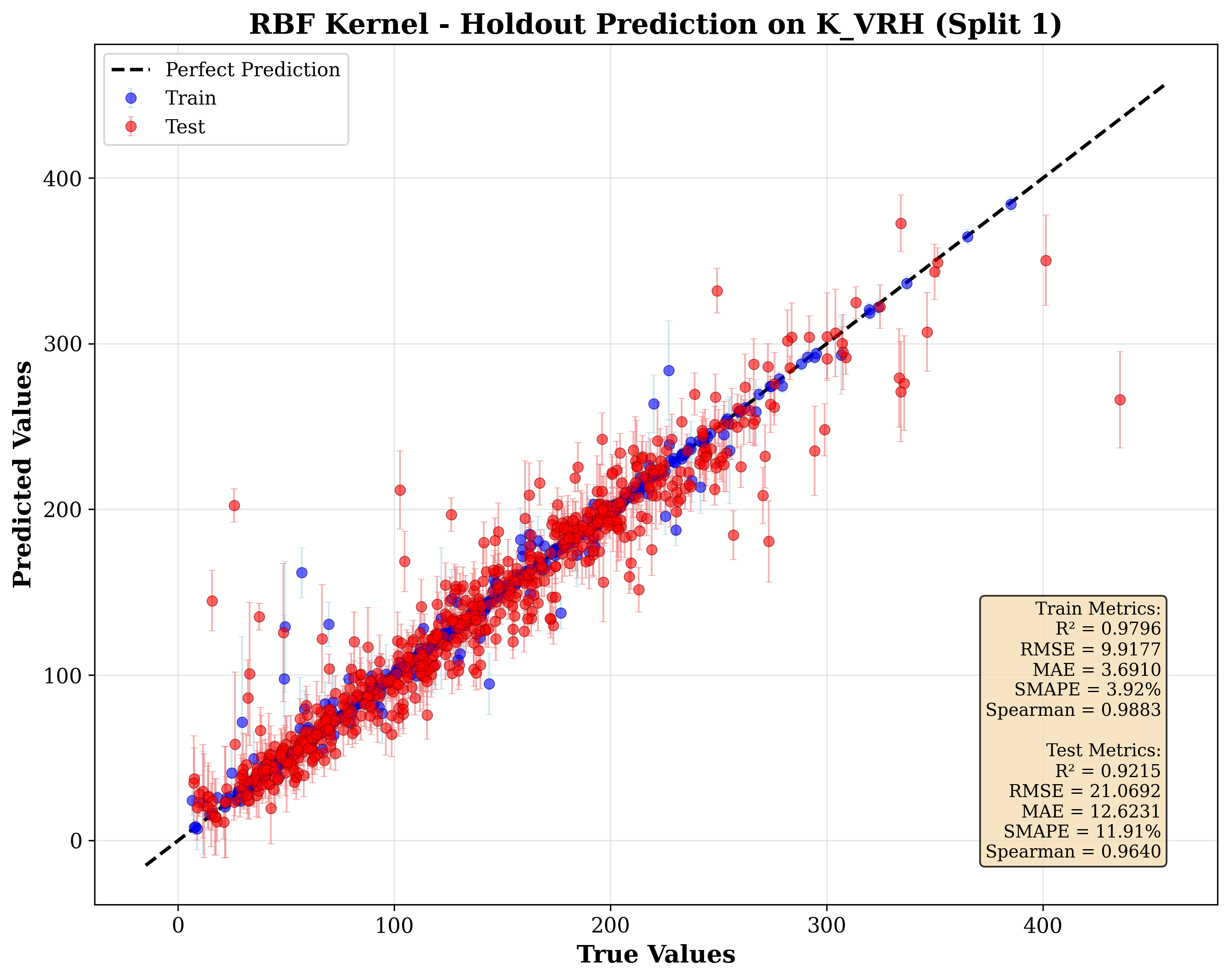}
        \caption{Elastic: $K_{\text{VRH}}$ ($R^{2} = 0.922$)}
    \end{subfigure}
    \hfill
    \begin{subfigure}[t]{0.32\textwidth}
        \centering
        \includegraphics[width=\textwidth]{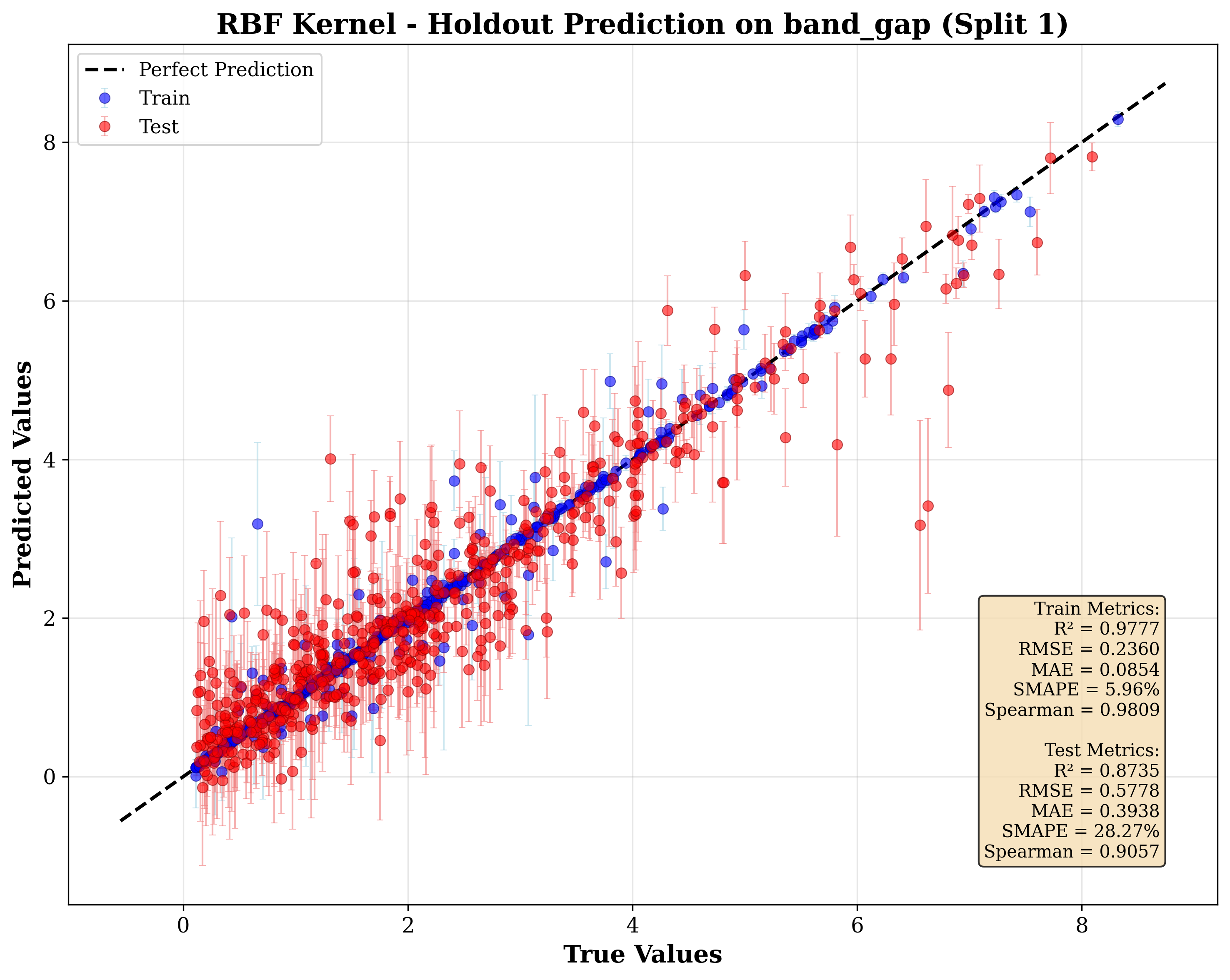}
        \caption{Dielectric: band gap ($R^{2} = 0.852$)}
    \end{subfigure}
    \hfill
    \begin{subfigure}[t]{0.32\textwidth}
        \centering
        \includegraphics[width=\textwidth]{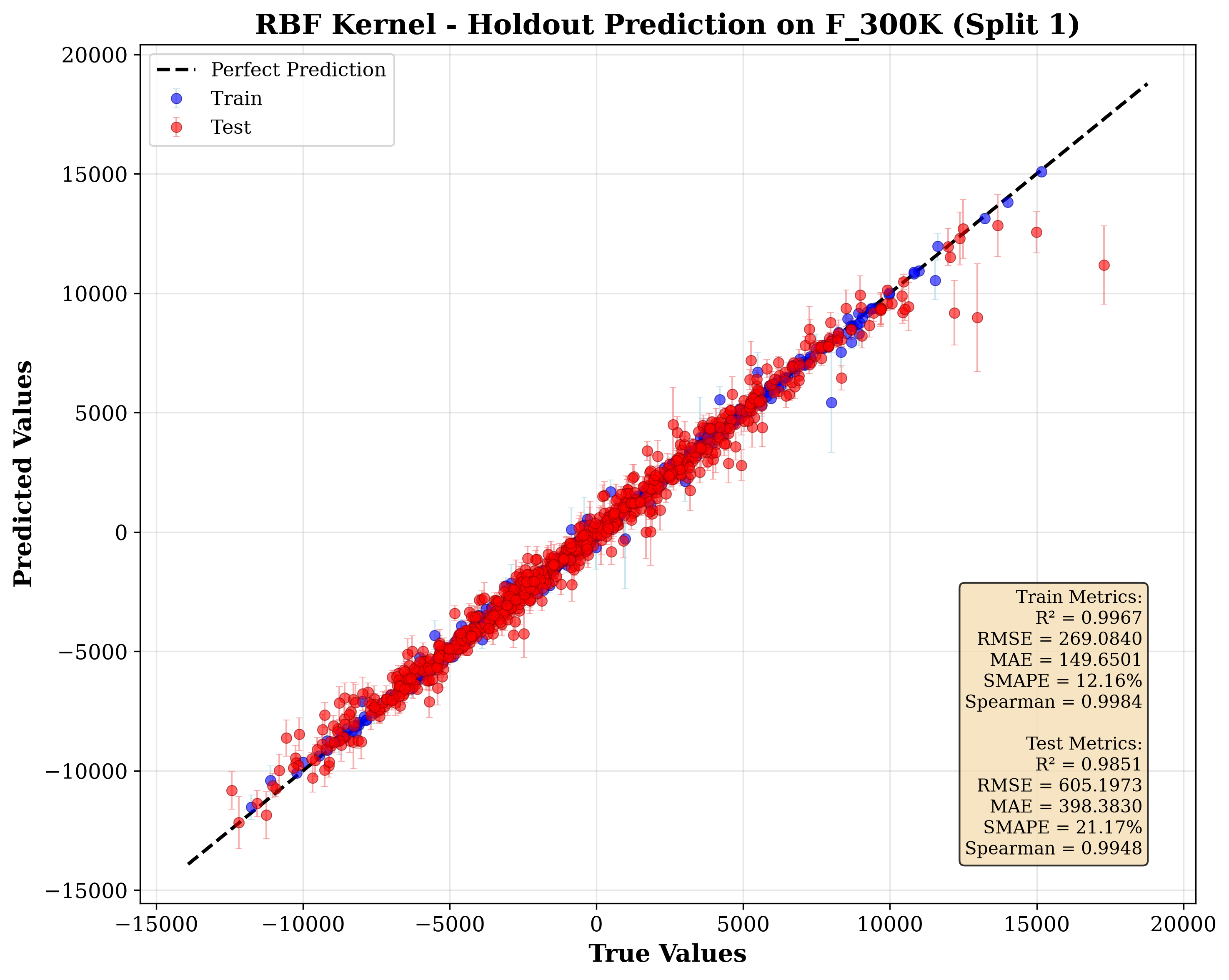}
        \caption{Phonon: free energy $F$ (300\,K) ($R^{2} = 0.984$)}
    \end{subfigure}
    \caption{Representative parity plots for the best-predicted property in each of the three benchmark datasets (ORB\,+\,GP, PCA\,=\,50, $\ntrain = 500$). Each panel shows the split-1 hold-out scatter; the $R^{2}$ quoted in the subcaptions is the 5-split mean. All three show tight clustering around the diagonal with well-calibrated uncertainty bands.}
    \label{si:fig:parity_examples}
\end{figure}

\begin{figure}[H]
    \centering
    \begin{subfigure}[t]{0.48\textwidth}
        \centering
        \includegraphics[width=\textwidth]{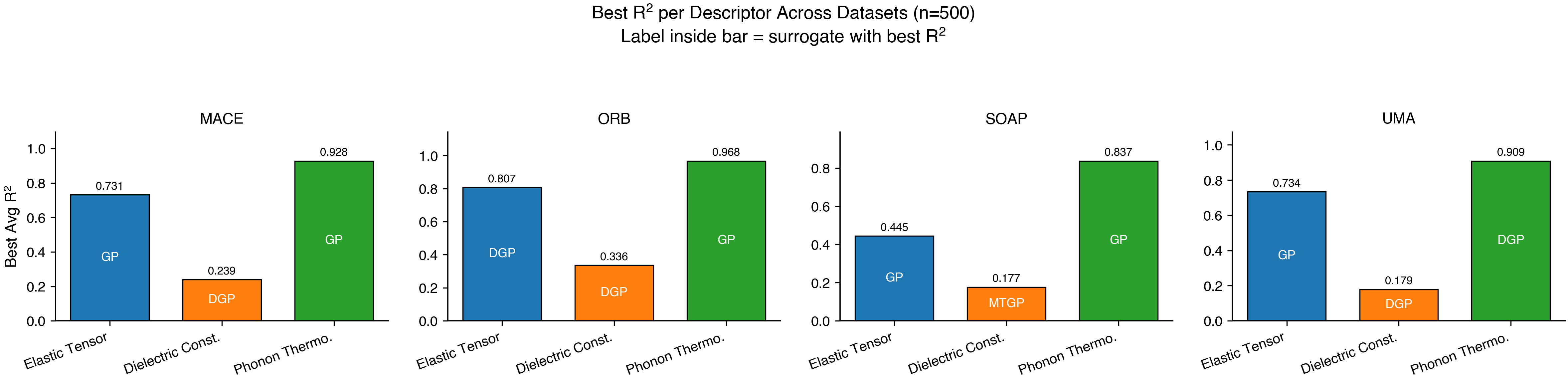}
        \caption{Best \Rtwo{} per descriptor (label = best surrogate)}
        \label{si:fig:cross_desc_r2}
    \end{subfigure}
    \hfill
    \begin{subfigure}[t]{0.48\textwidth}
        \centering
        \includegraphics[width=\textwidth]{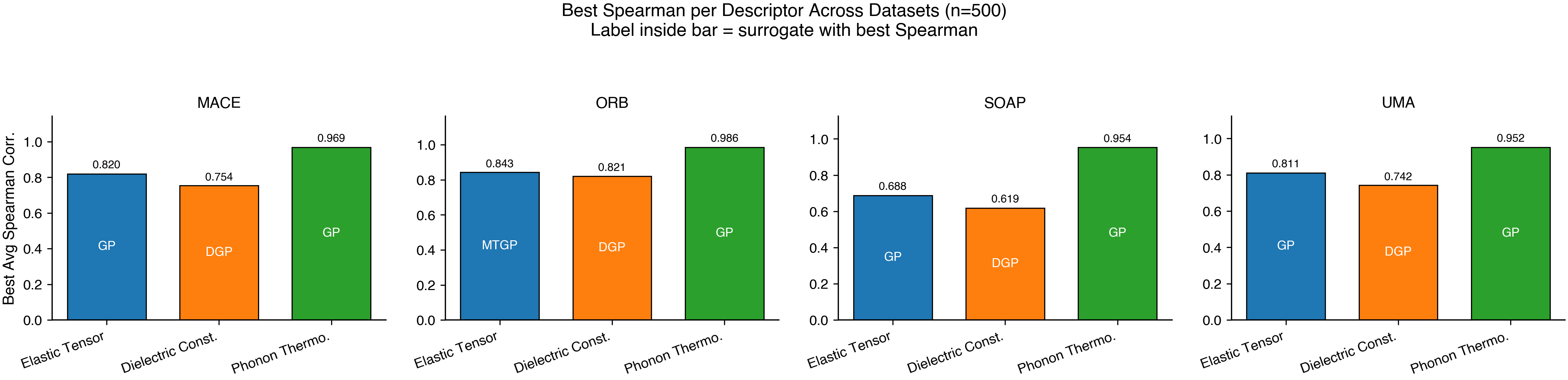}
        \caption{Best Spearman~$\rho$ per descriptor (label = best surrogate)}
        \label{si:fig:cross_desc_spearman}
    \end{subfigure}
    \caption{Cross-dataset comparison of best achievable performance per descriptor at $\ntrain = 500$. Each bar shows the best \Rtwo{} (a) or Spearman~$\rho$ (b) across all surrogate--PCA combinations, with the winning surrogate name overlaid. ORB achieves the highest \Rtwo{} on all three datasets, and its Spearman correlations stay above 0.7 across datasets.}
    \label{si:fig:cross_descriptor}
\end{figure}

\begin{figure}[H]
    \centering
    \includegraphics[width=0.75\textwidth]{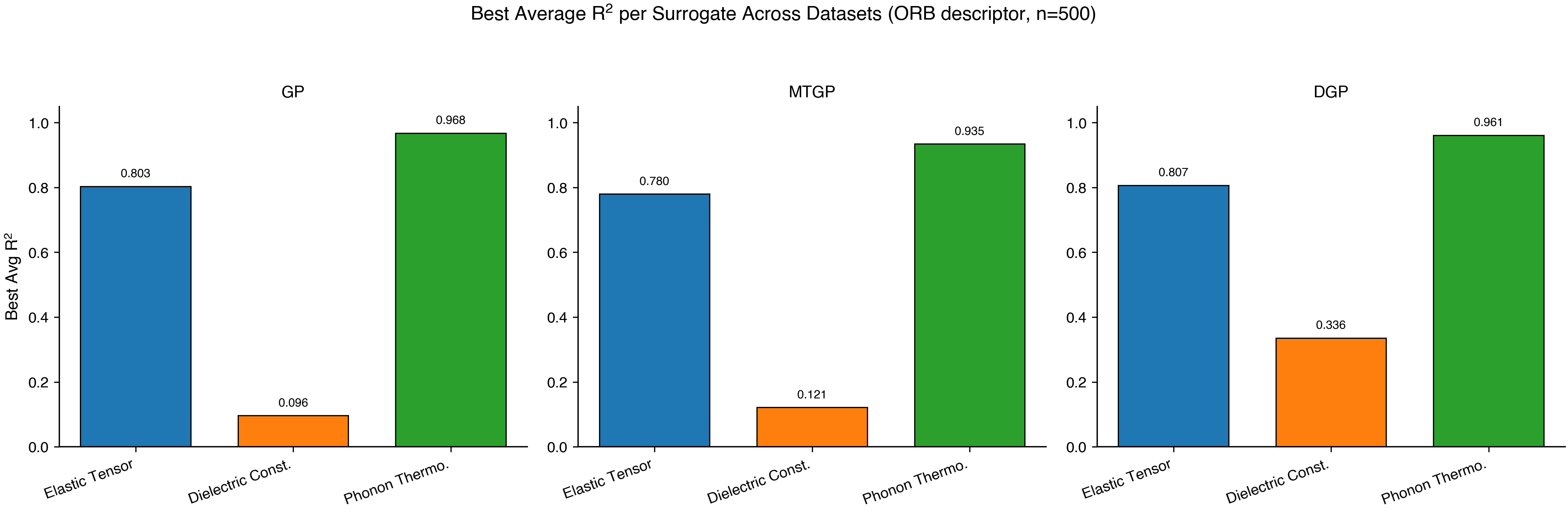}
    \caption{Best averaged \Rtwo{} per surrogate across the three datasets (ORB descriptor, $\ntrain = 500$, best PCA per surrogate). The DGP gives the highest \Rtwo{} on the elastic and dielectric datasets and the GP leads on phonon, which is why we use the DGP for non-linear targets and the GP as a default.}
    \label{si:fig:cross_surrogate}
\end{figure}

Figure~\ref{si:fig:parity_examples} shows parity plots for the best-predicted property in each dataset, and Figure~\ref{si:fig:cross_descriptor} compares the best achievable \Rtwo{} and $\rho$ per descriptor across datasets. Figure~\ref{si:fig:cross_surrogate} summarizes the best \Rtwo{} per surrogate, the basis for using the DGP on non-linear targets and the GP as a default.

\FloatBarrier
\section{Large-dataset benchmark: Pheasy phonon thermodynamics (11{,}818 materials)}
\label{si:pheasy}

\begin{figure}[H]
    \centering
    \includegraphics[width=0.72\textwidth]{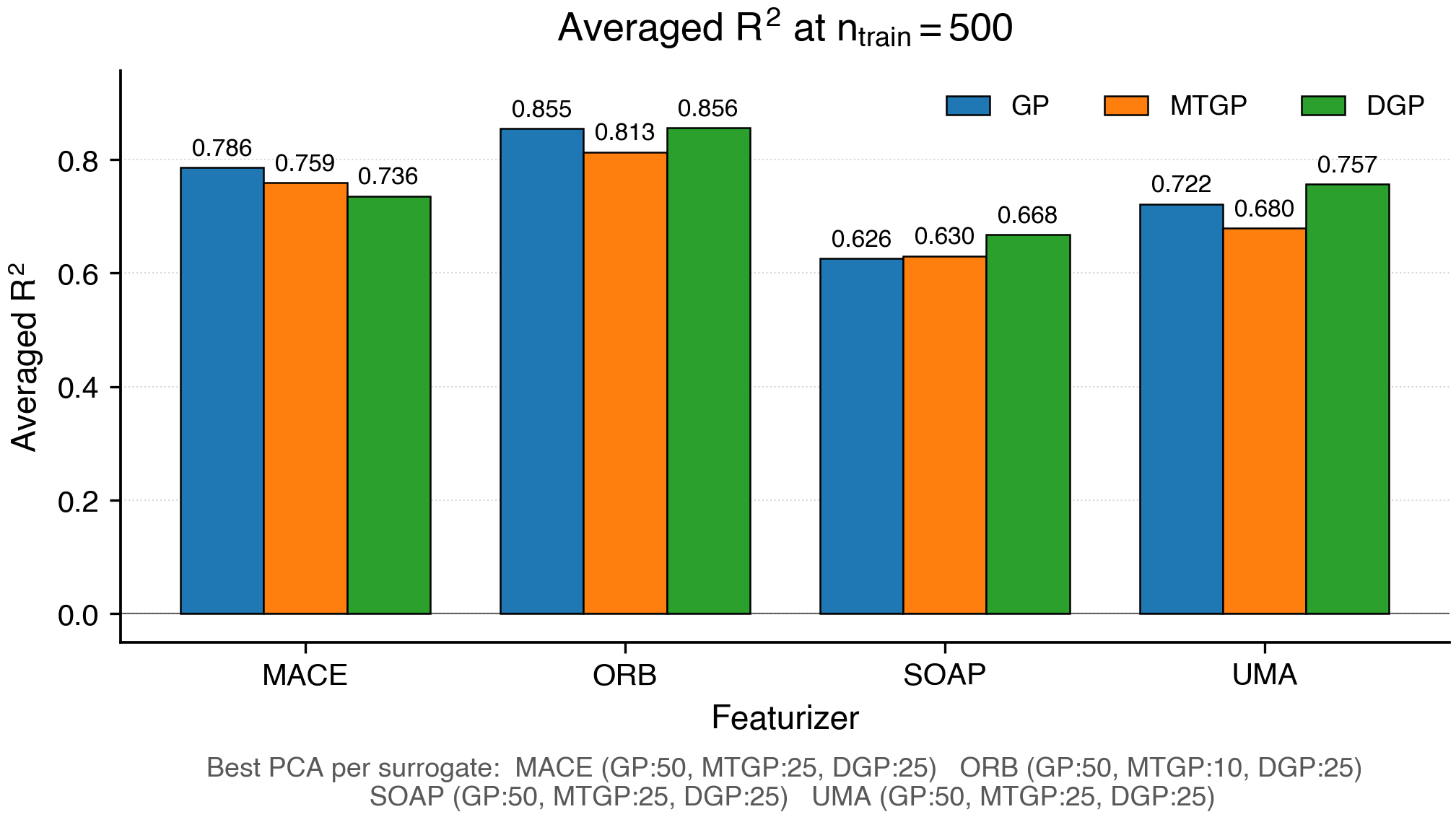}
    \caption{Averaged \Rtwo{} at $\ntrain = 500$ on the Pheasy phonon-thermodynamics dataset (11{,}818 materials), grouped by featurizer, for GP, MTGP, and DGP (best PCA per featurizer--surrogate). ORB is the strongest descriptor across all three surrogates and SOAP the weakest, matching the ordering on the smaller DFPT phonon set.}
    \label{si:fig:pheasy_bar}
\end{figure}

\begin{figure}[H]
    \centering
    \begin{subfigure}[t]{0.48\textwidth}
        \centering
        \includegraphics[width=\textwidth]{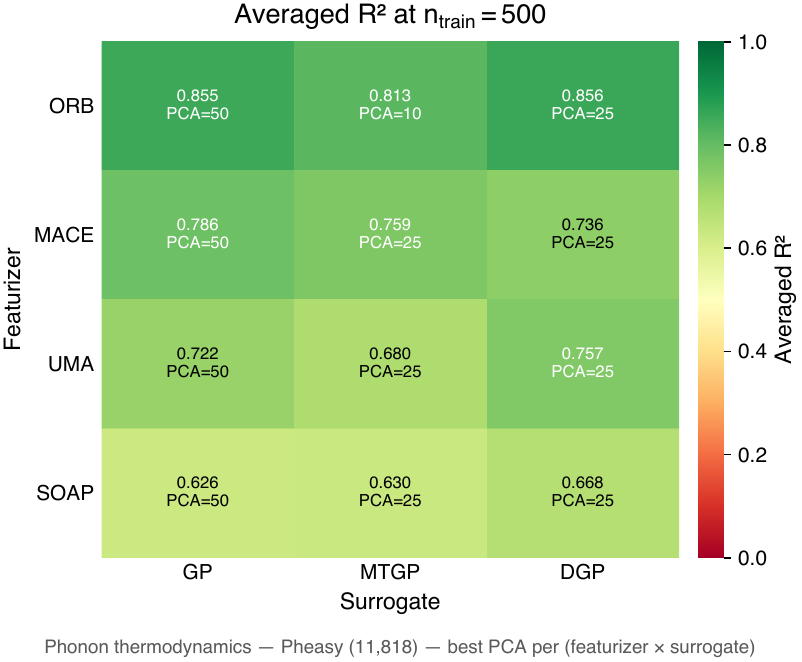}
        \caption{\Rtwo{} heatmap (featurizer $\times$ surrogate)}
        \label{si:fig:pheasy_heatmap}
    \end{subfigure}
    \hfill
    \begin{subfigure}[t]{0.48\textwidth}
        \centering
        \includegraphics[width=\textwidth]{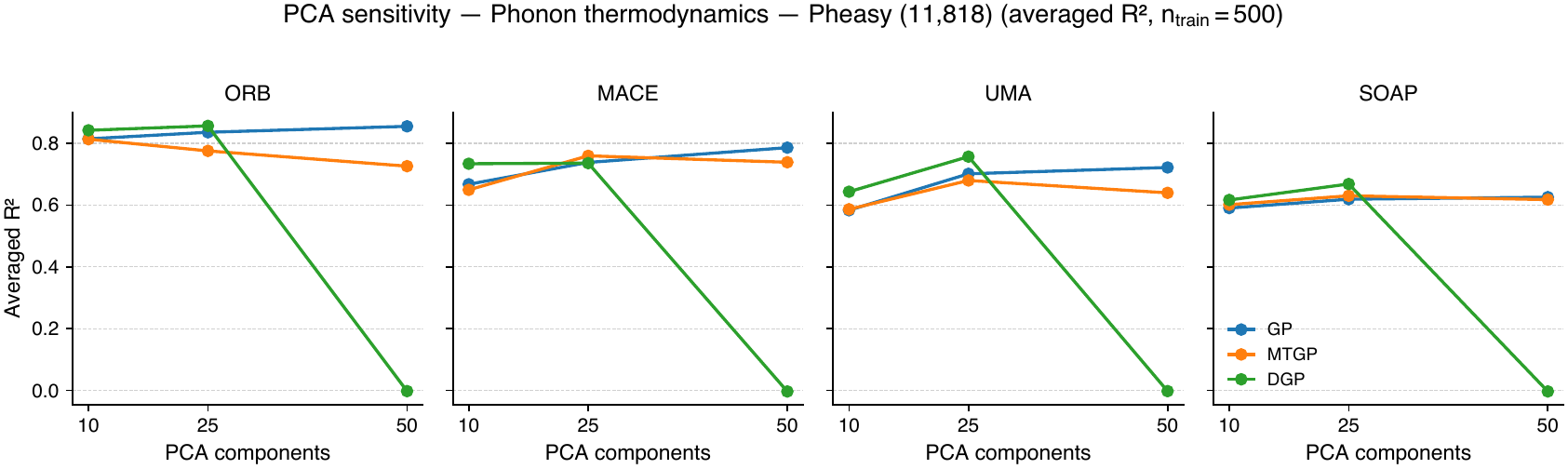}
        \caption{PCA sensitivity}
        \label{si:fig:pheasy_pca}
    \end{subfigure}
    \caption{Surrogate behavior on the Pheasy dataset at $\ntrain = 500$. (a) Averaged \Rtwo{} for each featurizer--surrogate combination at the best PCA. (b) Dependence on PCA dimensionality: the GP and MTGP improve or plateau toward PCA\,=\,50, while the DGP peaks at PCA\,$\leq$\,25 and collapses at PCA\,=\,50, the same behavior seen on the three smaller datasets (Fig.~\ref{si:fig:pca_all}).}
    \label{si:fig:pheasy_heatmap_pca}
\end{figure}

The three benchmark datasets in the main text each contain on the order of $10^{3}$ labeled materials. To check that the surrogate behavior reported there carries over to a substantially larger pool, we repeat the full static benchmark on the Pheasy phonon-thermodynamics dataset (11{,}818 materials), using the same four vibrational-thermodynamic targets as the DFPT phonon set ($C_V$, $S$, and $F$ at 300\,K, and the maximum phonon frequency) and the same protocol, with the training-set sweep extended to $\ntrain = 1000$ (DGP and MTGP) and $\ntrain = 2000$ (GP). The qualitative picture is unchanged. ORB is the strongest descriptor on every surrogate (Figs.~\ref{si:fig:pheasy_bar} and~\ref{si:fig:pheasy_heatmap_pca}), reaching averaged \Rtwo{}\,$\approx 0.855$ (GP) and $0.856$ (DGP) at $\ntrain = 500$, with SOAP the weakest. No single surrogate dominates: the DGP leads on ORB and UMA, the GP on MACE, and the MTGP on SOAP, and the descriptor-averaged best-PCA \Rtwo{} values are close (DGP $0.765$, GP $0.752$, MTGP $0.725$). The DGP collapse at PCA\,=\,50 seen on the three smaller datasets persists on this larger pool (Fig.~\ref{si:fig:pheasy_heatmap_pca}b), so the best-PCA summaries use the DGP at PCA\,=\,25; this is consistent with the over-parameterization reading in Section~\ref{si:surrogate} rather than a small-sample effect, since it recurs at $\ntrain \leq 1000$ regardless of the larger candidate pool. The learning curves (Fig.~\ref{si:fig:pheasy_lc}) are stable to slightly improving in both \Rtwo{} and Spearman~$\rho$, and the gain from $\ntrain = 500$ to $1000$--$2000$ is small, indicating that the embeddings reach most of their achievable accuracy on these targets within a few hundred training points.

\begin{figure}[H]
    \centering
    \begin{subfigure}[t]{0.48\textwidth}
        \centering
        \includegraphics[width=\textwidth]{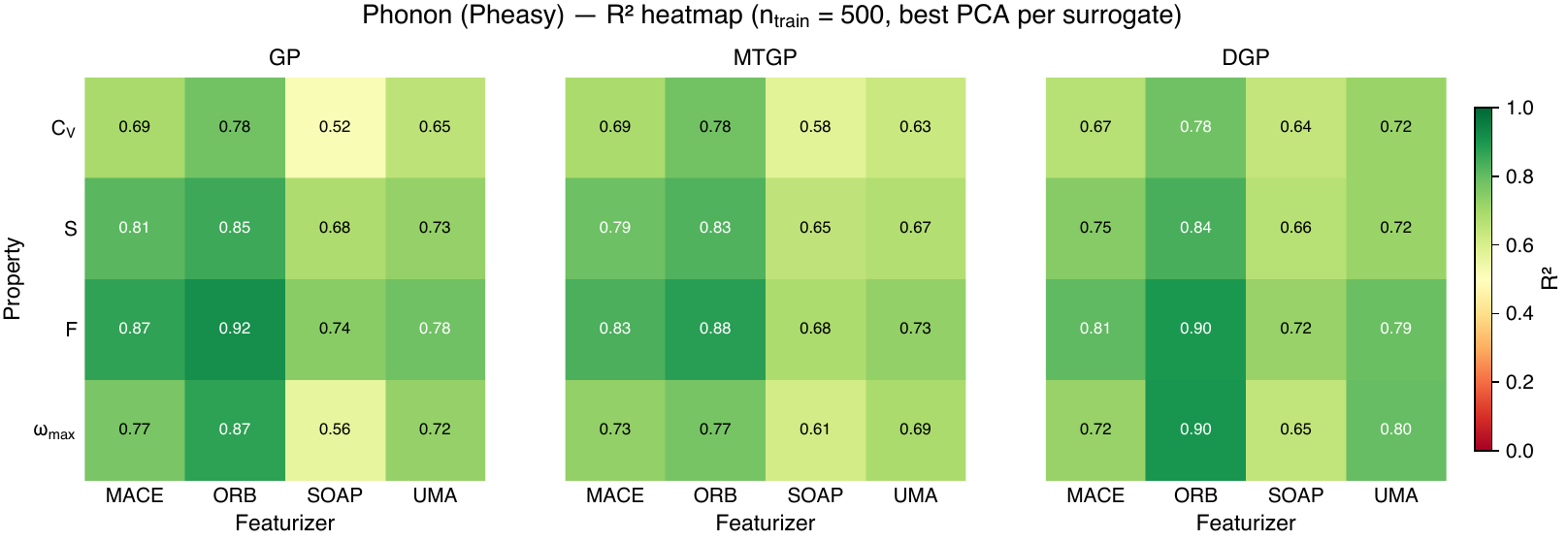}
        \caption{Per-property \Rtwo{} (per surrogate)}
        \label{si:fig:pheasy_difficulty}
    \end{subfigure}
    \hfill
    \begin{subfigure}[t]{0.48\textwidth}
        \centering
        \includegraphics[width=\textwidth]{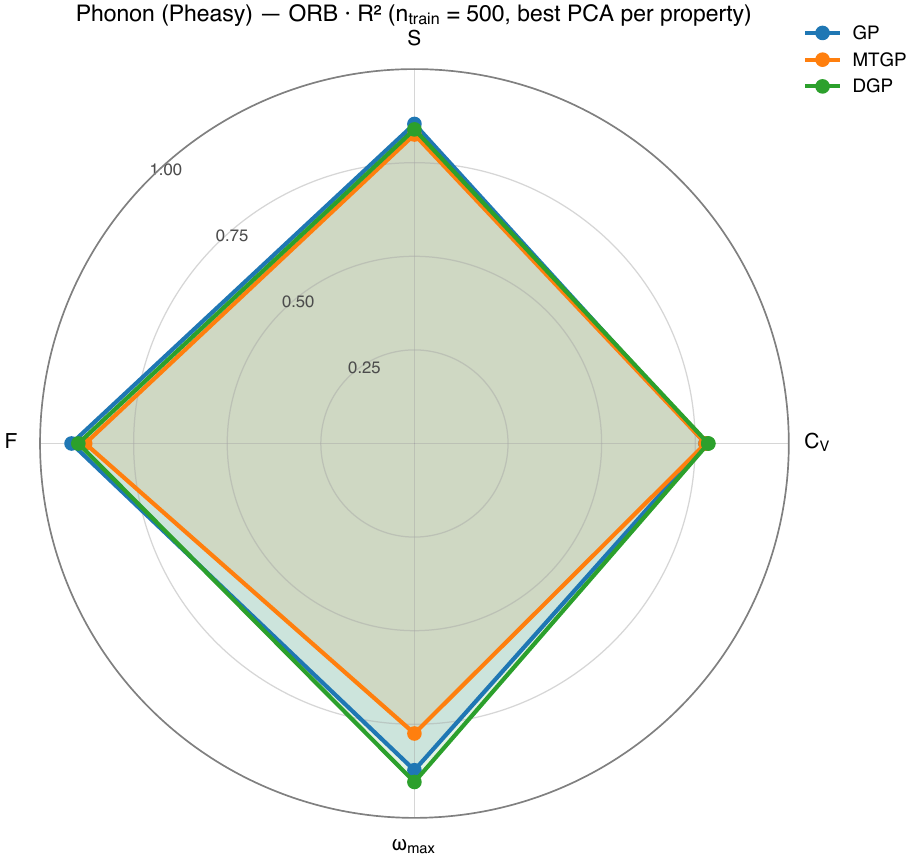}
        \caption{ORB per-property \Rtwo{} profile}
        \label{si:fig:pheasy_radar}
    \end{subfigure}
    \caption{Per-property difficulty on the Pheasy dataset at $\ntrain = 500$. (a) Best-PCA \Rtwo{} for each (property, featurizer) under GP, MTGP, and DGP; the free energy $F$ is the easiest target and $C_V$ the hardest, with ORB strongest throughout. (b) ORB's per-property \Rtwo{} profile across the three surrogates.}
    \label{si:fig:pheasy_difficulty_radar}
\end{figure}

\begin{figure}[H]
    \centering
    \begin{subfigure}[t]{0.48\textwidth}
        \centering
        \includegraphics[width=\textwidth]{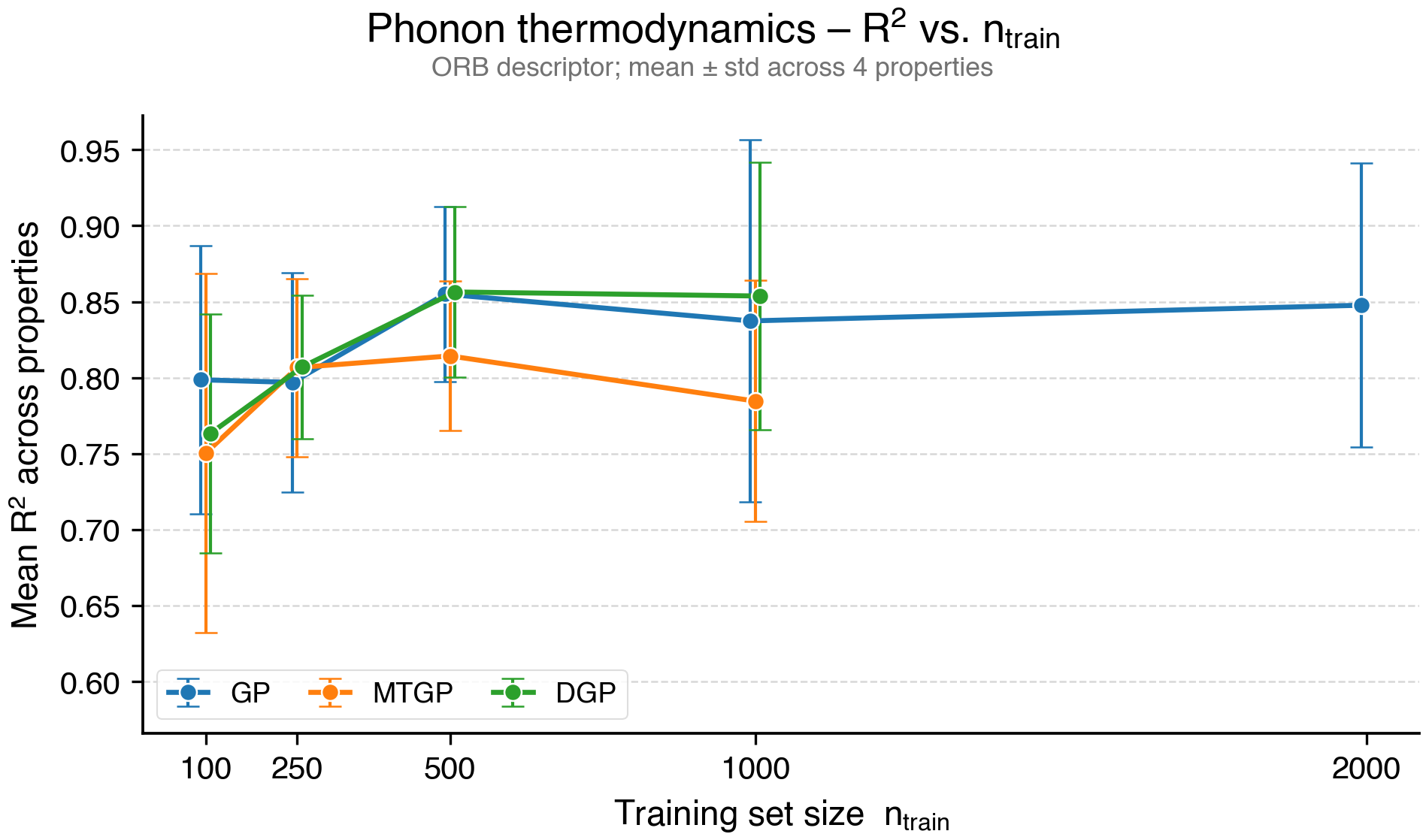}
        \caption{Averaged \Rtwo{}}
        \label{si:fig:pheasy_lc_r2}
    \end{subfigure}
    \hfill
    \begin{subfigure}[t]{0.48\textwidth}
        \centering
        \includegraphics[width=\textwidth]{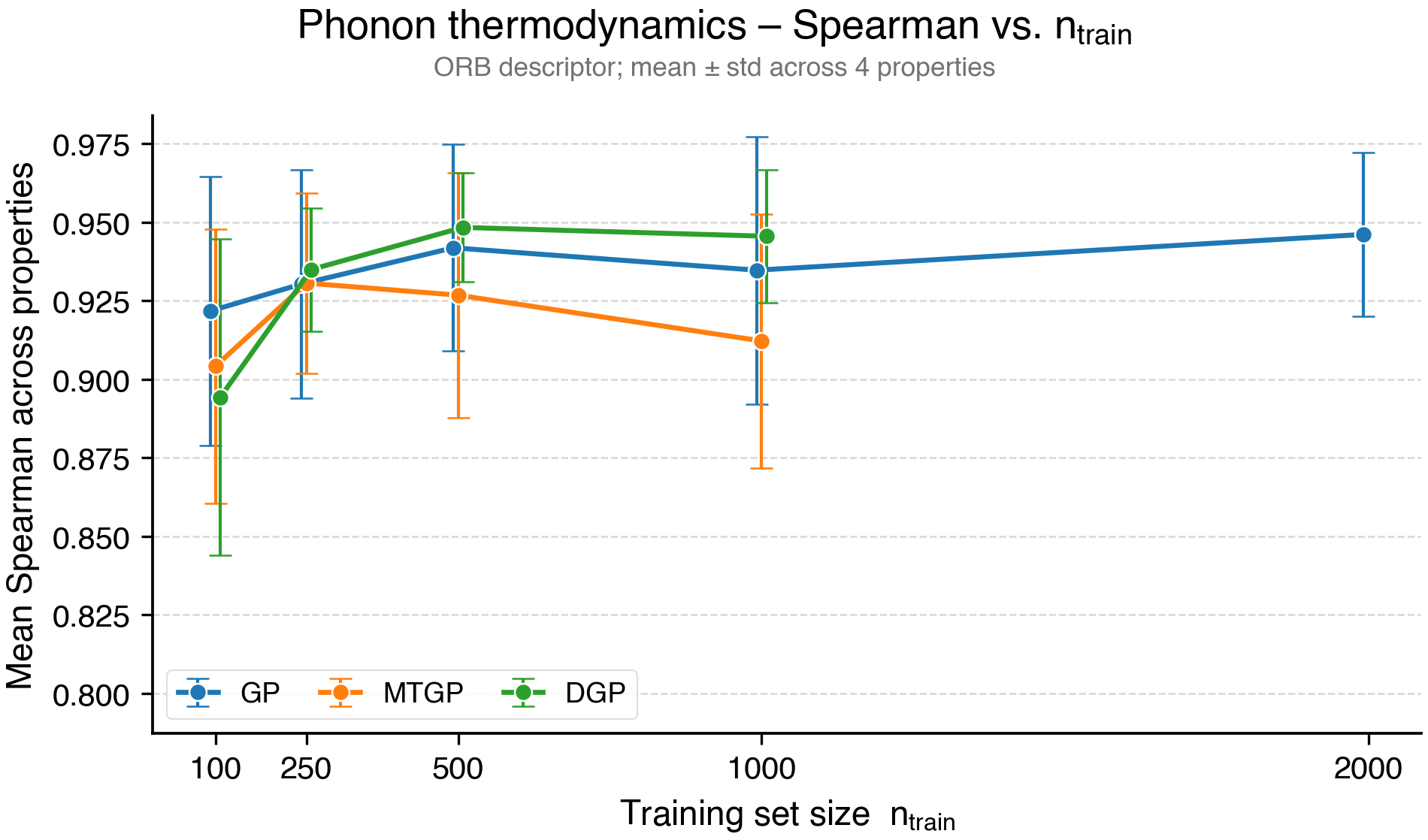}
        \caption{Averaged Spearman $\rho$}
        \label{si:fig:pheasy_lc_spearman}
    \end{subfigure}
    \caption{Learning curves on the Pheasy dataset (ORB descriptor, best PCA per surrogate), with $\ntrain$ extended to 2000 (GP) and 1000 (MTGP and DGP). Both \Rtwo{} and Spearman~$\rho$ are stable to slightly improving with training size, and the gain beyond $\ntrain = 500$ is small.}
    \label{si:fig:pheasy_lc}
\end{figure}

\FloatBarrier
\section{Synthesizability scoring details}
\label{si:synth}

We tuned the ORB-PU synthesizability classifier (main-text Methods) with Optuna across eight feature-set $\times$ base-learner configurations and selected ORB+Magpie with gradient-boosted trees by cross-validated area under the precision--recall curve (AUPRC). Table~\ref{si:tab:apu_leaderboard} gives the leaderboard on the held-out Materials-Project test split. Table~\ref{si:tab:apu_cgnf} reports the selected model and pretrained CGNF on the same split; these numbers favor ORB-PU because it is in-distribution whereas CGNF is applied zero-shot, so they are not a statement of method superiority. The comparison on equal footing is on the generated structures, reported in the main text.

\begin{table}[H]
    \centering
    \caption{Optuna-tuned ORB-PU configurations on the held-out Materials-Project test split, ordered by test AUPRC. CV~AUPRC is the cross-validated selection metric; TPR is recall on held-out positives at the 0.5 threshold; ECE is the expected calibration error; coverage is the fraction of test structures not abstained by the out-of-distribution criterion. The selected model is ORB+Magpie with gradient-boosted trees (XGB).}
    \label{si:tab:apu_leaderboard}
    \footnotesize
    \begin{tabular}{@{}llcccccc@{}}
        \toprule
        \textbf{Features} & \textbf{Base} & \textbf{CV AUPRC} & \textbf{AUPRC} & \textbf{AUROC} & \textbf{TPR} & \textbf{ECE} & \textbf{Cov.} \\
        \midrule
        ORB+Magpie       & XGB & 0.956 & 0.961 & 0.967 & 0.892 & 0.024 & 0.72 \\
        ORB+Magpie+stab  & XGB & 0.955 & 0.960 & 0.966 & 0.891 & 0.010 & 0.71 \\
        Magpie           & XGB & 0.948 & 0.949 & 0.956 & 0.872 & 0.007 & 0.72 \\
        ORB+Magpie       & RF  & 0.941 & 0.945 & 0.956 & 0.892 & 0.054 & 0.65 \\
        ORB+Magpie+stab  & RF  & 0.941 & 0.944 & 0.956 & 0.893 & 0.053 & 0.66 \\
        Magpie           & RF  & 0.935 & 0.937 & 0.948 & 0.875 & 0.033 & 0.68 \\
        ORB              & XGB & 0.919 & 0.929 & 0.948 & 0.873 & 0.034 & 0.71 \\
        ORB              & RF  & 0.904 & 0.912 & 0.936 & 0.861 & 0.082 & 0.60 \\
        \bottomrule
    \end{tabular}
\end{table}

\begin{table}[H]
    \centering
    \caption{Selected ORB-PU model versus pretrained CGNF on the same held-out Materials-Project test split (9{,}857 held-out positives, 12{,}000 unlabeled). TPR is recall on held-out positives at the 0.5 threshold; pessimistic precision counts unlabeled predicted-positives as false positives (a lower bound, since the unlabeled pool hides true positives); $\beta$ is each model's estimated positive fraction in the unlabeled pool. These metrics favor the in-distribution ORB-PU model and are not a method-superiority claim; the comparison on equal footing is on the generated structures (main text).}
    \label{si:tab:apu_cgnf}
    \footnotesize
    \begin{tabular}{@{}lcccccc@{}}
        \toprule
        \textbf{Model} & \textbf{TPR} & \textbf{Pess.\ prec.} & \textbf{AUPRC} & \textbf{AUROC} & \textbf{ECE} & $\beta$ \\
        \midrule
        CGNF (pretrained, zero-shot) & 0.973 & 0.574 & 0.805 & 0.853 & 0.241 & 0.623 \\
        ORB-PU (ORB+Magpie, XGB)     & 0.892 & 0.887 & 0.961 & 0.967 & 0.024 & 0.137 \\
        \bottomrule
    \end{tabular}
\end{table}

\FloatBarrier
\section{DFT validation: per-structure bulk moduli}
\label{si:dft_bm}

Table~\ref{si:tab:esen_vs_dft} lists the per-structure oracle-parity DFT bulk moduli of the 15 validated $K_{\text{VRH}}$ discoveries, alongside the eSEN oracle value and the Birch--Murnaghan pressure derivative $B_0'$ of each fit. The main-text summary statistics for oracle fidelity (MAE 8.5~GPa, MAPE 2.5\%, Spearman $\rho = 0.87$, bias $-2.3$~GPa) are computed from these rows. Repeated formulae (CoIrOs$_2$, FeB$_2$MoW) are distinct polymorphs recovered in different runs. Every fit is in-window with a physical pressure derivative $B_0' \in [4.1, 5.1]$.

\begin{table}[H]
    \centering
    \caption{Oracle-parity DFT versus eSEN-oracle bulk modulus $K_0$ (GPa) for the 15 DFT-validated closed-loop $K_{\text{VRH}}$ discoveries. Rank is the closed-loop leaderboard position; policy is BASE (ungated) or ACC (gated); $\Delta = $ eSEN $-$ DFT; $B_0'$ is the fitted Birch--Murnaghan pressure derivative.}
    \label{si:tab:esen_vs_dft}
    \small
    \begin{tabular}{@{}lccccc r@{}}
        \toprule
        \textbf{Structure} & \textbf{Rank} & \textbf{Policy} & \textbf{DFT $K_0$} & \textbf{eSEN $K_0$} & \textbf{$\Delta$} & \textbf{$B_0'$} \\
        \midrule
        MoN            & 1  & ACC  & 353.6 & 375.2 & $+21.6$ & 4.52 \\
        MoC            & 2  & BASE & 348.1 & 351.6 & $+3.5$  & 4.36 \\
        Os$_5$W$_3$    & 3  & ACC  & 355.9 & 339.6 & $-16.3$ & 4.66 \\
        CoIrOs$_2$     & 5  & ACC  & 346.0 & 338.0 & $-8.0$  & 4.93 \\
        CoIrOs$_2$     & 6  & ACC  & 346.0 & 338.0 & $-8.0$  & 4.93 \\
        CoIrOs$_2$     & 7  & ACC  & 348.1 & 337.6 & $-10.5$ & 4.85 \\
        Re$_5$W        & 8  & ACC  & 351.7 & 336.5 & $-15.2$ & 4.46 \\
        VIr$_7$        & 9  & BASE & 334.6 & 325.0 & $-9.6$  & 5.06 \\
        TiVN$_2$       & 10 & ACC  & 296.3 & 313.1 & $+16.8$ & 4.34 \\
        TaB$_2$W       & 11 & ACC  & 315.2 & 313.0 & $-2.2$  & 4.13 \\
        FeB$_2$MoW     & 13 & BASE & 306.9 & 306.2 & $-0.7$  & 4.32 \\
        FeB$_2$MoW     & 14 & BASE & 306.9 & 306.1 & $-0.8$  & 4.32 \\
        Mn(BMo)$_3$    & 16 & BASE & 300.5 & 298.4 & $-2.1$  & 4.46 \\
        FeRe(PW)$_2$   & 17 & ACC  & 293.4 & 297.8 & $+4.4$  & 4.59 \\
        ReIr$_2$Rh$_5$ & 19 & ACC  & 299.5 & 292.2 & $-7.3$  & 5.10 \\
        \bottomrule
    \end{tabular}
\end{table}

\FloatBarrier
\section{DFT energy above the convex hull of the generated structures}
\label{si:ehull}

As a thermodynamic complement to the bulk-modulus validation, we estimate the 0~K energy above the convex hull, $E_\mathrm{hull}$, for the top generated structures. We use the standard Materials-Project-compatible procedure implemented in pymatgen rather than a bespoke scheme, so that the values are directly comparable to the Materials Project convex hull; we do not introduce a new stability method. The DFT software and settings are those cited in the main-text Methods.

\subsection{Protocol}

\textbf{Energy.} The oracle-parity equation-of-state relaxation (main-text Methods; ENCUT~$=680$~eV, no Hubbard~$U$) is not on the Materials Project energy scale, so for each structure we run a separate single-point static calculation with the Materials Project input set (\texttt{MPStaticSet} in pymatgen: ENCUT~$=520$~eV, tetrahedron smearing, projector augmented-wave potentials, the PBE functional, and the Hubbard-$U$ values the Materials Project applies) on the DFT-relaxed cell, evaluated with VASP. Magnetic moments are seeded from the relaxed cell.

\textbf{Reference hull and corrections.} For every chemical sub-system of a structure we retrieve the Materials Project entries on the GGA/GGA$+$U thermodynamic scale, assemble them into a phase diagram with pymatgen, and read $E_\mathrm{hull}$ from the resulting hull. The Materials Project 2020 compatibility corrections (anion and GGA/GGA$+$U mixing) are applied to the generated entry through pymatgen's \texttt{MaterialsProject2020Compatibility}, so that the generated entry and the reference entries carry the same corrections. For the five anion-bearing structures in our set the relevant term is the standard anion correction ($-0.687$~eV per O atom and $-0.361$~eV per N atom); none of their cations triggers the Hubbard-$U$ scheme, which the Materials Project applies only to oxygen- or fluorine-coordinated \{Co, Cr, Fe, Mn, Mo, Ni, V, W\}, so the generated static calculations are GGA and sit on the same rung as the reference entries. We obtain values for 28 of the 29 structures; the self-consistent field of FeRe$_9$Os$_2$ did not converge and it is omitted.

\textbf{Consistency of the correction.} An earlier version of this calculation placed the uncorrected static energy on the corrected reference hull. Because the anion correction is one-signed and was absent from our entry, this inflated only the anion-bearing structures while leaving the 23 anion-free structures unchanged (their correction is exactly zero). Applying the correction to both sides lowers the five anion structures (Figure~\ref{si:fig:ehull}a) and gives the values we report; the uncorrected numbers are upper bounds for those five and are listed alongside in Table~\ref{si:tab:ehull}. The ``upper bound'' statement is exact for the anion phases, since the anion correction can only lower the energy, and is not claimed in general.

\subsection{Results}

\begin{figure}[H]
    \centering
    \includegraphics[width=\textwidth]{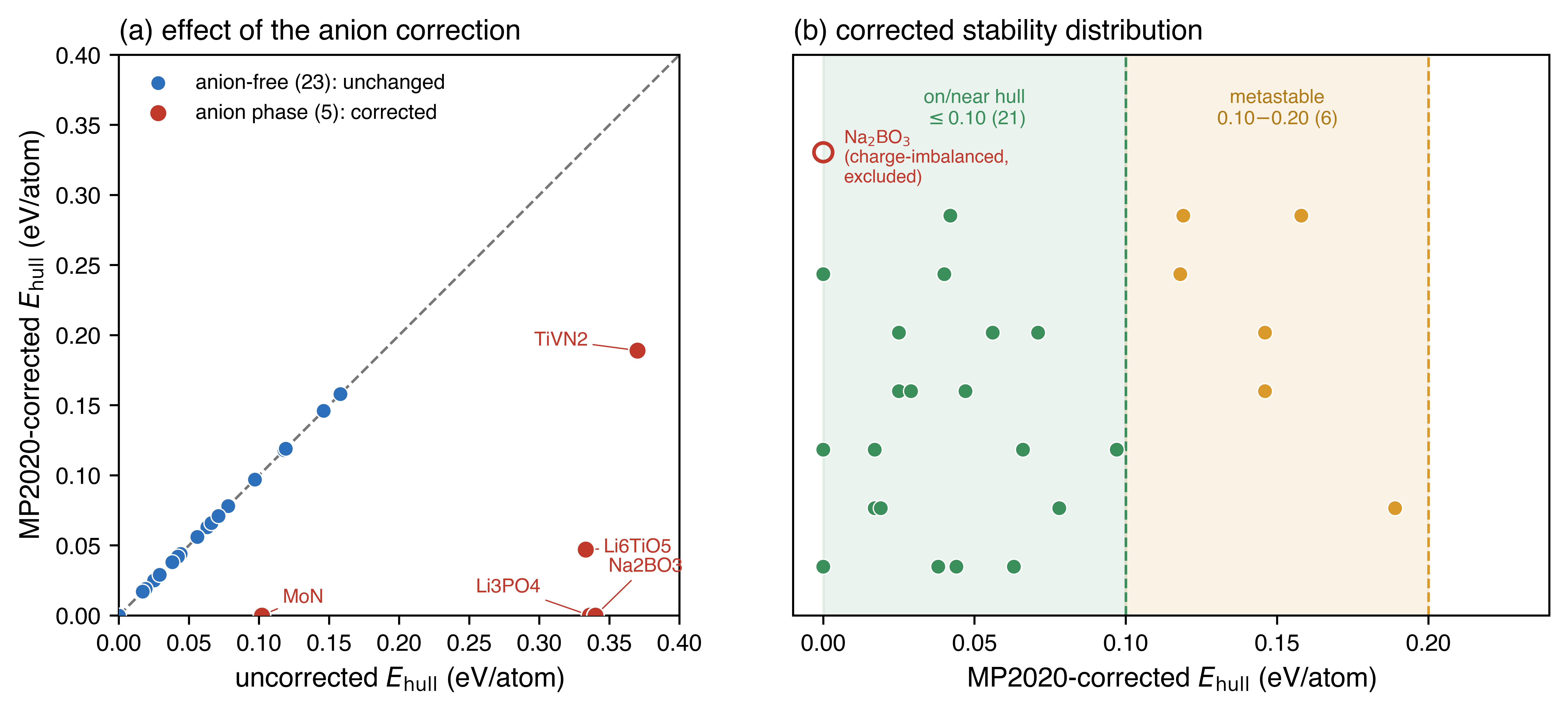}
    \caption{DFT energy above the Materials Project convex hull for the generated structures (Materials Project 2020 corrected). (a)~Uncorrected against corrected $E_\mathrm{hull}$: the anion correction is one-signed and moves only the five anion-bearing structures, while the 23 anion-free structures lie on the diagonal (correction exactly zero). (b)~Distribution of the corrected $E_\mathrm{hull}$ against the on/near-hull ($\leq 0.10$~eV/atom) and metastable-ceiling ($0.20$~eV/atom) references; the charge-imbalanced Na$_2$BO$_3$ is flagged and excluded from the tally.}
    \label{si:fig:ehull}
\end{figure}

After correction, of the 27 charge-balanced structures (the charge-imbalanced Na$_2$BO$_3$ is excluded; see Caveats), 21 lie on or near the hull of known phases ($E_\mathrm{hull} \leq 0.10$~eV/atom) and 6 are metastable ($0.10$--$0.20$~eV/atom); none exceeds $0.20$~eV/atom (Figure~\ref{si:fig:ehull}b). The largest value is TiVN$_2$ at $0.19$~eV/atom, reduced from $0.37$ by the nitrogen correction. Given a settings-and-correction uncertainty of about $\pm 0.05$~eV/atom, these are best read as a distribution lying within $0.20$~eV/atom of the known hull rather than as sharply binned counts; tier assignments within roughly $0.05$~eV/atom of the $0.10$ and $0.20$ boundaries, such as Os$_5$W$_3$ at $0.097$, are not firmly resolved.

\subsection{Caveats}

Three limits accompany these numbers. First, for the novel refractory chemistries the Materials Project hull is built from few entries (the $n_\mathrm{MP}$ column of Table~\ref{si:tab:ehull}, as low as 15 for Os$_5$W$_3$), so $E_\mathrm{hull}$ there is a lower bound and a value near zero means ``lowest among known phases at that composition'' rather than a proven ground state; for those structures we rely on the CALPHAD and oracle cross-checks rather than on $E_\mathrm{hull}$ alone. Second, the energy is a Materials-Project-INCAR single point on the oracle-parity-relaxed cell, not a relaxation under the Materials Project settings; because the cell is not re-relaxed this can only raise $E_\mathrm{hull}$, so the bias is conservative. Third, the composition of Na$_2$BO$_3$ is not charge-balanced, so a composition-level correction applied to it is not meaningful and its corrected value near zero is an artifact; we report it as a generation caught by the charge-balance pre-screen, not as a stable structure.

\begin{table}[H]
    \centering
    \caption{Uncorrected and Materials Project 2020 corrected energy above the hull, $E_\mathrm{hull}$ (eV/atom), for the generated structures, ordered by corrected value. $n_\mathrm{MP}$ is the number of Materials Project reference entries forming the hull of that chemical system; small values indicate a sparsely sampled hull, for which $E_\mathrm{hull}$ is a lower bound. The correction is nonzero only for the five anion-bearing structures and equals the standard anion term; the 23 anion-free structures are unchanged. FeRe$_9$Os$_2$ is omitted (SCF unconverged). $^{\dagger}$Na$_2$BO$_3$ is a charge-imbalanced generation; its value is not a stability estimate.}
    \label{si:tab:ehull}
    \small
    \begin{tabular}{@{}llccc@{}}
        \toprule
        \textbf{Structure} & \textbf{anion} & \textbf{$E_\mathrm{hull}$ uncorr.} & \textbf{$E_\mathrm{hull}$ corr.} & \textbf{$n_\mathrm{MP}$} \\
        \midrule
        MoN                  & N & 0.102 & 0.000 & 43  \\
        TaB$_2$W             &   & 0.000 & 0.000 & 55  \\
        Mn(BMo)$_3$          &   & 0.000 & 0.000 & 57  \\
        Li$_3$PO$_4$         & O & 0.336 & 0.000 & 108 \\
        Na$_2$BO$_3^{\dagger}$ & O & 0.340 & 0.000 & 125 \\
        MoC                  &   & 0.017 & 0.017 & 81  \\
        Li$_4$Mg             &   & 0.017 & 0.017 & 92  \\
        Li$_{13}$Mg$_3$      &   & 0.019 & 0.019 & 92  \\
        FeB$_2$MoW           &   & 0.025 & 0.025 & 100 \\
        FeB$_2$MoW           &   & 0.025 & 0.025 & 100 \\
        Li$_4$Mg             &   & 0.029 & 0.029 & 92  \\
        Li$_7$Mg$_3$         &   & 0.038 & 0.038 & 92  \\
        Li$_{13}$Mg$_3$Al$_2$ &  & 0.040 & 0.040 & 155 \\
        NaLi$_2$             &   & 0.042 & 0.042 & 25  \\
        VIr$_7$              &   & 0.044 & 0.044 & 19  \\
        Li$_6$TiO$_5$        & O & 0.333 & 0.047 & 263 \\
        NaLi$_5$             &   & 0.056 & 0.056 & 25  \\
        ReIr$_2$Rh$_5$       &   & 0.063 & 0.063 & 27  \\
        LiMg                 &   & 0.066 & 0.066 & 92  \\
        LiMg$_2$             &   & 0.071 & 0.071 & 92  \\
        Li$_5$CaMg$_2$       &   & 0.078 & 0.078 & 190 \\
        Os$_5$W$_3$          &   & 0.097 & 0.097 & 15  \\
        CoIrOs$_2$           &   & 0.118 & 0.118 & 27  \\
        Re$_5$W              &   & 0.119 & 0.119 & 20  \\
        CoIrOs$_2$           &   & 0.146 & 0.146 & 27  \\
        CoIrOs$_2$           &   & 0.146 & 0.146 & 27  \\
        FeRe(PW)$_2$         &   & 0.158 & 0.158 & 77  \\
        TiVN$_2$             & N & 0.370 & 0.189 & 62  \\
        \bottomrule
    \end{tabular}
\end{table}